\documentclass[conference]{IEEEtran}
\usepackage{xspace}
\usepackage{pifont}

\usepackage{booktabs}      
\usepackage[table]{xcolor} 
\usepackage{graphicx}      
\usepackage{subcaption} 
\usepackage{algorithm}
\usepackage{algpseudocode}
\usepackage{amsmath}     
\usepackage{amsfonts}
\usepackage{mathtools}   
\usepackage{bm}          
\usepackage{multirow}
\usepackage[normalem]{ulem}
\usepackage{url}
\usepackage{xurl} 
\usepackage{hyperref}

\usepackage{tikz}
\usepackage{xcolor}
\usepackage{tcolorbox}
\usetikzlibrary{positioning, calc, arrows.meta}

\newcommand{\tok}[1]{%
    \tikz[baseline=(t.base)]{%
        \node[draw, rounded corners=1.5pt, inner sep=2pt, 
              font=\ttfamily\scriptsize, thick] (t) {#1};%
    }%
}

\newcommand{\Toketive}{\texttt{Toketive}\xspace}
\newcommand{\crossmark}{\ding{51}}
\newcommand{\xmark}{\ding{55}}
\newcommand{\kstar}{k^{\ast}}

\usepackage{colortbl}
\usepackage[T1]{fontenc}

\definecolor{darkgreen}{rgb}{0.0, 0.45, 0.0}
\definecolor{darkorange}{rgb}{0.95,0.45,0}

\newcommand{\modelstyle}[2]{{\textbf{\textcolor{#1}{#2}}}}

\newcommand{\LlamaThree}{\modelstyle{darkgreen}{Llama3}\xspace}
\newcommand{\LlamaThreeOne}{\modelstyle{blue}{Llama3.1}\xspace}
\newcommand{\TuluThree}{\modelstyle{darkorange}{Tulu3}\xspace}
\newcommand{\TuluThreeOne}{\modelstyle{violet}{Tulu3.1}\xspace}
\newcommand{\OlmoTwo}{\modelstyle{red}{OLMo2}\xspace}

\algnewcommand\Input{\item[\textbf{Input:}]}
\algnewcommand\Output{\item[\textbf{Output:}]}

\ifCLASSINFOpdf
\else
\fi
\begin{document}
%
\title{The Tokens Remember: When Tokenization Bypasses Knowledge Editing and Unlearning}

\author{
\IEEEauthorblockN{
Manit Baser$^{\dagger}$,
Aditya Nawal$^{\dagger}$,
Dinil Mon Divakaran$^{\ddagger}$,
Mohan Gurusamy$^{\dagger}$
}
\IEEEauthorblockA{
$^{\dagger}$National University of Singapore
\quad
$^{\ddagger}$A*STAR Institute of Advanced Intelligence and Computing, Singapore
}
}
	

%


\maketitle

\begin{abstract}

Open-weight LLMs give downstream users control over the inference stack, but this flexibility can undermine post-release guarantees that sensitive knowledge has been modified or removed. Model editing and machine unlearning are used to modify or remove targeted knowledge without retraining models from scratch. However, existing security evaluations of these techniques face two critical limitations. First, they typically require access to either the original pre-edit/unlearning model or auxiliary classifiers to detect modifications or reconstruct pre-edit behavior. Second, they evaluate modifications under the canonical tokenization of an input, implicitly treating tokenization as a benign preprocessing step. We show that this assumption creates a security gap: the same input string can be represented by alternative valid tokenizations that induce different computational trajectories, allowing an adversary to bypass localized modifications and recover information intended to be suppressed.

We introduce \Toketive, a simple yet powerful reference-free attack that exploits the tokenization-based side channel to (i)~detect modified knowledge and (ii)~reconstruct the corresponding pre-edit response. \Toketive operates solely on the released model and requires neither the pre-edit model, training data, shadow models, nor auxiliary classifiers. Across five LLMs, six datasets, and six state-of-the-art editing and unlearning techniques, we find that 38.6\% of alternative tokenizations bypass the modification and recover the pre-edit response. \Toketive detects modified facts with an F1 score of 84.2\%, a 26.2\% relative gain over the strongest baseline, and reconstructs pre-edit responses with 74.5\% top-5 accuracy, 21.7\% higher than the best baseline. Our results reveal a fundamental gap between canonical-tokenization evaluation and the post-release security guarantees expected from model editing and unlearning, showing that localized modifications should not be treated as robust knowledge-control boundaries without adversarial evaluation over alternative representations.

\end{abstract}


%
\IEEEpeerreviewmaketitle

\section{Introduction}
\label{sec:intro}

The rapid progress of large language models (LLMs) has been accompanied by the widespread release of high-performance open-weight models, facilitating democratized access to state-of-the-art capabilities. Models such as Llama~\cite{grattafiori2024llama}, Qwen~\cite{yang2025qwen3}, Mistral~\cite{liu2026ministral}, Gemma~\cite{team2026gemma}, DeepSeek~\cite{xu2026deepseek}, GLM~\cite{zeng2026glm}, and OLMo~\cite{olmo2025olmo} now achieve performance comparable to many proprietary models while remaining publicly accessible. These have accelerated innovation by enabling model customization, domain adaptation, and scientific reproducibility~\cite{calvi2026shieldstral, vijayantares, kassianik2025llama, elzemity2026small}. As of August 2026, Hugging Face hosts over 395,000 open-weight text-generation checkpoints~\cite{huggingface_text_generation_2026}. The growing importance of open-weight models is also reflected by concerns over sovereignty, as recent restrictions (temporarily) limited access to some frontier models~\cite{anthropic2026fablemythosaccess, kwok2026opensource, wang2025shiftingcompute}.

As open-weight models become increasingly prevalent, their accessibility fundamentally changes the security assumptions under which they operate~\cite{makiej2026opensource, pang2025paladin, xu2025mark}. Unlike hosted LLM APIs, once an open-weight model is released, its developer no longer controls the inference stack, allowing downstream users, including malicious users, to inspect, modify, and execute the model under arbitrary local inference settings. Consequently, developers must ensure that released models remain safe, accurate, and compliant as regulations evolve~\cite{futureoflife2024aiacttimeline, jonesday2025gpai}, new vulnerabilities are discovered~\cite{russinovich2026oneprompt, lyu2024keeping}, factual knowledge changes~\cite{chen-etal-2024-lifelong}, and harmful memorized contents are identified~\cite{kassem2023preserving}. Model editing and machine unlearning provide efficient mechanisms for performing such targeted modifications without resorting to the prohibitively expensive approach of retraining models from scratch~\cite{meng2022mass, 10.1145/3749987, hu2024duty, batorski2026grom}. These modifications may involve removing personally identifiable information~\cite{cheng2025effective}, correcting outdated or erroneous knowledge~\cite{loth2026industrialized}, or mitigating unsafe behaviors~\cite{llmalignment}.

Editing and unlearning aim to ensure that modified knowledge cannot be reliably recovered during subsequent inference, including under adversarial attempts to recover the edited or unlearned information. These techniques seek precise, localized modifications and often follow a ``locate-then-edit'' (or ``unlearn'') paradigm: first identifying the internal parameters or representations associated with a target fact, and then applying a localized update~\cite{meng2022locating, jung-etal-2025-come, liang2024locate}. Alternative approaches to knowledge modification are discussed in Section~\ref{sec:related}.
\begin{figure}
\centering

    \includegraphics[width=\linewidth]{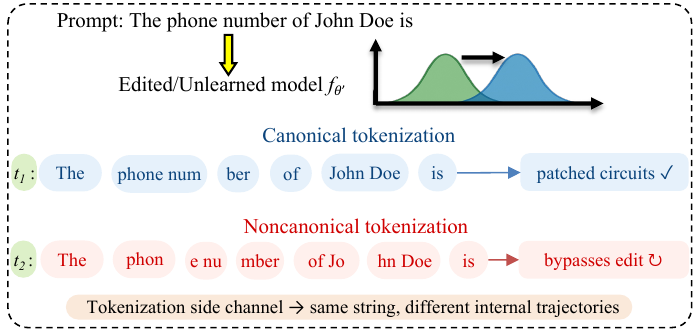}
    \caption{Different tokenizations of an input string ($t_1$ and $t_2$) induce different internal computational trajectories. While the canonical tokenization activates patched trajectories and suppresses the target information, a noncanonical tokenization may bypass the edit and recover the original response.}
    \label{fig:canonical}
\end{figure}

Beyond the primary objective editing and unlearning, however, unintended effects on overall model behavior have emerged as an important concern. In particular, their impact on model alignment has received significant attention~\cite{10.1145/3698590, spohn2025alignthenunlearn, jiang-etal-2024-learning}. Existing works that study the security implications of editing and unlearning have been mostly limited to exposing vulnerabilities such as membership inference~\cite{naderloui2025rectifying}, training-data extraction~\cite{cheng2025effective}, reverse engineering of updates~\cite{sun2026reverse}, auditing~\cite{du2025textual}, and multi-step knowledge recovery~\cite{baserthinkeval}. Furthermore, these approaches share two key limitations: (i)~they require either training auxiliary classifiers and shadow models or access to the original pre-edit model to reconstruct pre-edit behavior; and (ii)~they assume that the attacker queries the model with the same canonical tokenization used during editing or unlearning~\cite{zheng2025broken}. This assumption is implicit in existing threat models, which does not account for variation in how an input may be represented at the token level. We show that this assumption is fundamentally flawed (Section~\ref{sec:experiments}).

Existing editing and unlearning techniques assume that modifying parameters identified using the canonical tokenization is sufficient to update the corresponding fact across all inputs that express it. However, alternative tokenizations~\cite{geh2025adversarial} of the same string (noncanonical tokenizations) can induce different internal computational paths, exposing tokenization-based side channels that can bypass alignment mechanisms optimized for the canonical representation. Consequently, any residual factual association accessible through such alternative paths constitutes an information leak. As illustrated in Fig.~\ref{fig:canonical}, we find that an adversary can exploit this by querying the updated model with a noncanonical tokenization of the same input, thereby recovering the unlearned or edited knowledge without requiring access to the original model, an attack surface that existing threat models overlook.

In this work, we introduce \textbf{\Toketive}, a reference-free adversarial attack that operationalizes tokenization-based side channels to detect and recover edited or unlearned knowledge. Given only a query and an edited model (i.e., without access to the original model, external reference outputs or any auxiliary training), \Toketive performs a budget-aware search over alternate tokenizations to (i) detect whether a fact has been modified, and (ii) reconstruct the pre-edit response. \Toketive exploits the residual factual associations that editing leaves intact along unpatched computational paths, which is a tokenization-based side channel. To the best of our knowledge, \Toketive is the first attack to explicitly leverage tokenization variance as a mechanism for breaking the security assumptions of model editing and machine unlearning\footnote{Source code is available at: \url{https://anonymous.4open.science/r/Toketive}.}.

Across five models, six datasets, and six editing and unlearning techniques, \Toketive achieves an average edit/unlearning detection F1 score of 84.2\%, a 26.2\% relative gain over the strongest baseline. It also recovers pre-update responses with a top-5 accuracy of 74.5\%, 21.7\% higher than the best baseline. These results demonstrate that suppressed knowledge remains recoverable under tokenization-aware attacks like \Toketive, revealing a gap between current adversarial evaluation practices and the security properties they aim to capture. We summarize our contributions below.

\noindent\ding{172} \textbf{Tokenization as an attack surface.}
We identify attacker-controlled tokenization as a previously overlooked attack surface for post-release knowledge control, showing that state-of-the-art localized editing and unlearning techniques fail to generalize across alternative representations of the same input. This vulnerability is particularly relevant to open-weight models, where users can freely construct queries with alternative tokenizations (Section~\ref{sec:threatmodel}).

\noindent\ding{173} \textbf{\Toketive: a reference-free adversarial attack.}
We develop \Toketive, a tokenization-aware attack that detects vulnerable modifications and reconstructs pre-edit responses using only the released model, without pre-edit models, training data, shadow models, auxiliary classifiers, or external reference outputs~(Section~\ref{sec:toketive}).

\noindent\ding{174} \textbf{Comprehensive evaluation.}
We evaluate \Toketive across five models, six datasets, and six editing and unlearning techniques, showing that it achieves strong detection performance and reliably reconstructs pre-edit responses, significantly outperforming existing baselines~(Section~\ref{sec:experiments}).

\noindent\ding{175} \textbf{Defense analysis and robustness-locality trade-off.}
We investigate a tokenization-aware adaptive editing strategy that incorporates discovered bypass tokenizations into subsequent updates. While adaptive editing reduces tokenization-based bypasses, we find that increased robustness comes at the cost of broader changes to neighboring knowledge, revealing a fundamental robustness-locality trade-off~(\hyperref[app:adaptive_editing]{Appendix~A}).

\section{Preliminaries}
\subsection{Model Editing and Machine Unlearning}

Model editing aims to update specific factual knowledge stored in a pretrained language model without retraining from scratch~\cite{meng2022locating,Fang2025_AlphaEdit}. Machine 
unlearning pursues the complementary goal of removing the influence of specific training data or factual associations from a model, such that it behaves as if it was never trained on that information~\cite{li2026editing}. A common formulation represents factual knowledge as a triple 
$(s, r, o)$, where $s$ denotes a subject, $r$ a relation, and $o$ the object. Given a natural language prompt $p(s,r)$ (e.g., ``The president of the United States is''), an autoregressive language model $f_\theta$ predicts a distribution over possible continuations, ideally assigning high probability to ($o$).

\textit{Model editing.}
The goal of editing is to replace an existing association $(s,r,o)$ with a new target $(s,r,o^*)$ while preserving unrelated behavior. Formally, given an editing request $e = (s,r,o,o^*)$, an editing procedure $\mathcal{E}$ produces a model,
\[
f_{\theta'} = \mathcal{E}(f_\theta, e),
\]
such that: (i)~Reliability: $f_{\theta'}(p(s,r))$ assigns high probability to $o^*$; (ii)~Generalization: the new model recalls $o^*$ under paraphrased prompts describing the same fact; and (iii)~Locality: predictions for unrelated prompts remain largely unchanged.

\textit{Machine unlearning.}
With unlearning, the goal is to suppress a target association $(s, r, o)$ entirely, such that the model no longer recalls $o$ from $p(s,r)$ and behaves as if it was never trained on that fact. Formally, given an unlearning request over a forget set 
$\mathcal{D}_f$, an unlearning procedure $\mathcal{U}$ produces an updated model,
\[
f_{\theta'} = \mathcal{U}(f_\theta, \mathcal{D}_f),
\]
such that: (i)~Efficacy: $f_{\theta'}$ no longer assigns high probability to $o$ under $p(s,r)$ or its paraphrases; (ii)~Utility: predictions for facts outside $\mathcal{D}_f$ remain unchanged; and (iii)~Privacy: the membership of a forget-set fact should be indistinguishable from a never-trained fact by an adversary querying $f_{\theta'}$.

\textit{The locate-then-edit/unlearn paradigm.}
Both model editing and localization-based unlearning follow a common two-stage procedure~\cite{liang2024locate,li2026editing}. First, they locate internal model components responsible for storing the target fact, typically identifying specific feed-forward MLP layers and subject token positions via causal tracing. Second, they apply targeted parameter updates restricted to those components, modifying weights so that the model retrieves the new association or suppresses the target association, while minimizing collateral damage to unrelated behavior.

\subsection{Noncanonical tokenization}

Let $x = (x_1, \dots, x_n)$ denote a string of characters and let $\mathcal{V}$ be a subword vocabulary constructed via Byte-Pair Encoding (BPE)~\cite{geh2025adversarial, zheng2025broken}. A tokenization of $x$ with respect to $\mathcal{V}$ is a sequence of tokens $v = (v_1, \dots, v_m)$ such that,

\[
v_1 \oplus v_2 \oplus \dots \oplus v_m = x,
\]

where $\oplus$ denotes string concatenation and $v_i \in \mathcal{V}$.

LLM pipelines employ a deterministic tokenizer $k^\ast$ to map each string $x$ to a unique canonical tokenization $v^\ast = k^\ast(x)$, obtained by greedily applying merge rules in the order they were learned. However, as $v^\ast$ is only one element of the full tokenization set,

\[
\mathcal{T}_{\mathcal{V}}(x) = \{ v : v \text{ is a valid tokenization of } x \}.
\]

In general, $|\mathcal{T}_{\mathcal{V}}(x)|$ grows exponentially with $|x|$, even under BPE vocabularies. Consider the string $x =$ \texttt{"The phone number of John Doe is"}. As illustrated in Fig.~\ref{fig:canonical}, canonical tokenization $v^\ast$ may segment this string into subwords such as \texttt{$t_1$ = ["The", " phone num", "ber", " of", " John Doe", " is"]}. However, alternative tokenizations in $\mathcal{T}_{\mathcal{V}}(x)$ can split the same string differently, e.g., \texttt{$t_2$ = ["The", " phon", "e nu", " mber", " of Jo", "hn Doe", " is"]}, while still satisfying $v_1 \oplus \dots \oplus v_m = x$. Although both correspond to the same character string, they induce different token sequences and thus different internal computational model trajectories.

We refer to any $v \in \mathcal{T}_{\mathcal{V}}(x)$ such that $v \neq k^\ast(x)$ as a noncanonical tokenization. Although LLMs are trained exclusively on canonical tokenizations, noncanonical tokenizations can preserve substantial semantic information about the underlying input, as demonstrated by prior work~\cite{geh2025adversarial}. This implies that the conditional distribution may remain semantically meaningful even when $v \neq k^\ast(x)$, while potentially activating different internal trajectories.

\subsection{Representational Entanglement}
\label{subsec:clare}
LLMs encode factual knowledge within high-dimensional hidden representations distributed across transformer layers. 
Let $p_i = p(s_i, r_i)$ denote a prompt corresponding to a factual triple $(s_i, r_i, o_i)$. 
As the input tokens propagate through a model's layers, their hidden representations evolve according to the residual, attention, and feed-forward transformations:
\[
h_i^{(l)} = h_i^{(l-1)} + a_i^{(l)} + m_i^{(l)},
\]
where $a_i^{(l)}$ and $m_i^{(l)}$ denote the attention and MLP contributions at layer $l$. Prior works~\cite{meng2022locating, meng2022mass, baser2026clare} show that factual associations are often mediated by a subset of intermediate MLP layers, sometimes referred to as critical layers, where subject tokens elicit key-like activation patterns that shape the final prediction. Representations extracted at or near the last such critical layer provide a compact yet informative snapshot of how a fact is encoded before downstream mixing and decoding constraints are applied.

{\textbf{Definition (Representational Entanglement).}}
Let $h_i^{(\mathcal{L})}$ and $h_j^{(\mathcal{L})}$ denote hidden representations at a chosen probe layer $\mathcal{L}$, for prompts $p_i$ and $p_j$, respectively. 
The representational entanglement~\cite{baser2026clare} between the corresponding facts is defined as the cosine similarity between their representations at $\mathcal{L}$:
\[
\mathcal{S}(i, j) = \cos(h_i^{(\mathcal{L})}, h_j^{(\mathcal{L})}).
\]

A high entanglement score indicates that two prompts induce similar representations at the probe layer. In \Toketive, this similarity serves as the primary signal for selecting alternative tokenizations. Tokenizations that are too close to the canonical representation are more likely to preserve the target factual association but also to get affected by the edit or unlearning update, whereas tokenizations that are too distant may bypass the update but lose the relevant factual association. \Toketive therefore searches for tokenizations within an intermediate similarity range that balances these two properties. This target range forms the basis of the adaptive tokenization sampler described in Section~\ref{sec:toketive}.


\begin{table*}
\centering
\caption{Comparison of \Toketive with existing methods for security and privacy in editing and unlearning.}
\label{tab:comparison}
\resizebox{.75\textwidth}{!}{ 
\begin{tabular}{c|ccc}

\toprule

\rowcolor{gray!25}
\textbf{Method} & \textbf{Pre-edit model required} & \textbf{Requires training auxiliary models} & \textbf{Reconstructs pre-edit output} \\

\midrule

DEED~\cite{youssef2025has} 
& \textcolor{red}{\crossmark\ Required} 
& \textcolor{red}{\crossmark\ Trains AdaBoost classifier} 
& \textcolor{orange}{$\sim$ Indirectly} \\

FUMA~\cite{deepak2025identifying} 
& \textcolor{green!60!black}{\xmark\ Not required} 
& \textcolor{green!60!black}{\xmark\ No training required} 
& \textcolor{red}{\xmark\ No} \\

KSTER~\cite{sun2026reverse} 
& \textcolor{red}{\crossmark\ Required} 
& \textcolor{green!60!black}{\xmark\ No training} 
& \textcolor{orange}{$\sim$ Indirectly} \\

RULI~\cite{naderloui2025rectifying} 
& \textcolor{green!60!black}{\xmark\ Not required} 
& \textcolor{red}{\crossmark\ Trains shadow models} 
& \textcolor{red}{\xmark\ No} \\

U-LiRA~\cite{hayes2025inexact} 
& \textcolor{green!60!black}{\xmark\ Not required} 
& \textcolor{red}{\crossmark\ Trains shadow models} 
& \textcolor{red}{\xmark\ No} \\

TULA-DR~\cite{du2025textual} 
& \textcolor{red}{\crossmark\ Required} 
& \textcolor{green!60!black}{\xmark\ No training required} 
& \textcolor{green!60!black}{\crossmark\ Yes (via weight diff)} \\

\rowcolor{brown!20}
\textbf{\Toketive (ours)} 
& \textcolor{green!60!black}{\xmark\ Not required} 
& \textcolor{green!60!black}{\xmark\ No training required} 
& \textcolor{green!60!black}{\crossmark\ Yes} \\

\bottomrule

\end{tabular}
}
\end{table*}

\section{Threat model}
\label{sec:threatmodel}

\textbf{Setting.} We consider a scenario where a model $f_\theta$ is edited or unlearned to produce $f_{\theta'}$, which is subsequently released. We assume access to the released model's parameters and internal activations. This reflects the intended use of open-weight foundation models, where downstream users can inspect and execute the model locally. Consequently, recent works have leveraged access to model parameters and internal representations to recover edited or unlearned information~\cite{youssef2025has, deepak2025identifying, sun2026reverse}. Moreover, open-weight models are increasingly adopted because they avoid the recurring inference costs of proprietary APIs while enabling local execution with competitive performance~\cite{10.1145/3630106.3658966}. Advances in quantization, compression, and hardware-aware optimization further allow modern open-weight models to run efficiently on commodity hardware, maintaining strong performance in resource-constrained settings~\cite{10.1145/3803798, ni2025large}.

\textbf{In-scope and out-of-scope.} We focus on inexact editing and unlearning procedure, particularly those utilizing the locate-then-edit paradigm~\cite{Fang2025_AlphaEdit, meng2022mass, gu2401model}, which modifies a targeted subset of parameters for a given prompt. Recent works~\cite{liang2024locate, li2026editing} conceptualize unlearning as a special case of the locate-then-edit paradigm. We exclude fine-tuning-based approaches~\cite{sinha2025unstar, 3692070.3693215, yang2026fine} because they primarily suppress target outputs rather than remove the underlying knowledge from the model, while also introducing substantial collateral changes to unrelated knowledge and capabilities~\cite{batorski2026grom, betley2025emergent, hong2024dissecting}. Although fine-tuning more effectively suppresses noncanonical tokenizations, our experiments in \hyperref[app:finetuning]{Appendix~B} show that this comes at a substantial cost to locality, with fine-tuning producing $5.2\times$ larger ripple effects than editing techniques.

\noindent\textbf{Attacker's Capabilities.} An attacker in our threat model can:
\begin{enumerate}
    \item Query $f_{\theta'}$ with arbitrary token sequences, including noncanonical tokenizations of any string;
    \item Read internal activations of $f_{\theta'}$ at any layer and position.
\end{enumerate}

\noindent The attacker does not have access to:
\begin{itemize}

    \item the pre-edit model $f_\theta$ or any of its outputs;
    \item the pre-edit response $o$ for any target fact;
    \item any training or fine-tuning data for $f_\theta$;
    \item any information about which editing or unlearning technique was applied;
    \item shadow models, auxiliary classifiers or training facilities.
\end{itemize}

This access scenario captures a realistic and practically relevant adversary. For example, a malicious actor attempting to recover PII that was supposed to have been deleted from a publicly released model, or a regulator auditing a deployed model for compliance with a deletion request.

\noindent\textbf{Attack Goal.}
The attacker pursues two coupled objectives:

    \ding{172} Edit/Unlearning Detection: Given $f_{\theta'}$ and a prompt $p(s, r)$, determine whether the fact $(s, r, \cdot)$ was modified post-training.
    
    \ding{173} Adversarial Reconstruction: If the fact was modified, recover the original object from the pre-edit association $(s, r, o)$.

These goals are intentionally coupled, as detection without reconstruction only establishes that an edit occurred, while reconstruction without detection would require knowing in advance which facts to target. \Toketive addresses both simultaneously, using a unified search procedure over the tokenization space. Once an open-weight model is released, the model owner no longer controls how it is deployed or queried. An adversary can locally invoke the model with raw token IDs and inspect hidden activations, bypassing any canonicalization that a hosted API might enforce. Existing evaluations overestimate suppression by assuming canonical tokenization during inference. \Toketive identifies a previously overlooked attack surface under this open-weight setting.

\section{\Toketive: A New Adversarial Attack on Editing \& Unlearning}
\label{sec:toketive}

In this section, we identify limitations of existing detection methods, privacy attacks, and evaluation frameworks in editing and unlearning. Subsequently, we introduce a novel adversarial attack that overcomes these limitations, while unifying the detection and reconstruction of edited or unlearned knowledge.

\subsection{Limitations of Existing Methods}
\label{sec:limitations}

Existing approaches share three structural limitations that restrict their applicability in practical scenarios. Table~\ref{tab:comparison} summarizes their capabilities and limitations.

\textbf{Limitation I: Dependence on the Pre-edit Model.}
Many approaches analyzing model editing and machine unlearning require access to the original pre-edit model $f_\theta$. DEED~\cite{youssef2025has} assumes access to the original model for obtaining its training set. KSTER~\cite{sun2026reverse} performs spectral analysis on the weight difference $\theta' - \theta$ to recover the editing subject. TULA-DR~\cite{du2025textual} solves a constrained optimization over the weight differential to reconstruct unlearned text. In each case, the pre-edit model is an input to the procedure. This assumption is strong and often unrealistic. In operation, the original model is typically unavailable once an edit or unlearning procedure is applied, particularly when the edit is motivated by a right-to-be-forgotten request under GDPR~\cite{pmlr-v238-oesterling24a} or similar regulation. An auditor or adversary operating after the modification has access only to the released post-edit model.

\begin{figure*}
\centering

    \centering
    \includegraphics[width=0.9\linewidth]{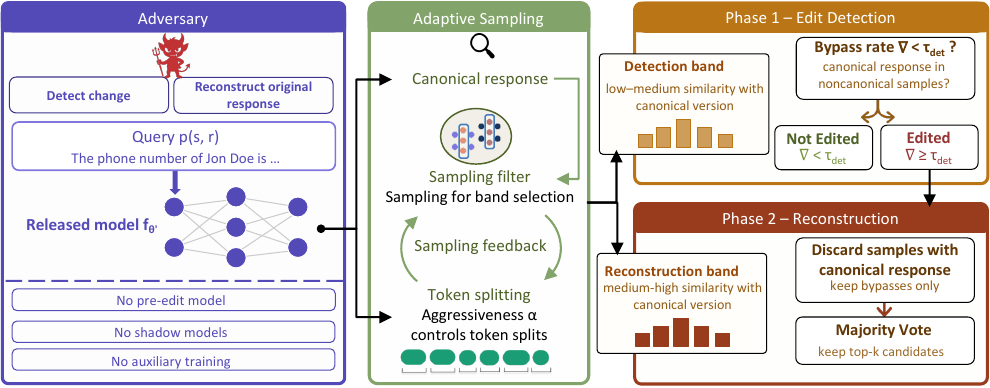}

\caption{Overview of the \Toketive adversarial attack. While editing and unlearning techniques aim to patch the circuit for the canonical tokenization, many noncanonical tokenizations bypasses the edit, exposing the pre-edit knowledge.}
\label{fig:attack_pipeline}
\end{figure*}

\textbf{Limitation II: Dependence on Shadow Models or Auxiliary Training.}
Other methods that avoid the pre-edit model requirement instead require training shadow models or auxiliary classifiers to function. U-LiRA~\cite{hayes2025inexact} trains multiple shadow models to construct membership likelihood ratios, while RULI~\cite{naderloui2025rectifying} requires training multiple shadow models alongside their unlearned variants to audit privacy leakage. DEED~\cite{youssef2025has} also trains an AdaBoost classifier on labeled examples of edited and unedited facts. In these approaches, the process imposes substantial training overhead; they also require knowledge of the training and unlearning algorithms as well as access to the underlying data distribution, which a post-hoc auditor or an adversary operating on a released model is unlikely to possess.

\textbf{Limitation III: Methods Without Pre-edit Model Access Cannot Reconstruct Pre-edit Responses.}
Among the methods that do not require the pre-edit model (U-LiRA, RULI, and FUMA~\cite{deepak2025identifying}), none are capable of reconstructing the pre-edit response. U-LiRA and RULI are membership inference methods, as they determine whether a data point was part of the training set, but produce no information about what the model's pre-edit response was. FUMA identifies what was unlearned via gradient signals but similarly does not recover the pre-edit output. For an auditor verifying compliance with a deletion request, knowing that an edit occurred is necessary but not sufficient. The auditor must also verify what was suppressed and whether it remains accessible post-modification.

\subsection{The \Toketive Attack}
\label{sec:attack}

We design \Toketive around the fact that alternative tokenizations of the same input string induce distinct internal computational trajectories. Our key observation is that many alternate trajectories remain unaffected even after the model is edited or unlearned, and continue to retrieve the original association. We exploit this asymmetry with a simple yet powerful tokenization-based side channel,  where \Toketive searches over noncanonical tokenizations to identify and recover the pre-edit response. \Toketive preserves the exact input string and varies only its valid tokenization, isolating tokenization-dependent computation.

\Toketive takes as input the post-edit model $f_{\theta'}$ and a target prompt $p(s, r)$, and produces two outputs: a detection label $\ell \in \{\texttt{edited}, \texttt{unedited}\}$ and, if $\ell = \texttt{edited}$, a reconstructed pre-edit response $\hat{o}$. \Toketive consists of an adaptive tokenization sampler as a shared upstream component, as its output is routed to two downstream phases: \ding{172} edit detection and \ding{173} pre-edit reconstruction. As its filtering signal, the sampler uses representational entanglement~(Section~\ref{subsec:clare}), a technique originally proposed to predict ripple effects between different facts, which we repurpose here as a bypass prediction signal within the tokenization space of a single prompt. We evaluate alternative filtering strategies in Section~\ref{sec:experiments}. Algorithm~\ref{alg:spectre} presents the complete procedure.

\textbf{Canonical reference.}
Before sampling begins, \Toketive performs a single forward pass on the canonical tokenization $v^* = \kstar(p(s,r))$ through $f_{\theta'}$, recording two quantities: the canonical output $\hat{y}_{v^*}$ (the post-edit response under $v^*$), and the canonical reference vector $h^{(\mathcal{L})}_{v^*}$, the hidden state at the final subject token position at the last critical layer $\mathcal{L}$~(Algo.~\ref{alg:spectre}, line~no.~17-24). All sampling scores computed during the search are measured relative to $h^{(\mathcal{L})}_{v^*}$.

\begin{algorithm}[t!]
\caption{\small\Toketive: Tokenization-Based Adversarial Attack.} 
\label{alg:spectre}
\small
\begin{algorithmic}[1]
\Input Post-edit model $f_{\theta'}$, prompt $p(s,r)$, subject $s$,
       sample budgets $n_{\text{det}}, n_{\text{rec}}$,
       detection threshold $\tau_{\text{det}}$,
       band bounds $\beta^{\text{det}}, \beta^{\text{rec}}$
\Output Detection label $\ell$; reconstructed pre-edit response $\hat{o}$
        (if $\ell = \texttt{edited}$)
\State \textbf{Initialize:} $\mathcal{C}_{\text{det}},\, \mathcal{C}_{\text{rec}},\, \mathcal{C}_{\text{byp}}$
\vspace{2pt}
\State $(o^*,\, h^*) \leftarrow \textsc{CanonicalRef}(f_{\theta'},\, p,\, s)$
\vspace{4pt}
\State {\color{blue!70!black} $>$ \textit{Phase \ding{172}: edit detection}}
\State $\mathcal{C}_{\text{det}} \leftarrow \textsc{Sample}(f_{\theta'},\, p,\, s,\, h^*,\,
       \beta^{\text{det}},\, n_{\text{det}})$
\State $\mathcal{r} \leftarrow \dfrac{1}{|\mathcal{C}_{\text{det}}|}
       \displaystyle\sum_{u \in \mathcal{C}_{\text{det}}}
       \mathbf{1}\bigl[o^* \in f_{\theta'}(u)\bigr]$
\If{$\mathcal{r} \geq \tau_{\text{det}}$}
    \State \Return $(\texttt{unedited},\; \bot)$
\EndIf
\vspace{4pt}
\State {\color{blue!70!black} $>$ \textit{Phase \ding{173}: pre-edit reconstruction}}
\State $\mathcal{C}_{\text{rec}} \leftarrow \textsc{Sample}(f_{\theta'},\, p,\, s,\, h^*,\,
       \beta^{\text{rec}},\, n_{\text{rec}})$
\State $\mathcal{C}_{\text{byp}} \leftarrow
       \{\, u \in \mathcal{C}_{\text{rec}} \mid o^* \notin f_{\theta'}(u) \,\}$
\If{$\mathcal{C}_{\text{byp}} = \emptyset$}
    \State \Return $(\texttt{edited},\; \bot)$
\EndIf
\State $\hat{o} \leftarrow \arg\max_{c}\;
       \displaystyle\sum_{u \in \mathcal{C}_{\text{byp}}}
       \mathbf{1}\bigl[f_{\theta'}(u)[|p|{:}] = c\bigr]$
\State \Return $(\texttt{edited},\; \hat{o})$
\vspace{4pt}
\State {\color{blue!70!black} $>$ \textit{canonical reference}}
\Procedure{CanonicalRef}{$f_{\theta'},\, p,\, s$}
    \State $\mathcal{L} \leftarrow \lfloor \mathrm{depth}(f_{\theta'}) / 3 \rfloor$
    \State $v^* \leftarrow \kstar(p)$
    \State $o^* \leftarrow f_{\theta'}(v^*)[|p|{:}]$
    \State $h^* \leftarrow \textsc{Hidden}(f_{\theta'},\, v^*,\, \mathcal{L},\, s)$
    \State \Return $(o^*,\, h^*)$
\EndProcedure
\vspace{4pt}
\State {\color{blue!70!black} $>$ \textit{adaptive tokenization sampling}}
\Procedure{Sample}{$f_{\theta'},\, p,\, s,\, h^*,\, [\beta_l, \beta_h),\, n$}
    \State $\mathcal{C} \leftarrow \emptyset$
    \State $\mathcal{L} \leftarrow \lfloor \mathrm{depth}(f_{\theta'}) / 3 \rfloor$
    \While{$|\mathcal{C}| < n$} \Comment{$\alpha$ adapted by proportional controller}
        \State $u \leftarrow \textsc{Tokenize}(p,\, \alpha)$
        \If{$\textsc{Concat}(u) = p$}
            \State $h_u \leftarrow \textsc{Hidden}(f_{\theta'},\, u,\, \mathcal{L},\, s)$
            \State $e \leftarrow \cos(h_u,\, h^*)$
            \If{$\beta_l \leq e < \beta_h$}
                \State $\mathcal{C} \leftarrow \mathcal{C} \cup \{\,u\,\}$
            \EndIf
        \EndIf
    \EndWhile
    \State \Return $\mathcal{C}$
\EndProcedure
\end{algorithmic}
\end{algorithm}

\subsubsection{Adaptive Tokenization Sampler}
\label{sec:sampler}

The sampler is controlled by a single aggressiveness parameter, $\alpha \in [0,1]$, which is initialized to 0.5. It governs how aggressively the prompt string is fragmented into smaller subword units (Algorithm~\ref{alg:spectre}, lines~25-39). A higher $\alpha$ produces more fragmented tokenizations, whereas lower values generate segmentations closer to the canonical form. As illustrated in Fig.~\ref{fig:canonical}, increasing $\alpha$ results in finer-grained tokenizations (e.g., $t_2$ versus $t_1$), inducing alternative internal computational trajectories for the same input string. During search, $\alpha$ is automatically adjusted using a proportional controller driven by the rolling mean similarity over the previous ten sampled tokenizations. Once the controller reaches the target similarity band, a small random perturbation is applied to $\alpha$ to encourage local exploration and avoid stagnation. For each candidate tokenization $u$, the sampler performs a forward pass through $f_{\theta'}$ and computes the entanglement score (Section~\ref{subsec:clare}):
\[
\mathcal{S}(u, v^*) = \cos\!\left(h_u^{(\mathcal{L})},\; 
h_{v^*}^{(\mathcal{L})}\right).
\]
A candidate $u$ is accepted only if its concatenation recovers the original string $p$ exactly, ensuring the underlying character sequence is unchanged. Based on this score (between $0$ and $1$), each candidate is routed to one of three destinations:
\begin{itemize}
    \item Detection band $[\beta^{\text{det}}_l,
    \beta^{\text{det}}_h)$: forwarded to Phase~\ding{172}.
    \item Reconstruction band $[\beta^{\text{rec}}_l,
    \beta^{\text{rec}}_h)$: forwarded to 
    Phase~\ding{173}.
    \item Out-of-band: discarded without querying for completion.
\end{itemize}

For each candidate tokenization, \Toketive first re-identifies the subject span after tokenization and extracts the hidden representation of the subject's final token at probe layer $\mathcal{L}=\lfloor\text{depth}/3\rfloor$. Because subject boundaries are recomputed independently for every tokenization, this procedure naturally accommodates subjects that are split into multiple tokens, byte-level tokens, or tokenizer-specific segmentations. We adopt $\mathcal{L}=\lfloor\text{depth}/3\rfloor$, where depth is the total number of layers in the given language model, following prior work~\cite{baser2026clare}, which reports that intermediate-layer representations provide a substantially more reliable indicator of representational entanglement than gradient-based alternatives~\cite{qin2024does}. Thus, representational entanglement serves as an empirically validated routing heuristic that prioritizes tokenizations likely to bypass localized edits. Consequently, \Toketive does not depend on any particular theory of how knowledge is localized within the model. While a formal characterization of representational entanglement remains an open research problem, our objective is algorithmic rather than mechanistic: we evaluate this heuristic solely by its ability to improve attack success, rather than as evidence for a causal theory of knowledge storage.

We treat the tokenizer as a black-box mapping from strings to token sequences with a fixed vocabulary $\mathcal{V}$. While some methods use BPE~\cite{geh2025adversarial}, \Toketive does not rely on any specific tokenization scheme. Instead, it operates over the set of valid tokenizations $\mathcal{T}_{\mathcal{V}}(x)$ that decode to the same string.

\textbf{Proportional controller.}
The sampler maintains a diversity buffer, which is a rolling window of the most recent entanglement scores across all attempts, including discarded ones~(Algo.~\ref{alg:spectre}, line~no.~28). After each iteration, $\alpha$ is updated based on how the recent mean score compares to the active target band. If scores are consistently too high, $\alpha$ is increased to produce more divergent tokenizations; if scores are too low, $\alpha$ is decreased to pull candidates back toward the target band. A small random perturbation is applied when scores are already in band, preventing the search from stagnating at a fixed aggressiveness. Recording scores for all attempts and not just accepted samples ensures the controller has a complete view of the sampling distribution and recovers quickly from iterations with low acceptance rate.

\subsubsection{Phase \ding{172} (Edit/Unlearning Detection)}
\label{sec:detect}

This phase processes samples routed to the detection band. This band captures tokenizations that are significantly different from the canonical tokenization in terms of their internal activations of the model, i.e., different enough to potentially bypass patched circuits, but not so different that semantic signal is lost~(Algo.~\ref{alg:spectre}, line~no.~3-8).

The sampler accumulates $n_{\text{det}}$ detection-band candidates $\mathcal{C}_{\text{det}}$. For each $u \in \mathcal{C}_{\text{det}}$, \Toketive queries $f_{\theta'}$ to obtain the completion $\hat{y}_u$ and checks whether it contains the post-modification object $o^*$. For editing, $o^*$ corresponds to the updated target answer; for unlearning, $o^*$ is the deterministic replacement response (e.g., "unknown") specified by the unlearning procedure and produced by the canonical prompt after unlearning.
The canonical-response bypass rate is:
\[
\mathcal{r} = \frac{1}{|\mathcal{C}_{\text{det}}|}
\sum_{u \in \mathcal{C}_{\text{det}}}
\mathbf{1}\!\left[o^* \in \hat{y}_u\right].
\]

where $\hat{y}_u = f_{\theta'}(u)$ is the model's completion under noncanonical tokenization $u$, and $o^*$ is  extracted from the canonical response $\hat{y}_{v^*}$. If the edit/unlearning generalizes across tokenizations, alternative tokenizations in this band will produce the post-modification object $o^*$, yielding $\mathcal{r} \approx 1$. If the edit/unlearning is superficial, many alternatives will bypass it and recover the pre-update response rather than $o^*$, yielding low $\mathcal{r}$. Thus, the detection verdict $\ell$ becomes:
\[
\ell =
\begin{cases}
\texttt{edited (or unlearned)}   & \text{if } \mathcal{r} < \tau_{\text{det}} \\
\texttt{unedited (or not unlearned)} & \text{otherwise,}
\end{cases}
\]
where $\tau_{\text{det}}$ is the detection threshold, and is treated as a hyperparameter. If $\ell = \texttt{unedited (or not unlearned}$, \Toketive halts and returns $(\texttt{unedited (or not unlearned},\; \bot)$.

\subsubsection{Phase \ding{173} (Pre-Edit Reconstruction)}
\label{sec:recon}

This phase runs only if $\ell = \texttt{edited (or unlearned}$. It consumes samples routed to the reconstruction band~(Algo.~\ref{alg:spectre}, line~no.~9-16). This band captures tokenizations that remain internally similar to the canonical representation, likely drawing on the same underlying factual knowledge, but just different enough that their key vectors are not fully aligned with $k^*$, allowing them to escape the patched circuits. The sampler accumulates $n_{\text{rec}}$ reconstruction-band candidates $\mathcal{C}_{\text{rec}}$.

\textbf{Bypass filtering.}
For each $u \in \mathcal{C}_{\text{rec}}$, \Toketive generates the completion $\hat{y}_u$ and discards any sample for which the post-update answer $o^* \in \hat{y}_u$. The retained bypass continuations are:
\[
\mathcal{C}_{\text{byp}} =
\left\{ u \in \mathcal{C}_{\text{rec}} : o^* \notin \hat{y}_u 
\right\}.
\]

\textbf{Majority vote reconstruction.}
\Toketive extracts the predicted object span from each generated output $\hat{y}_u$:
\[
c_u = \hat{y}_u[|p(s,r)|{:}].
\]

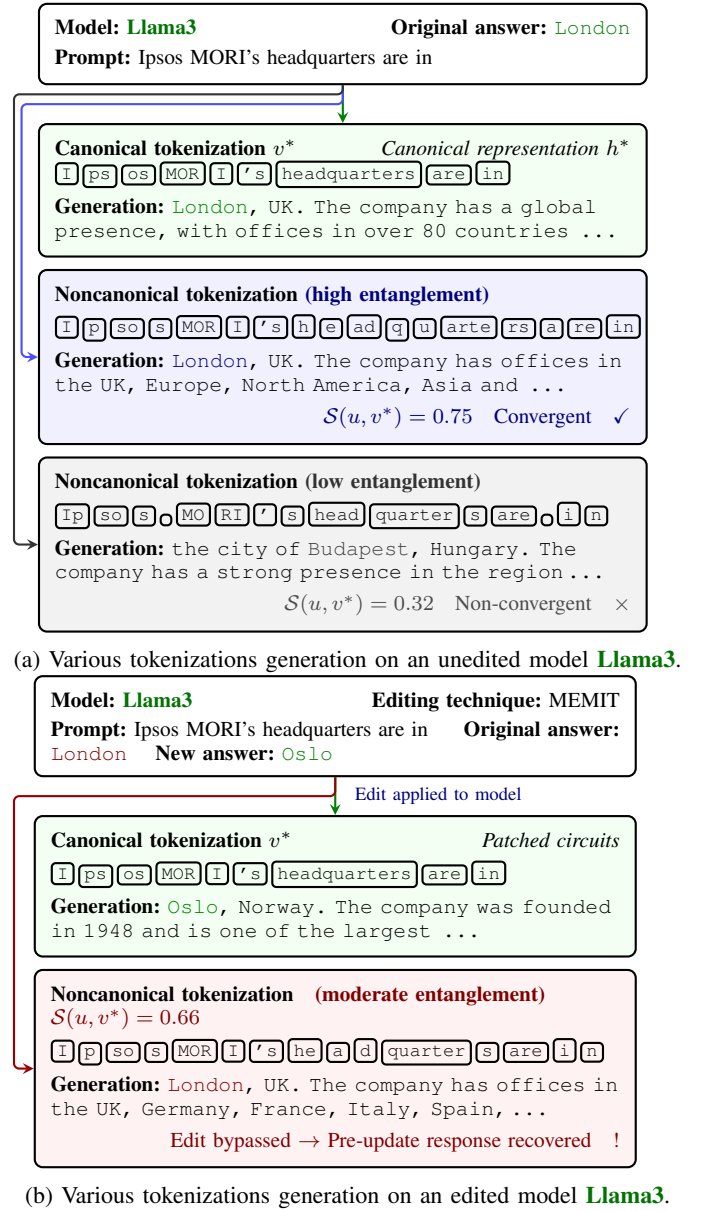
\begin{figure}[t]
\centering

\begin{subfigure}{\linewidth}
\begin{tikzpicture}[
    font=\footnotesize,
    box/.style={
        rectangle, rounded corners=3pt,
        draw, thick,
        text width=0.86\linewidth,
        align=left,
        inner sep=6pt
    },
    arrow/.style={->, >=stealth, thick},
]

\node[box, fill=white!10] (prompt) at (0,0) {
    \textbf{Model:} \LlamaThree \hfill
    \textbf{Original answer:} \texttt{\textcolor{green!50!black}{London}}\\[2pt]
    \textbf{Prompt:} Ipsos MORI's headquarters are in
};

\node[box, fill=green!5, below=0.5cm of prompt] (canon) {
    \textbf{Canonical tokenization $v^*$} \hfill\textit{Canonical representation $h^*$}
    \tok{I}\,\tok{ps}\,\tok{os}\,\tok{ MOR}\,\tok{I}\,%
    \tok{'s}\,\tok{ headquarters}\,\tok{ are}\,\tok{ in}\\[4pt]
    \textbf{Generation:}
    \texttt{\textcolor{green!50!black}{London}, UK. The company has a global presence, with offices in over 80 countries \,\ldots}\\[3pt]
};

\node[box, fill=blue!6, below=0.15cm of canon] (high) {
    \textbf{Noncanonical tokenization}
    \textcolor{blue!50!black}{\textbf{(high entanglement)}}\\[3pt]
    \tok{I}\,\tok{p}\,\tok{so}\,\tok{s}\,\tok{ MOR}\,\tok{I}\,%
    \tok{'s}\,\tok{ h}\,\tok{e}\,\tok{ad}\,\tok{q}\,\tok{u}\,%
    \tok{arte}\,\tok{rs}\,\tok{ a}\,\tok{re}\,\tok{ in}\\[4pt]
    \textbf{Generation:}
    \texttt{\textcolor{blue!50!black}{London}, UK. The company has offices in the UK, Europe, North America, Asia and \,\ldots}\\[3pt]
    \hfill\textcolor{blue!50!black}{
        $\mathcal{S}(u, v^*) = 0.75$\quad
        Convergent\quad\checkmark}
};

\node[box, fill=gray!10, below=0.15cm of high] (low) {
    \textbf{Noncanonical tokenization}
    \textcolor{gray!40!black}{\textbf{(low entanglement)}}\\[3pt]
    \tok{Ip}\,\tok{so}\,\tok{s}\,\tok{ }\,\tok{MO}\,\tok{RI}\,%
    \tok{'}\,\tok{s}\,\tok{ head}\,\tok{quarter}\,\tok{s}\,%
    \tok{ are}\,\tok{ }\,\tok{i}\,\tok{n}\\[4pt]
    \textbf{Generation:}
    \texttt{the city of \textcolor{gray!50!black}{Budapest}, Hungary. The company has a strong presence in the region\,\ldots}\\[3pt]
    \hfill\textcolor{gray!60!black}{
        $\mathcal{S}(u, v^*) = 0.32$\quad
        Non-convergent\quad$\times$}
};

\tikzset{ arrow/.style={->, >=stealth, thick, rounded corners=2pt} }

\draw[arrow, green!50!black] (prompt.south) -- (canon.north);

\draw[arrow, blue!70] 
    (prompt.south) -- ++(0,-0.25) -- 
    ++(-4.25, 0) |- (high.west);

\draw[arrow, gray!40!black]
    (prompt.south) -- ++(0,-0.12) --
    ++(-4.35, 0) |- (low.west);

\end{tikzpicture}
\caption{Various tokenizations generation on an unedited model \LlamaThree.}
\label{fig:example_convergence}
\end{subfigure}

\begin{subfigure}{\linewidth}
\begin{tikzpicture}[
    font=\footnotesize,
    box/.style={
        rectangle, rounded corners=3pt,
        draw, thick,
        text width=0.85\linewidth,
        align=left,
        inner sep=6pt
    },
    arrow/.style={->, >=stealth, thick},
]

\node[box, fill=white!10] (prompt) at (0,0) {
    \textbf{Model:} \LlamaThree \hfill
    \textbf{Editing technique:} MEMIT\\[2pt]
    \textbf{Prompt:} Ipsos MORI's headquarters are in \hfill
    \textbf{Original answer:} \texttt{\textcolor{red!50!black}{London}}
    \quad\textbf{New answer:} \texttt{\textcolor{green!50!black}{Oslo}}
};

\node[box, fill=green!5, below=0.5cm of prompt] (canon) {
    \textbf{Canonical tokenization $v^*$} \hfill
    \textit{Patched circuits}\\[3pt]
    \tok{I}\,\tok{ps}\,\tok{os}\,\tok{ MOR}\,\tok{I}\,%
    \tok{'s}\,\tok{ headquarters}\,\tok{ are}\,\tok{ in}\\[4pt]
    \textbf{Generation:}
    \texttt{\textcolor{green!50!black}{Oslo}, Norway. The company was founded in 1948 and is one of the largest \,\ldots}
};

\node[box, fill=red!5, below=0.15cm of canon] (bypass) {
    \textbf{Noncanonical tokenization}\quad
    \textcolor{red!50!black}{\textbf{(moderate entanglement)}}
    \hfill
    \textcolor{red!50!black}{$\mathcal{S}(u, v^*) = 0.66$}\\[3pt]
    \tok{I}\,\tok{p}\,\tok{so}\,\tok{s}\,\tok{ MOR}\,\tok{I}\,%
    \tok{'s}\,\tok{ he}\,\tok{a}\,\tok{d}\,\tok{quarter}\,%
    \tok{s}\,\tok{ are}\,\tok{ i}\,\tok{n}\\[4pt]
    \textbf{Generation:}
    \texttt{\textcolor{red!50!black}{London}, UK. The company 
    has offices in the UK, Germany, France, Italy, 
    Spain,\,\ldots}\\[3pt]
    \hfill\textcolor{red!50!black}{
        Edit bypassed $\rightarrow$ Pre-update response recovered\quad !}
};

\tikzset{arrow/.style={->, >=stealth, thick, rounded corners=2pt}}
\draw[arrow, green!50!black] (prompt.south) -- (canon.north);
\draw[arrow, red!60!black]
    (prompt.south) -- ++(0,-0.25) --
    ++(-4.25, 0) |- (bypass.west);

\node[font=\scriptsize, text=blue!50!black, right=4pt]
    at ($(prompt.south)!0.5!(canon.north)$)
    {Edit applied to model};

\end{tikzpicture}

\caption{Various tokenizations generation on an edited model \LlamaThree.}
\label{fig:example_bypass}
\end{subfigure}
\caption{Observed tokenization convergence and bypass in \LlamaThree. \textbf{(a)} On an unedited model, representational entanglement predicts whether a noncanonical tokenization retrieves the same answer as the canonical one. \textbf{(b)} After MEMIT edits the model, canonical tokenization produces the new answer. A noncanonical tokenization with moderate entanglement bypasses the patched circuits and recovers the pre-update answer.}
\label{fig:example_combined}
\end{figure}

Given $\mathcal{C}_{\text{byp}}$, \Toketive aggregates these object candidates and reconstructs the pre-edit response via plurality voting:
\[
\hat{o} = \arg\max_{c} \sum_{u \in \mathcal{C}_{\text{byp}}}
\mathbf{1}[c_u = c].
\]

For evaluation, we rank candidate object spans by frequency. The most frequent candidate is reported as the top-1 prediction $\hat{o}$, while top-$k$ accuracy measures whether the ground-truth pre-edit object appears among the $k$ most frequent candidates.

Fig.~\ref{fig:example_combined} illustrates the core intuition behind \Toketive. Panel~(a) shows that representational entanglement predicts whether a noncanonical tokenization retrieves the same response as the canonical one in an unedited model, motivating its use as a routing signal. Panel~(b) shows that after MEMIT edits the model, a noncanonical tokenization with moderate entanglement bypasses the patched circuits and recovers the pre-edit answer, demonstrating the tokenization side channel that \Toketive exploits.

We highlight that \Toketive does not exhaustively enumerate the tokenization space. Instead, it performs adaptive sampling over valid token segmentations under a fixed maximum-attempt budget, yielding runtime bounded by the sampling budget rather than the exponentially large tokenization space. Starting from the canonical tokenization, the sampler proposes alternative local segmentations according to the aggressiveness parameter $\alpha$, retaining only candidates whose decode–reencode round-trip exactly reconstructs the original substring. This validation automatically handles byte-level tokens, leading-space markers, Unicode boundaries, and tokenizer-specific artifacts; invalid candidates are discarded, duplicate tokenizations are removed, and special tokens are never introduced.
\hyperref[app:sampler]{Appendix~C} provides further implementation details and complete sampler pseudocode. \hyperref[app:rq3]{Appendix~D} runtime complexity and search-cost analysis for \Toketive. \hyperref[app:hyperparams]{Appendix~E} presents hyperparameter calibration.

\section{Experiments}
\label{sec:experiments}

All experiments are conducted on a cluster of NVIDIA H200 GPUs with 141 GB memory each. We propose and investigate the following research questions:

\noindent \textit{Q1} [\textit{\textbf{Predictability}}]: Are there measurable signals that predict whether alternative tokenizations of a prompt will converge to the canonical response in an unedited model?

\noindent\textit \textit{Q2} [\textit{\textbf{Tokenization Invariance}}]: To what extent do current editing and unlearning techniques produce updates that generalize across alternative tokenizations of the same semantic prompt?

\noindent \textit{Q3} [\textit{\textbf{Adversarial Efficacy}}]: Can an adversary identify whether a fact has been modified, and if so, reconstruct the pre-edit response from an edited model without access to the pre-edit model or response?

We first describe the common experimental setup used across all of our experiments. The subsequent subsections discuss and address \textit{Q1}, \textit{Q2}, and \textit{Q3}, respectively.

\subsection{Experimental Setup}
\label{sec:setup}

\textbf{Datasets.}
We evaluate on six benchmarks spanning factual editing and machine unlearning.
    
    \ding{172} CounterFact~\cite{meng2022locating}: a benchmark for factual knowledge editing that constructs prompts from $(s, r, o)$ triples and counterfactual variants to test whether edits correctly update specific facts while preserving unrelated knowledge. It is widely used to evaluate edit efficacy, generalization to paraphrased prompts, and specificity in LLMs.
    
    \ding{173} RippleEdits~\cite{cohen2024rippleedits}: a diagnostic benchmark for measuring the ripple effects of editing a single fact, evaluating how changes propagate across related relations and compositional queries, and whether updates remain appropriately localized.
    
    \ding{174} MQuAKE~\cite{zhong2023mquake}: a benchmark of multi-hop questions to test whether edited models correctly update answers implied by the edits. It includes counterfactual and temporal subsets, with examples constructed from chains of interdependent facts to assess the propagation of edits through reasoning.
    
    \ding{175} Known-1000~\cite{meng2022locating}: a question-answering dataset derived from structured knowledge base triples, where each fact is paired with a question formulation probing the same relation, enabling analysis of how models store and retrieve facts.
    
    \ding{176} TOFU – World Facts~\cite{maini2025tofu}: a split of the TOFU unlearning benchmark consisting of general Q\&A pairs, used to evaluate whether unlearning methods preserve performance on broad real-world knowledge without causing collateral degradation.
    
    \ding{177} TOFU – Real Authors~\cite{maini2025tofu}: a split of the TOFU unlearning benchmark containing Q\&A pairs about real-world authors, used to assess whether unlearning fictitious author data preserves knowledge about actual individuals and avoids unintended interference with semantically related entities.

For CounterFact, we use the first 1000 samples for all our experiments. For MQuAKE and RippleEdits, we extract 1000 factual triples from each dataset along with their corresponding prompt templates. For the TOFU datasets, we convert Q\&A pairs into prompt-based formats compatible with the editing and unlearning framework. This standardization enables consistent comparison across datasets with differing formats. To ensure reliable ground truth, we restrict evaluation to instances where the canonical tokenization produces the correct answer for the models in all the experiments. To eliminate stochastic variation and isolate the effect of tokenization from sampling noise, we generate model outputs using deterministic decoding (greedy decoding with no sampling) in all our experiments.

\begin{table*}
\centering
\caption{Predictability of tokenization convergence across models and datasets. Bold indicates the better signal per metric.}
\label{tab:predictability}
\resizebox{0.85\textwidth}{!}{
\begin{tabular}{ll|cc|cc|cc|cc|cc|cc}
\toprule
\multirow{2}{*}{\centering Model} & 
\multirow{2}{*}{\centering Method} & 
\multicolumn{2}{c}{Real Authors} & 
\multicolumn{2}{c}{CounterFact} & 
\multicolumn{2}{c}{Known-1000} & 
\multicolumn{2}{c}{MQuAKE} & 
\multicolumn{2}{c}{RippleEdits} & 
\multicolumn{2}{c}{World Facts} \\
\cmidrule(lr){3-14}
& & 
AUC & $|r_b|$ & 
AUC & $|r_b|$ & 
AUC & $|r_b|$ & 
AUC & $|r_b|$ & 
AUC & $|r_b|$ & 
AUC & $|r_b|$ \\
\midrule
\multirow[c]{2}{*}{\LlamaThree} & Edit distance & 0.662 & 0.323 & 0.632 & 0.264 & 0.621 & 0.242 & 0.614 & 0.228 & 0.565 & 0.130 & 0.628 & 0.256 \\
 & Repr. entanglement & \textbf{0.803} & \textbf{0.606} & \textbf{0.688} & \textbf{0.375} & \textbf{0.774} & \textbf{0.548} & \textbf{0.797} & \textbf{0.594} & \textbf{0.702} & \textbf{0.403} & \textbf{0.752} & \textbf{0.503} \\
\cline{1-14}
\multirow[c]{2}{*}{\LlamaThreeOne} & Edit distance & 0.662 & 0.324 & 0.666 & 0.331 & 0.580 & 0.160 & 0.622 & 0.244 & 0.533 & 0.066 & 0.595 & 0.190 \\
 & Repr. entanglement & \textbf{0.828} & \textbf{0.656} & \textbf{0.721} & \textbf{0.442} & \textbf{0.761} & \textbf{0.523} & \textbf{0.800} & \textbf{0.601} & \textbf{0.730} & \textbf{0.460} & \textbf{0.738} & \textbf{0.475} \\
\cline{1-14}
\multirow[c]{2}{*}{\OlmoTwo} & Edit distance & 0.669 & 0.338 & 0.634 & 0.267 & 0.563 & 0.127 & 0.674 & 0.348 & 0.567 & 0.133 & 0.658 & 0.317 \\
 & Repr. entanglement & \textbf{0.819} & \textbf{0.638} & \textbf{0.684} & \textbf{0.368} & \textbf{0.784} & \textbf{0.568} & \textbf{0.815} & \textbf{0.629} & \textbf{0.735} & \textbf{0.470} & \textbf{0.799} & \textbf{0.599} \\
\cline{1-14}
\multirow[c]{2}{*}{\TuluThree} & Edit distance & 0.647 & 0.293 & 0.611 & 0.223 & 0.579 & 0.157 & 0.629 & 0.258 & 0.526 & 0.052 & 0.590 & 0.180 \\
 & Repr. entanglement & \textbf{0.850} & \textbf{0.699} & \textbf{0.723} & \textbf{0.446} & \textbf{0.749} & \textbf{0.499} & \textbf{0.828} & \textbf{0.657} & \textbf{0.712} & \textbf{0.424} & \textbf{0.769} & \textbf{0.538} \\
\cline{1-14}
\multirow[c]{2}{*}{\TuluThreeOne} & Edit distance & 0.661 & 0.322 & 0.636 & 0.273 & 0.583 & 0.167 & 0.632 & 0.264 & 0.528 & 0.055 & 0.588 & 0.177 \\
 & Repr. entanglement & \textbf{0.845} & \textbf{0.690} & \textbf{0.704} & \textbf{0.408} & \textbf{0.764} & \textbf{0.527} & \textbf{0.827} & \textbf{0.654} & \textbf{0.716} & \textbf{0.431} & \textbf{0.763} & \textbf{0.527} \\
\bottomrule
\end{tabular}
}
\end{table*}

\textbf{Models and techniques.}
We evaluate on five language models, consistent with the scenario of Section~\ref{sec:threatmodel}: \LlamaThree~(8B, Instruct), \LlamaThreeOne~(8B, Instruct), \OlmoTwo~(13B, Instruct), \TuluThree~(8B) and \TuluThreeOne~(8B). We evaluate the following six techniques for our research questions.

\ding{172} AlphaEdit~\cite{Fang2025_AlphaEdit} projects editing perturbations onto the null space of preserved knowledge, reducing interference with existing information. It relies on accurate null-space estimation, and avoids explicit trade-offs between update and preservation, improving sequential editing stability.

\ding{173} RECT~\cite{gu2401model} mitigates degradation of general abilities by regularizing weight updates based on relative parameter changes. This reduces overfitting to edited facts, but may constrain the flexibility of large edits.

\ding{174} MEMIT~\cite{meng2022mass} enables large-scale insertion of factual knowledge by directly modifying key-value memory structures in transformer layers. While it supports thousands of edits with strong generalization, performance can degrade under extensive sequential editing.

\ding{175} PRUNE~\cite{ma2024perturbation} constrains the condition number of edited weight matrices to limit perturbation during sequential edits. This helps preserve general abilities, though it may restrict the magnitude of effective updates.

\ding{176} CoME~\cite{jung-etal-2025-come} integrates unlearning with editing by removing outdated knowledge and inserting new information. This reduces knowledge conflicts, but selective unlearning may risk discarding useful shared representations.

\ding{177} Locate-then-Unlearn (LTU)~\cite{liang2024locate} identifies task-specific neurons and selectively unlearns them to reduce task interference. This improves multi-task adaptation.

\subsection{Q1: Predictability of Tokenization Convergence}
\label{sec:rq1}

Not all noncanonical tokenizations behave identically, as even in unedited models, some diverge from the canonical response while others preserve it.
Whether this divergence is predictable from internal model signals is the central question for predictability of tokenization convergence. We construct sets of alternative tokenizations of the same prompt string, which differ in segmentation but decode to identical character sequences. For each input prompt $x$, we generate a canonical tokenization $v^*$ and a set of alternative tokenizations $\{u_i\}$. We define the canonical output $\hat{y}_{v^*}$ as the model's response under the canonical tokenization, and $\hat{y}_{u}$ as the response under an alternative tokenization.

\begin{figure}
    \centering
    \begin{subfigure}{0.49\linewidth}
        \includegraphics[width=\linewidth]{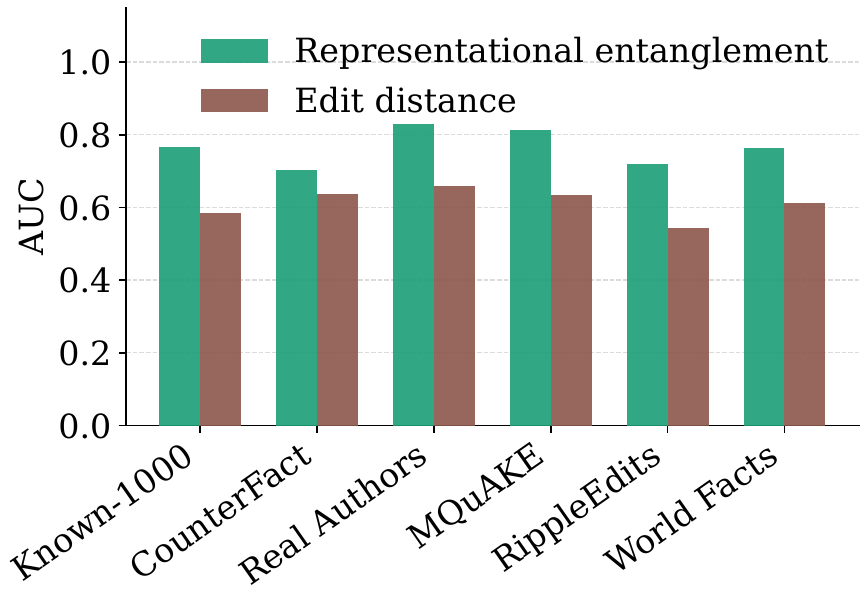}
        \caption{AUC}
        \label{fig:rq1_auc}
    \end{subfigure}
    \hfill
    \begin{subfigure}{0.49\linewidth}
        \includegraphics[width=\linewidth]{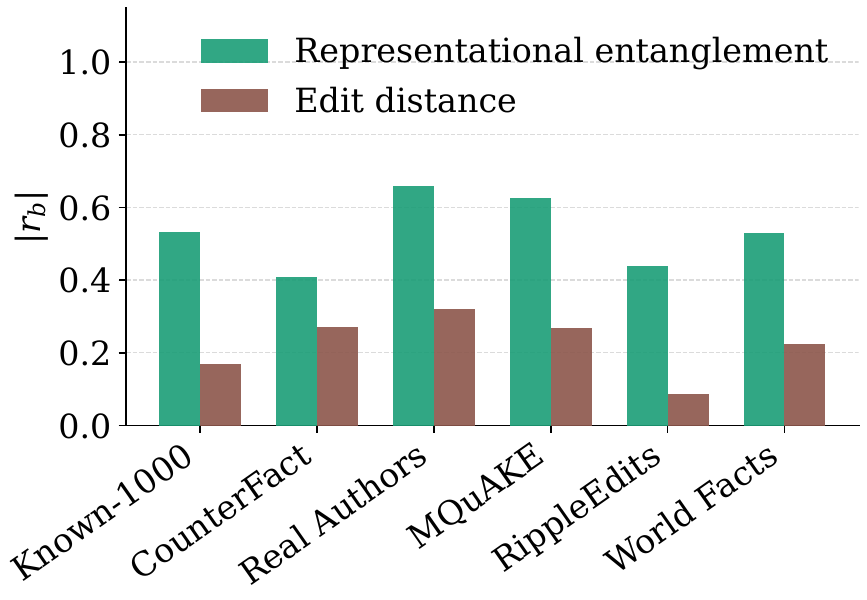}
        \caption{Rank-Biserial $|r_b|$}
        \label{fig:rq1_rb}
    \end{subfigure}
    \caption{Predictability of tokenization convergence averaged across all five models. Representational entanglement consistently outperforms edit distance as a predictor of whether a noncanonical tokenization will retrieve the same 
    factual object, achieving higher AUC 
    and effect size $|r_b|$ across all datasets.}
    \label{fig:rq1_signals}
\end{figure}

{\textbf{Definition (Convergence)}.}
A noncanonical tokenization $u$ is convergent if the 
object $o$ (from the unedited triple $(s,r,o)$) appears in the generated continuation $\hat{y}_u$. Formally,
\[
\text{convergence}(u) = \mathbf{1}\bigl[o \in \hat{y}_u\bigr],
\quad \text{given } o \in \hat{y}_{v^*}.
\]

For each sample prompt $p(s, r)$ in a given dataset, we generate 30 random noncanonical tokenizations, yielding a diverse coverage of the tokenization space $\mathcal{T}_{\mathcal{V}}(x)$.

{\textbf{Predictive Signals.}}
For each $u$, we compute the following signals as candidate predictors of convergence:
\begin{enumerate}
    \item Representational entanglement: representation-level similarity~(Section~\ref{subsec:clare}) between hidden states of $u$ and $v^*$.
    \item Edit distance: normalized Levenshtein distance between token sequences, serving as a lexical baseline.
\end{enumerate}

For evaluation, we assess predictability using (i) area under the ROC curve (AUC: measures discrimination between converging and non-converging tokenizations) and (ii) rank-biserial correlation ($|r_b|$: quantifies effect size between the two score distributions). Additional analyses including mean score analysis, calibration metrics and distributional visualizations are provided in \hyperref[app:rq1_means]{Appendix~F}, \hyperref[app:rq1_dist]{G}, and \hyperref[sec:calibration]{H} respectively.

Table~\ref{tab:predictability} presents AUC and $|r_b|$ across all model-dataset combinations. Representational entanglement achieves AUC between 0.684 and 0.850 and $|r_b|$ between 0.368 and 0.699, indicating it carries meaningful signal about whether a noncanonical tokenization will retrieve the same factual object as the canonical one. Edit distance achieves AUC between 0.526 and 0.674 and $|r_b|$ between 0.052 and 0.348, consistently lagging behind representational entanglement. Strikingly, even the worst representational entanglement result (AUC\,=\,0.684) exceeds the best edit distance result (AUC\,=\,0.674), suggesting a consistent separation between the two signals across settings. The gap is most pronounced on RippleEdits, where edit distance approaches random performance across all models while representational entanglement remains reliable. This highlights a fundamental limitation of lexical metrics, that surface-level token similarity is a poor proxy for internal computation. Fig.~\ref{fig:rq1_signals} summarizes AUC and $|r_b|$ averaged across all five models.

\begin{figure}
    \centering
    \begin{subfigure}[b]{0.49\linewidth}
        \includegraphics[width=\linewidth]{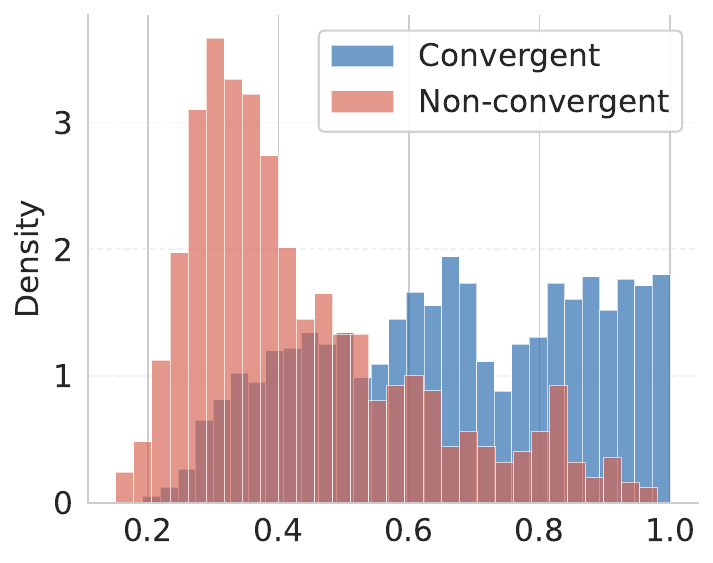}
        \caption{Repr. entanglement}
    \end{subfigure}
    \hfill
    \begin{subfigure}[b]{0.49\linewidth}
        \includegraphics[width=\linewidth]{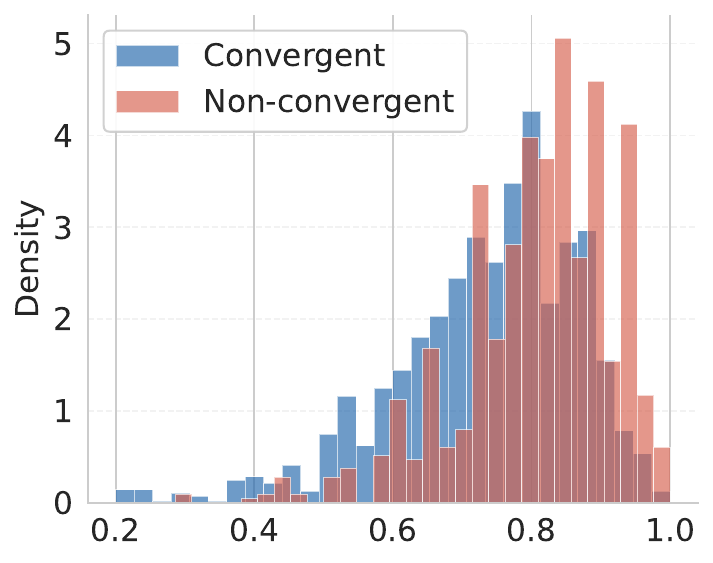}
        \caption{Edit distance}
    \end{subfigure}
    \centering
    \begin{subfigure}[b]{0.49\linewidth}
        \includegraphics[width=\linewidth]{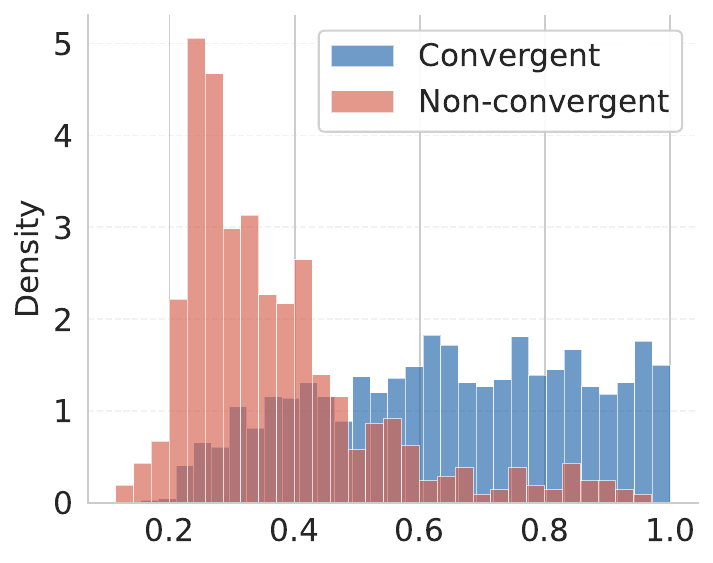}
        \caption{Repr. entanglement}
    \end{subfigure}
    \hfill
    \begin{subfigure}[b]{0.49\linewidth}
        \includegraphics[width=\linewidth]{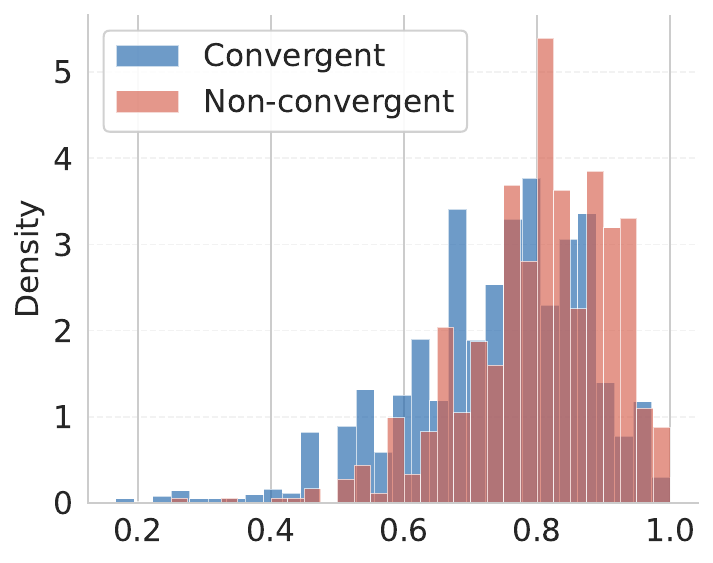}
        \caption{Edit distance}
    \end{subfigure}
    \caption{Score distributions for convergent and non-convergent noncanonical tokenizations for \LlamaThreeOne ((a), (b)) and \TuluThree ((c), (d)) for Real Authors. Representational entanglement shows a better separation between the two groups, while edit distance distributions overlap substantially.}
    \label{fig:rq1_histograms}
\end{figure}

\begin{table*}
\centering
\caption{Bypass rates (\%) for noncanonical tokenizations across models, techniques, and datasets. Each cell reports the percentage of noncanonical tokenizations that evade the edit and recover the pre-update response. Color intensity for each model reflects relative bypass rate, with darker shades indicating higher bypass rates.}
\label{tab:bypass_rates}
\resizebox{0.65\textwidth}{!}{
\begin{tabular}{llcccccc}
\toprule
 Model & Technique & Known-1000 & CounterFact & Real Authors & MQuAKE & RippleEdits & World Facts \\
\midrule
\multirow{6}{*}{\LlamaThree} & MEMIT & \cellcolor{darkgreen!14}{36.5} & \cellcolor{darkgreen!33}{50.3} & \cellcolor{darkgreen!13}{43.8} & \cellcolor{darkgreen!31}{46.9} & \cellcolor{darkgreen!19}{37.5} & \cellcolor{darkgreen!6}{27.0} \\
\multirow[t]{5}{*}{} & RECT & \cellcolor{darkgreen!19}{39.1} & \cellcolor{darkgreen!35}{51.6} & \cellcolor{darkgreen!16}{46.4} & \cellcolor{darkgreen!35}{49.7} & \cellcolor{darkgreen!26}{42.3} & \cellcolor{darkgreen!9}{30.1} \\
 & PRUNE & \cellcolor{darkgreen!12}{35.8} & \cellcolor{darkgreen!31}{49.3} & \cellcolor{darkgreen!11}{42.6} & \cellcolor{darkgreen!34}{49.5} & \cellcolor{darkgreen!19}{37.7} & \cellcolor{darkgreen!8}{28.7} \\
 & AlphaEdit & \cellcolor{darkgreen!8}{33.8} & \cellcolor{darkgreen!28}{47.3} & \cellcolor{darkgreen!11}{42.6} & \cellcolor{darkgreen!27}{43.1} & \cellcolor{darkgreen!16}{35.6} & \cellcolor{darkgreen!2}{23.2} \\
 & CoME & \cellcolor{darkgreen!6}{32.9} & \cellcolor{darkgreen!30}{48.3} & \cellcolor{darkgreen!13}{44.1} & \cellcolor{darkgreen!26}{42.7} & \cellcolor{darkgreen!12}{33.1} & \cellcolor{darkgreen!0}{21.1} \\
 & LTU & \cellcolor{darkgreen!5}{32.1} & \cellcolor{darkgreen!17}{39.1} & \cellcolor{darkgreen!32}{57.1} & \cellcolor{darkgreen!32}{47.5} & \cellcolor{darkgreen!2}{26.9} & \cellcolor{darkgreen!32}{51.8} \\
\cline{1-8}
\multirow{6}{*}{\LlamaThreeOne} & MEMIT & \cellcolor{blue!25}{42.0} & \cellcolor{blue!31}{49.1} & \cellcolor{blue!10}{42.1} & \cellcolor{blue!23}{39.7} & \cellcolor{blue!11}{32.3} & \cellcolor{blue!10}{30.9} \\
\multirow[t]{5}{*}{} & RECT & \cellcolor{blue!29}{43.9} & \cellcolor{blue!34}{51.5} & \cellcolor{blue!15}{45.2} & \cellcolor{blue!26}{42.0} & \cellcolor{blue!20}{38.3} & \cellcolor{blue!14}{34.5} \\
 & PRUNE & \cellcolor{blue!21}{39.9} & \cellcolor{blue!30}{48.3} & \cellcolor{blue!12}{43.4} & \cellcolor{blue!23}{39.5} & \cellcolor{blue!7}{30.2} & \cellcolor{blue!9}{29.7} \\
 & AlphaEdit & \cellcolor{blue!15}{36.9} & \cellcolor{blue!26}{45.8} & \cellcolor{blue!8}{40.5} & \cellcolor{blue!13}{31.2} & \cellcolor{blue!7}{30.2} & \cellcolor{blue!1}{22.1} \\
 & CoME & \cellcolor{blue!18}{38.3} & \cellcolor{blue!24}{44.3} & \cellcolor{blue!4}{37.8} & \cellcolor{blue!18}{35.2} & \cellcolor{blue!0}{25.3} & \cellcolor{blue!3}{23.9} \\
 & LTU & \cellcolor{blue!27}{42.8} & \cellcolor{blue!27}{46.0} & \cellcolor{blue!31}{56.4} & \cellcolor{blue!26}{42.2} & \cellcolor{blue!3}{27.8} & \cellcolor{blue!29}{48.6} \\
\cline{1-8}
\multirow{6}{*}{\OlmoTwo} & MEMIT & \cellcolor{red!13}{36.1} & \cellcolor{red!3}{29.0} & \cellcolor{red!5}{38.5} & \cellcolor{red!5}{24.6} & \cellcolor{red!16}{35.8} & \cellcolor{red!16}{36.6} \\
\multirow[t]{5}{*}{} & RECT & \cellcolor{red!17}{38.0} & \cellcolor{red!4}{30.0} & \cellcolor{red!5}{38.6} & \cellcolor{red!5}{24.2} & \cellcolor{red!17}{36.1} & \cellcolor{red!15}{35.7} \\
 & PRUNE & \cellcolor{red!13}{36.0} & \cellcolor{red!1}{27.8} & \cellcolor{red!4}{38.0} & \cellcolor{red!5}{24.0} & \cellcolor{red!15}{35.4} & \cellcolor{red!12}{32.9} \\
 & AlphaEdit & \cellcolor{red!0}{29.7} & \cellcolor{red!0}{26.7} & \cellcolor{red!3}{37.2} & \cellcolor{red!0}{19.7} & \cellcolor{red!10}{31.9} & \cellcolor{red!5}{25.6} \\
 & CoME & \cellcolor{red!8}{33.8} & \cellcolor{red!1}{27.7} & \cellcolor{red!3}{37.1} & \cellcolor{red!2}{21.7} & \cellcolor{red!15}{35.3} & \cellcolor{red!12}{32.3} \\
 & LTU & \cellcolor{red!18}{38.7} & \cellcolor{red!13}{36.5} & \cellcolor{red!4}{37.5} & \cellcolor{red!14}{31.9} & \cellcolor{red!22}{39.6} & \cellcolor{red!16}{36.9} \\
\cline{1-8}
\multirow{6}{*}{\TuluThree} & MEMIT & \cellcolor{darkorange!21}{39.8} & \cellcolor{darkorange!26}{45.8} & \cellcolor{darkorange!3}{37.1} & \cellcolor{darkorange!26}{42.2} & \cellcolor{darkorange!31}{45.5} & \cellcolor{darkorange!12}{32.3} \\
\multirow[t]{5}{*}{} & RECT & \cellcolor{darkorange!26}{42.1} & \cellcolor{darkorange!28}{47.3} & \cellcolor{darkorange!7}{39.8} & \cellcolor{darkorange!27}{43.1} & \cellcolor{darkorange!35}{47.4} & \cellcolor{darkorange!17}{37.4} \\
 & PRUNE & \cellcolor{darkorange!22}{40.3} & \cellcolor{darkorange!26}{45.5} & \cellcolor{darkorange!3}{36.8} & \cellcolor{darkorange!25}{41.4} & \cellcolor{darkorange!30}{44.8} & \cellcolor{darkorange!12}{32.4} \\
 & AlphaEdit & \cellcolor{darkorange!8}{33.8} & \cellcolor{darkorange!17}{39.2} & \cellcolor{darkorange!1}{35.7} & \cellcolor{darkorange!15}{33.0} & \cellcolor{darkorange!16}{35.7} & \cellcolor{darkorange!2}{23.5} \\
 & CoME & \cellcolor{darkorange!17}{38.0} & \cellcolor{darkorange!21}{42.1} & \cellcolor{darkorange!0}{34.6} & \cellcolor{darkorange!20}{37.2} & \cellcolor{darkorange!22}{39.8} & \cellcolor{darkorange!7}{28.0} \\
 & LTU & \cellcolor{darkorange!8}{33.8} & \cellcolor{darkorange!19}{40.5} & \cellcolor{darkorange!20}{48.7} & \cellcolor{darkorange!30}{46.0} & \cellcolor{darkorange!23}{40.1} & \cellcolor{darkorange!34}{53.9} \\
 \cline{1-8}
\multirow{6}{*}{\TuluThreeOne} & MEMIT & \cellcolor{violet!29}{43.5} & \cellcolor{violet!32}{49.7} & \cellcolor{violet!20}{49.0} & \cellcolor{violet!26}{42.8} & \cellcolor{violet!21}{38.6} & \cellcolor{violet!10}{30.5} \\
\multirow[t]{5}{*}{} & RECT & \cellcolor{violet!35}{46.3} & \cellcolor{violet!33}{50.5} & \cellcolor{violet!23}{51.1} & \cellcolor{violet!28}{44.2} & \cellcolor{violet!29}{43.7} & \cellcolor{violet!14}{35.0} \\
 & PRUNE & \cellcolor{violet!28}{43.2} & \cellcolor{violet!28}{47.2} & \cellcolor{violet!19}{47.8} & \cellcolor{violet!24}{40.6} & \cellcolor{violet!19}{37.6} & \cellcolor{violet!12}{32.3} \\
 & AlphaEdit & \cellcolor{violet!15}{37.0} & \cellcolor{violet!22}{43.0} & \cellcolor{violet!15}{45.3} & \cellcolor{violet!21}{38.4} & \cellcolor{violet!10}{32.2} & \cellcolor{violet!0}{20.8} \\
 & CoME & \cellcolor{violet!21}{40.1} & \cellcolor{violet!26}{45.5} & \cellcolor{violet!15}{45.5} & \cellcolor{violet!23}{39.7} & \cellcolor{violet!8}{30.7} & \cellcolor{violet!6}{26.6} \\
 & LTU & \cellcolor{violet!20}{39.6} & \cellcolor{violet!24}{44.0} & \cellcolor{violet!35}{58.9} & \cellcolor{violet!29}{45.1} & \cellcolor{violet!21}{39.0} & \cellcolor{violet!35}{54.0} \\
\cline{1-8}
\bottomrule
\end{tabular}
}
\end{table*}

Fig.~\ref{fig:rq1_histograms} shows the score distributions for convergent and non-convergent tokenizations for two representative models on the Real Authors dataset~\cite{maini2025tofu}. Convergent tokenizations consistently appear with higher representational entanglement values than non-convergent ones, indicating that internal representational similarity to the canonical tokenization is a reliable indicator of factual retrieval. No such ordering is observed for edit distance, where convergent and non-convergent tokenizations exhibit substantially overlapping score distributions, indicating that token-level surface similarity may carry little information about internal computational behavior. Representational entanglement directly reflects whether two tokenizations activate similar hidden representations, motivating its use as the routing criterion in \Toketive's adaptive sampler~(Section~\ref{sec:sampler}).

\subsection{Q2: Tokenization Invariance of Editing and Unlearning}
\label{sec:rq2}

Even when an edit or unlearning operation succeeds under the canonical tokenization, the update may fail to generalize to alternative tokenizations of the same input string. We measure this directly via the bypass rate (the fraction of noncanonical tokenizations that evade the update and recover the pre-update response). A tokenization-invariant update would yield a bypass rate of zero, as no alternative tokenization would be able to retrieve the suppressed knowledge. High bypass rates indicate that the update is superficial, aim to patch the canonical computational path while leaving residual factual associations intact along alternate trajectories.

For measuring the bypass rate, we utilise the same samples and their respective noncanonical tokenizations as in \textit{Q1}. For each fact, we apply the editing or unlearning technique. The bypass rate for a given model, technique, and dataset is:
\[
\text{bypass rate} =  
\frac{1}{K} \sum_{i=1}^{K} 
\mathbf{1}\bigl[o \in \hat{y}_{u_i}\bigr],
\]
where $o$ is the pre-update object, $u_i$ are the $K = 30$ noncanonical tokenizations of $p(s, r)$, and $\hat{y}_{u_i}$ is the model completion under $u_i$.

Table~\ref{tab:bypass_rates} reports bypass rates across all five models, six datasets, and six techniques. Several consistent patterns emerge. Bypass rates range from 19\% to 57\% with an average of 38.6\%, demonstrating that no technique achieves tokenization-invariant updates. Even AlphaEdit, the most robust technique as per these results, leaves a substantial fraction of noncanonical tokenizations capable of recovering the pre-update response. LTU (Locate-then-Unlearn) is the most vulnerable technique, as it achieves notably high bypass rates on Real Authors and World Facts across all models. Additional qualitative examples are provided in \hyperref[app:qualitative_examples]{Appendix~I}. A detailed breakdown of output distributions across all combinations of datasets, models, and techniques is provided in \hyperref[app:rq2]{Appendix~J}.

\begin{figure}
    \centering
    \includegraphics[width=\linewidth]{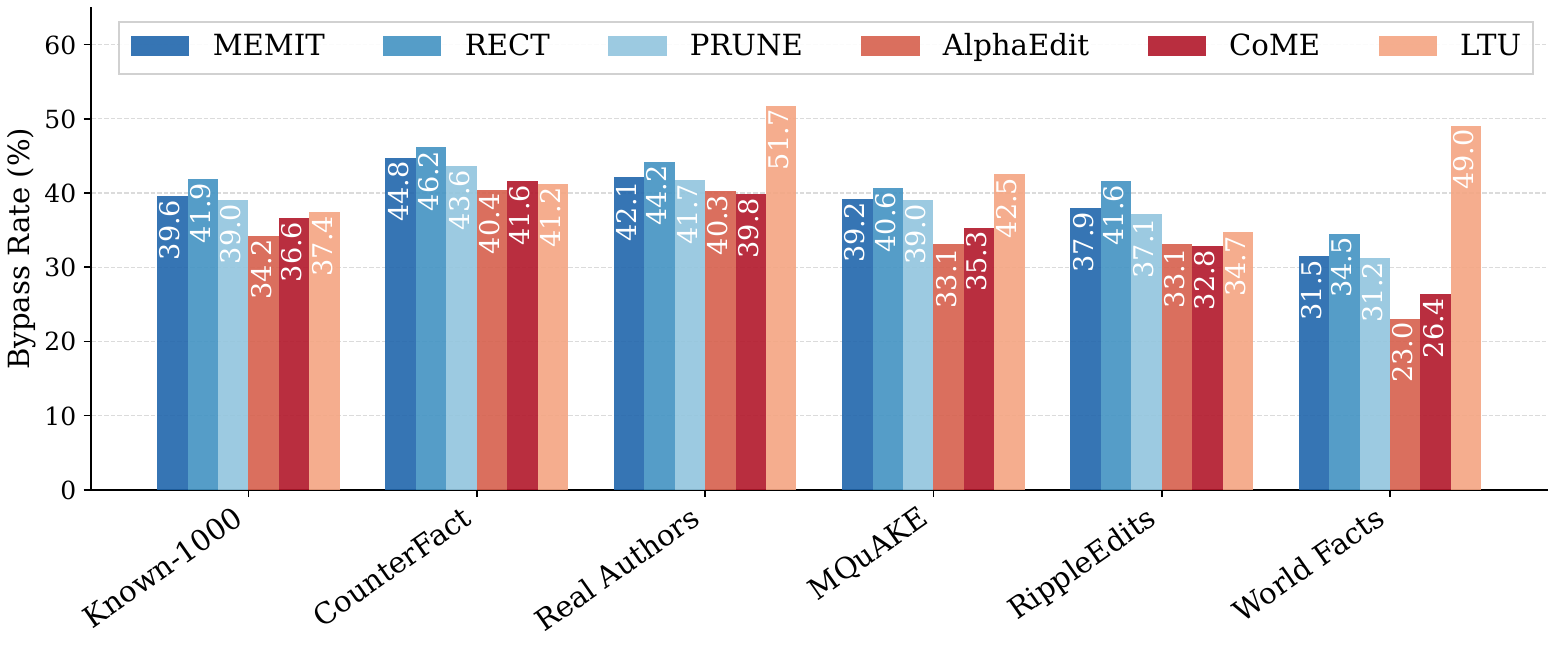}
    \caption{Bypass rates (\%) averaged across all five models per editing and unlearning technique and dataset. All techniques exhibit substantial bypass rates across all datasets, showing that they do not achieve tokenization-invariant updates.}
    \label{fig:rq2_bypass}
\end{figure}

Fig.~\ref{fig:rq2_analysis} illustrates the tokenization invariance failure for MEMIT on Real Authors. Before editing, 70.1\% of noncanonical tokenizations recover the old answer. After editing, this drops to 43.8\%, but a substantial fraction of tokenizations continue to bypass the edit, showing that MEMIT patches the canonical path without achieving tokenization-invariant suppression. Post-edit, 26.5\% of noncanonical tokenizations produce neither the old nor the new answer (Others). The entanglement histograms show that old-answer tokenizations cluster at higher similarity scores while new-answer tokenizations concentrate at lower scores, consistent with the band structure exploited by \Toketive~(Section~\ref{sec:attack}). Fig.~\ref{fig:rq2_bypass} summarizes bypass rates averaged across all five models per technique and dataset. These results show that current adversarial evaluations, which assess edit or unlearning success exclusively under canonical tokenizations, fail to capture an important failure mode of editing and unlearning techniques. A fact that appears successfully suppressed under standard metrics may remain recoverable by a tokenization-aware adversary.

\begin{table*}
\centering
\setlength{\tabcolsep}{2.5pt}
\caption{Edit detection performance of various methods averaged across editing and unlearning techniques for each model and dataset. True positive rate (TPR), false positive rate (FPR), Precision (Prec.), and F1 are reported as percentages. Color intensity within each model group reflects relative F1 magnitude. Best F1 scores are indicated in boldface.}
\label{tab:rq3_detection_main}
\resizebox{0.99\textwidth}{!}{
\begin{tabular}{c l cccc cccc cccc cccc cccc cccc}
\toprule
\multicolumn{2}{c}{} & \multicolumn{4}{c}{Known-1000} & \multicolumn{4}{c}{CounterFact} & \multicolumn{4}{c}{TOFU-Real Authors} & \multicolumn{4}{c}{MQuAKE} & \multicolumn{4}{c}{RippleEdits} & \multicolumn{4}{c}{TOFU-World Facts} \\
\cmidrule(lr){3-6} \cmidrule(lr){7-10} \cmidrule(lr){11-14} \cmidrule(lr){15-18} \cmidrule(lr){19-22} \cmidrule(lr){23-26}
Model & Method & TPR & FPR & Prec. & F1 & TPR & FPR & Prec. & F1 & TPR & FPR & Prec. & F1 & TPR & FPR & Prec. & F1 & TPR & FPR & Prec. & F1 & TPR & FPR & Prec. & F1 \\
\midrule
\multirow{4}{*}{\rotatebox{90}{\LlamaThree}} & Random-sampling & 40.9 & 0.0 & 100.0 & \cellcolor{darkgreen!6}{58.0} & 40.6 & 1.6 & 96.1 & \cellcolor{darkgreen!11}{57.0} & 24.4 & 0.0 & 100.0 & \cellcolor{darkgreen!2}{39.3} & 57.8 & 0.0 & 100.0 & \cellcolor{darkgreen!16}{73.2} & 34.1 & 26.8 & 56.0 & \cellcolor{darkgreen!4}{42.4} & 18.9 & 0.0 & 100.0 & \cellcolor{darkgreen!7}{31.8} \\
 & Edit Distance & 60.6 & 14.0 & 81.2 & \cellcolor{darkgreen!13}{69.4} & 62.2 & 11.9 & 83.9 & \cellcolor{darkgreen!17}{71.5} & 49.4 & 14.1 & 77.8 & \cellcolor{darkgreen!11}{60.5} & 70.0 & 18.9 & 78.7 & \cellcolor{darkgreen!16}{74.1} & 58.3 & 20.8 & 73.7 & \cellcolor{darkgreen!14}{65.1} & 29.4 & 12.8 & 69.7 & \cellcolor{darkgreen!10}{41.4} \\
 & FUMA-gradient & 100.0 & 100.0 & 50.0 & \cellcolor{darkgreen!11}{66.7} & 100.0 & 100.0 & 50.0 & \cellcolor{darkgreen!15}{66.7} & 96.1 & 90.0 & 51.6 & \cellcolor{darkgreen!14}{67.2} & 100.0 & 100.0 & 50.0 & \cellcolor{darkgreen!13}{66.7} & 100.0 & 100.0 & 50.0 & \cellcolor{darkgreen!15}{66.7} & 96.7 & 96.7 & 50.0 & \cellcolor{darkgreen!19}{65.9} \\
 & \Toketive (ours) & 81.9 & 15.5 & 84.0 & \cellcolor{darkgreen!20}{\textbf{82.9}} & 87.2 & 15.2 & 85.1 & \cellcolor{darkgreen!24}{\textbf{86.1}} & 82.8 & 9.5 & 89.8 & \cellcolor{darkgreen!22}{\textbf{86.1}} & 95.0 & 14.3 & 86.9 & \cellcolor{darkgreen!25}{\textbf{90.8}} & 90.9 & 15.3 & 85.6 & \cellcolor{darkgreen!24}{\textbf{88.2}} & 78.7 & 13.2 & 85.7 & \cellcolor{darkgreen!25}{\textbf{82.1}} \\
\cmidrule(lr){1-26}
\multirow{4}{*}{\rotatebox{90}{\LlamaThreeOne}} & Random-sampling & 37.8 & 0.0 & 100.0 & \cellcolor{blue!4}{54.8} & 34.4 & 0.5 & 98.4 & \cellcolor{blue!8}{51.0} & 24.4 & 10.0 & 71.0 & \cellcolor{blue!1}{36.4} & 53.3 & 5.0 & 91.4 & \cellcolor{blue!13}{67.4} & 32.2 & 5.0 & 86.6 & \cellcolor{blue!6}{47.0} & 18.9 & 2.8 & 87.3 & \cellcolor{blue!7}{31.1} \\
 & Edit Distance & 61.8 & 14.5 & 81.0 & \cellcolor{blue!13}{70.1} & 65.0 & 14.0 & 82.3 & \cellcolor{blue!18}{72.6} & 55.0 & 14.5 & 79.2 & \cellcolor{blue!13}{64.9} & 72.7 & 28.3 & 72.0 & \cellcolor{blue!15}{72.3} & 62.0 & 21.7 & 74.1 & \cellcolor{blue!15}{67.5} & 31.7 & 16.1 & 66.3 & \cellcolor{blue!11}{42.9} \\
 & FUMA-gradient & 99.5 & 96.7 & 50.7 & \cellcolor{blue!11}{67.2} & 100.0 & 100.0 & 50.0 & \cellcolor{blue!15}{66.7} & 100.0 & 100.0 & 50.0 & \cellcolor{blue!14}{66.7} & 96.1 & 96.7 & 49.9 & \cellcolor{blue!12}{65.7} & 100.0 & 100.0 & 50.0 & \cellcolor{blue!15}{66.7} & 100.0 & 100.0 & 50.0 & \cellcolor{blue!19}{66.7} \\
 & \Toketive (ours) & 83.0 & 12.6 & 86.8 & \cellcolor{blue!21}{\textbf{84.8}} & 89.4 & 13.5 & 86.9 & \cellcolor{blue!25}{\textbf{88.1}} & 83.9 & 15.0 & 84.8 & \cellcolor{blue!22}{\textbf{84.4}} & 93.9 & 15.2 & 86.1 & \cellcolor{blue!24}{\textbf{89.8}} & 83.6 & 15.8 & 84.1 & \cellcolor{blue!22}{\textbf{83.9}} & 76.3 & 12.6 & 85.8 & \cellcolor{blue!24}{\textbf{80.8}} \\
\cmidrule(lr){1-26}
\multirow{4}{*}{\rotatebox{90}{\OlmoTwo}} & Random-sampling & 65.0 & 0.0 & 100.0 & \cellcolor{red!18}{78.8} & 45.6 & 0.0 & 100.0 & \cellcolor{red!13}{62.6} & 82.2 & 1.1 & 98.7 & \cellcolor{red!24}{89.7} & 55.6 & 0.0 & 100.0 & \cellcolor{red!15}{71.4} & 78.9 & 2.2 & 97.2 & \cellcolor{red!23}{87.1} & 48.4 & 0.0 & 100.0 & \cellcolor{red!19}{65.2} \\
 & Edit Distance & 84.5 & 10.6 & 88.9 & \cellcolor{red!22}{86.6} & 71.8 & 13.3 & 84.3 & \cellcolor{red!20}{77.6} & 80.8 & 10.1 & 88.9 & \cellcolor{red!22}{84.6} & 73.1 & 15.0 & 83.0 & \cellcolor{red!18}{77.7} & 69.5 & 26.1 & 72.7 & \cellcolor{red!16}{71.0} & 51.7 & 7.9 & 86.8 & \cellcolor{red!18}{64.8} \\
 & FUMA-gradient & 50.0 & 50.0 & 50.0 & \cellcolor{red!2}{50.0} & 23.3 & 23.3 & 50.0 & \cellcolor{red!0}{31.8} & 100.0 & 100.0 & 50.0 & \cellcolor{red!14}{66.7} & 33.9 & 33.3 & 50.4 & \cellcolor{red!0}{40.5} & 23.3 & 23.3 & 50.0 & \cellcolor{red!0}{31.8} & 5.6 & 3.3 & 62.8 & \cellcolor{red!0}{10.2} \\
 & \Toketive (ours) & 92.8 & 12.8 & 87.9 & \cellcolor{red!25}{\textbf{90.3}} & 89.1 & 17.1 & 83.9 & \cellcolor{red!24}{\textbf{86.4}} & 93.6 & 12.2 & 88.5 & \cellcolor{red!25}{\textbf{90.9}} & 93.9 & 16.6 & 85.0 & \cellcolor{red!24}{\textbf{89.2}} & 95.4 & 17.3 & 84.6 & \cellcolor{red!25}{\textbf{89.7}} & 61.1 & 0.0 & 100.0 & \cellcolor{red!22}{\textbf{75.9}} \\
\cmidrule(lr){1-26}
\multirow{4}{*}{\rotatebox{90}{\TuluThree}} & Random-sampling & 33.9 & 0.0 & 100.0 & \cellcolor{darkorange!2}{50.6} & 42.8 & 1.1 & 97.5 & \cellcolor{darkorange!12}{59.5} & 19.4 & 0.0 & 100.0 & \cellcolor{darkorange!0}{32.5} & 41.2 & 0.0 & 100.0 & \cellcolor{darkorange!8}{58.4} & 43.3 & 4.4 & 90.7 & \cellcolor{darkorange!11}{58.6} & 15.0 & 0.0 & 100.0 & \cellcolor{darkorange!5}{26.1} \\
 & Edit Distance & 69.5 & 6.8 & 91.1 & \cellcolor{darkorange!18}{78.8} & 66.1 & 11.7 & 85.0 & \cellcolor{darkorange!18}{74.3} & 50.0 & 6.5 & 88.4 & \cellcolor{darkorange!13}{63.9} & 68.8 & 12.2 & 84.9 & \cellcolor{darkorange!17}{76.0} & 63.4 & 15.0 & 80.8 & \cellcolor{darkorange!16}{71.1} & 22.8 & 7.5 & 75.3 & \cellcolor{darkorange!8}{35.0} \\
 & FUMA-gradient & 100.0 & 100.0 & 50.0 & \cellcolor{darkorange!11}{66.7} & 100.0 & 100.0 & 50.0 & \cellcolor{darkorange!15}{66.7} & 88.9 & 83.3 & 51.6 & \cellcolor{darkorange!14}{65.3} & 99.5 & 100.0 & 49.9 & \cellcolor{darkorange!12}{66.4} & 96.7 & 96.7 & 50.0 & \cellcolor{darkorange!14}{65.9} & 98.3 & 96.7 & 50.4 & \cellcolor{darkorange!19}{\textbf{66.7}} \\
 & \Toketive (ours) & 87.2 & 8.3 & 91.3 & \cellcolor{darkorange!24}{\textbf{89.2}} & 88.2 & 15.5 & 85.1 & \cellcolor{darkorange!24}{\textbf{86.6}} & 79.7 & 8.9 & 90.0 & \cellcolor{darkorange!22}{\textbf{84.5}} & 92.3 & 13.5 & 87.3 & \cellcolor{darkorange!24}{\textbf{89.7}} & 81.6 & 10.7 & 88.4 & \cellcolor{darkorange!22}{\textbf{84.9}} & 40.6 & 0.5 & 98.7 & \cellcolor{darkorange!16}{57.5} \\
\cmidrule(lr){1-26}
\multirow{4}{*}{\rotatebox{90}{\TuluThreeOne}} & Random-sampling & 30.0 & 0.0 & 100.0 & \cellcolor{violet!0}{46.2} & 41.1 & 1.1 & 97.4 & \cellcolor{violet!11}{57.8} & 21.6 & 8.7 & 71.4 & \cellcolor{violet!0}{33.2} & 32.2 & 0.0 & 100.0 & \cellcolor{violet!4}{48.7} & 40.6 & 1.6 & 96.1 & \cellcolor{violet!10}{57.0} & 15.0 & 0.0 & 100.0 & \cellcolor{violet!5}{26.1} \\
 & Edit Distance & 61.6 & 4.4 & 93.3 & \cellcolor{violet!15}{74.2} & 66.5 & 11.7 & 85.1 & \cellcolor{violet!19}{74.7} & 52.2 & 7.1 & 88.1 & \cellcolor{violet!14}{65.6} & 61.1 & 10.0 & 85.9 & \cellcolor{violet!15}{71.4} & 60.2 & 19.4 & 75.6 & \cellcolor{violet!15}{67.0} & 21.7 & 5.4 & 80.0 & \cellcolor{violet!8}{34.1} \\
 & FUMA-gradient & 100.0 & 96.7 & 50.8 & \cellcolor{violet!12}{67.4} & 100.0 & 100.0 & 50.0 & \cellcolor{violet!15}{66.7} & 89.5 & 83.3 & 51.8 & \cellcolor{violet!14}{65.6} & 96.1 & 96.7 & 49.9 & \cellcolor{violet!12}{65.7} & 92.2 & 90.0 & 50.6 & \cellcolor{violet!14}{65.4} & 97.2 & 93.3 & 51.0 & \cellcolor{violet!19}{\textbf{66.9}} \\
 & \Toketive (ours) & 87.2 & 7.2 & 92.4 & \cellcolor{violet!24}{\textbf{89.7}} & 83.5 & 16.2 & 83.8 & \cellcolor{violet!23}{\textbf{83.7}} & 81.1 & 8.3 & 90.7 & \cellcolor{violet!22}{\textbf{85.6}} & 86.3 & 14.8 & 85.4 & \cellcolor{violet!22}{\textbf{85.8}} & 75.5 & 11.5 & 86.8 & \cellcolor{violet!21}{\textbf{80.8}} & 45.0 & 5.6 & 89.0 & \cellcolor{violet!17}{59.8} \\
\cmidrule(lr){1-26}
\bottomrule
\end{tabular}
}
\end{table*}

\subsection{Q3: Adversarial Efficacy of \Toketive}
\label{subsec:RQ3}

We evaluate whether \Toketive can (i) detect whether a fact has been modified, and (ii) reconstruct the original response, using only the post-edit model $f_{\theta'}$ and target prompt $p(s,r)$. To ensure reliable evaluation, we restrict our analysis to instances where edits are successful as per Efficacy Score~\cite{meng2022locating}, i.e., where $P$(new fact) > $P$(old fact). For edit detection, we compare \Toketive against three baselines:

    \ding{172} Random sampling: uniform random noncanonical tokenizations with no band selection.
    
    \ding{173} Edit distance: band routing based on normalized Levenshtein distance between token sequences, replacing representational entanglement as the filtering signal.
    
    \ding{174} FUMA-gradient: inspired by FUMA~\cite{deepak2025identifying}, we adapt a gradient-based detection signal that requires only the post-edit model $f_{\theta'}$, without access to the original model or any candidate set. For each prompt $p(s,r)$, we compute the L2 norm of the gradient of the token-average loss with respect to all model parameters as a continuous edit detection score.

For pre-modification response reconstruction, we compare \Toketive against the first two baselines, as gradient norms in FUMA-gradient provide no information about the content of the pre-edit response. We report top-1 and top-5 accuracy, where a fact is considered recovered if the pre-edit object $o$ appears among the top-$k$ most frequent bypass responses.

\textbf{Construction of negative samples.} Each fact is evaluated twice using the same detector: once on the edited (or unlearned) model and once on the corresponding unedited base model. For the unedited model, the detector measures whether noncanonical tokenizations continue to reproduce the canonical response, exactly as in the edited setting. Consequently, false positives arise solely from the intrinsic variability introduced by noncanonical tokenizations, rather than from separately constructed or mismatched negative samples.

\textbf{Edit Detection.}
Table~\ref{tab:rq3_detection_main} reports TPR, FPR, Precision, and F1-scores for edit detection averaged across editing and unlearning techniques. \Toketive consistently achieves the highest F1 across all models and datasets, with an average of 84.2\%, substantially outperforming all baselines. The second highest F1-score is achieved by Edit distance, with a value of 66.7\%. Random sampling achieves near-zero FPR and perfect precision in many settings, but at the cost of very low TPR, indicating that uniformly sampled noncanonical tokenizations rarely fall in the detection band and therefore fail to reliably flag edited facts. Edit distance improves recall over random sampling but consistently underperforms \Toketive across all settings, indicating that lexical token distance is an insufficient proxy for internal computational divergence. The FUMA-adapted gradient baseline achieves near-perfect TPR but also near-perfect FPR on most model-dataset combinations, collapsing to a precision of approximately 50\% and an average F1-score of 60.8\%. This indicates the gradient norm signal is non-discriminative for such a strict setting where the original unedited models are not available.

\begin{figure}[t]
    \centering
    \begin{subfigure}{0.8\linewidth}
        \includegraphics[width=\linewidth]{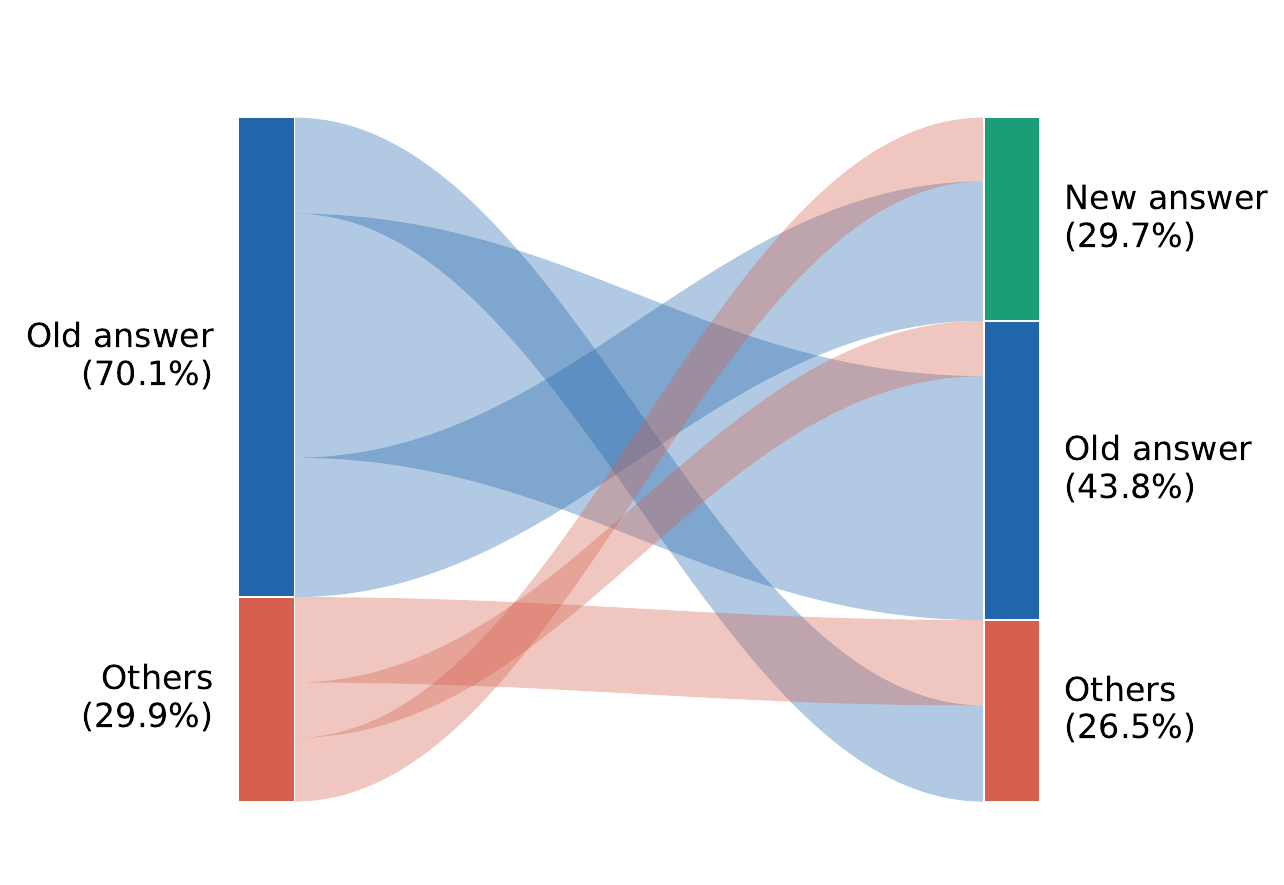}
        \caption{Response redistribution before and after editing.}
        \label{fig:rq2_sankey}
    \end{subfigure}
    \begin{subfigure}[b]{0.49\linewidth}
        \includegraphics[width=\linewidth]{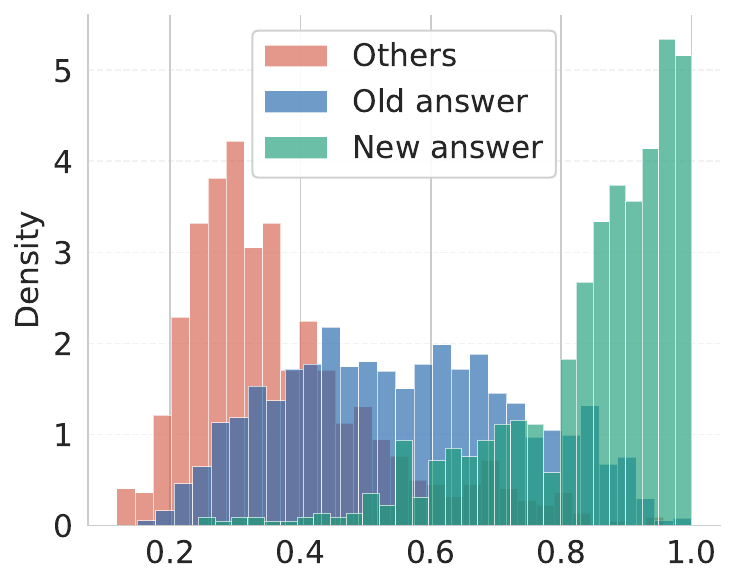}
        \caption{Repr. entanglement score distributions.}
        \label{fig:rq2_entanglement}
    \end{subfigure}
    \hfill
    \begin{subfigure}[b]{0.49\linewidth}
        \includegraphics[width=\linewidth]{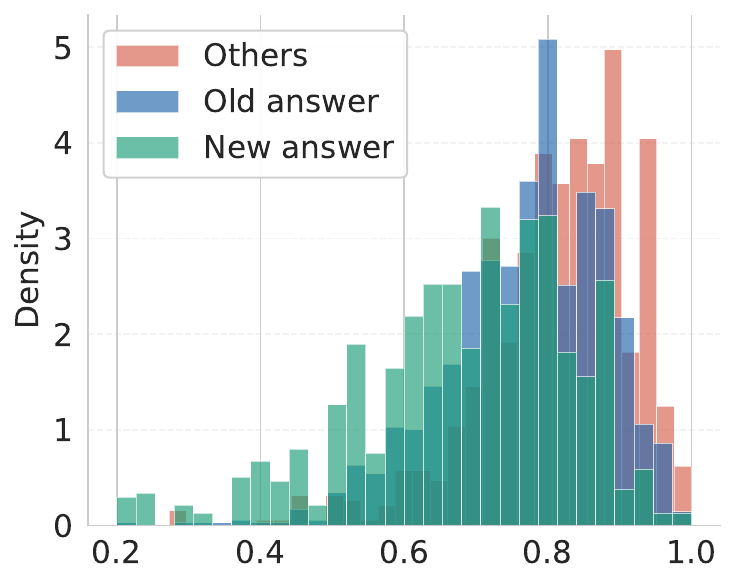}
        \caption{Edit distance score distributions.}
        \label{fig:rq2_edit_distance}
    \end{subfigure}
    \caption{Analysis of tokenization invariance for MEMIT on the Real Authors dataset (\LlamaThree). (a) Sankey diagram showing the redistribution of noncanonical tokenization responses before and after editing. (b,c) Score distributions of noncanonical tokenizations grouped by their post-edit response category.}
    \label{fig:rq2_analysis}
\end{figure}

\textbf{Pre-Edit Reconstruction.}
Table~\ref{tab:rq3_reconstruction_main} reports top-1 and top-5 accuracy averaged across techniques. \Toketive achieves an average top-1 accuracy of 60.9\% and top-5 accuracy of 74.5\%, consistently outperforming the other baselines by at least 21.7\%. Random sampling achieves substantially lower accuracy across 
all models and datasets with an average Top-1 and Top-5 accuracy of 30.7\% and 36.9\% respectively, demonstrating that unstructured tokenization search is insufficient for reliable reconstruction. Bypass continuations without band filtering are too diverse to concentrate on the pre-update object. Edit distance improves over random sampling and in some \OlmoTwo settings approaches over \Toketive, but underperforms across the majority of model-dataset combinations. The gap between \Toketive and edit distance is most pronounced on MQuAKE and CounterFact, showing that representational entanglement provides a stronger signal for identifying bypass tokenizations that encode factual associations.

\begin{table*}
\centering
\setlength{\tabcolsep}{3pt}
\caption{Pre-edit response reconstruction accuracy of various methods averaged across editing and unlearning techniques for each model and dataset. Top-1 Acc.~(\%) reports the fraction of facts for which the pre-edit object $o$ is the most frequent bypass continuation. Top-5 Acc.~(\%) reports the fraction of facts for which $o$ appears among the five most frequent bypass continuations. Color intensity within each model group reflects relative Top-k accuracy magnitude.}
\label{tab:rq3_reconstruction_main}
\resizebox{0.80\textwidth}{!}{
\begin{tabular}{l l cc cc cc cc cc cc}
\toprule
\multicolumn{2}{c}{} & \multicolumn{2}{c}{Known-1000} & \multicolumn{2}{c}{CounterFact} & \multicolumn{2}{c}{TOFU-Real Authors} & \multicolumn{2}{c}{MQuAKE} & \multicolumn{2}{c}{RippleEdits} & \multicolumn{2}{c}{TOFU-World Facts} \\
\cmidrule(lr){3-4} \cmidrule(lr){5-6} \cmidrule(lr){7-8} \cmidrule(lr){9-10} \cmidrule(lr){11-12} \cmidrule(lr){13-14}
Model & Method & Top-1\% & Top-5\% & Top-1\% & Top-5\% & Top-1\% & Top-5\% & Top-1\% & Top-5\% & Top-1\% & Top-5\% & Top-1\% & Top-5\% \\
\midrule
\multirow{3}{*}{\LlamaThree} & Random-sampling & \cellcolor{darkgreen!4}{33.7} & \cellcolor{darkgreen!4}{39.8} & \cellcolor{darkgreen!4}{36.7} & \cellcolor{darkgreen!2}{40.0} & \cellcolor{darkgreen!0}{21.1} & \cellcolor{darkgreen!0}{21.1} & \cellcolor{darkgreen!10}{41.7} & \cellcolor{darkgreen!10}{54.4} & \cellcolor{darkgreen!2}{20.2} & \cellcolor{darkgreen!2}{29.7} & \cellcolor{darkgreen!0}{13.9} & \cellcolor{darkgreen!1}{18.3} \\
 & Edit Distance & \cellcolor{darkgreen!13}{52.0} & \cellcolor{darkgreen!12}{59.9} & \cellcolor{darkgreen!16}{57.2} & \cellcolor{darkgreen!13}{61.7} & \cellcolor{darkgreen!9}{44.6} & \cellcolor{darkgreen!10}{49.4} & \cellcolor{darkgreen!16}{52.4} & \cellcolor{darkgreen!16}{68.5} & \cellcolor{darkgreen!11}{37.4} & \cellcolor{darkgreen!14}{58.0} & \cellcolor{darkgreen!4}{21.1} & \cellcolor{darkgreen!6}{29.2} \\
 & \Toketive (ours) & \cellcolor{darkgreen!15}{55.4} & \cellcolor{darkgreen!15}{66.8} & \cellcolor{darkgreen!24}{70.5} & \cellcolor{darkgreen!22}{79.5} & \cellcolor{darkgreen!17}{68.4} & \cellcolor{darkgreen!18}{73.4} & \cellcolor{darkgreen!25}{68.3} & \cellcolor{darkgreen!25}{86.7} & \cellcolor{darkgreen!18}{52.6} & \cellcolor{darkgreen!19}{71.5} & \cellcolor{darkgreen!21}{48.9} & \cellcolor{darkgreen!25}{68.5} \\
\cmidrule(lr){1-14}
\multirow{3}{*}{\LlamaThreeOne} & Random-sampling & \cellcolor{blue!0}{25.0} & \cellcolor{blue!3}{36.6} & \cellcolor{blue!0}{30.0} & \cellcolor{blue!0}{34.4} & \cellcolor{blue!1}{22.2} & \cellcolor{blue!1}{22.8} & \cellcolor{blue!2}{28.9} & \cellcolor{blue!7}{47.8} & \cellcolor{blue!0}{15.0} & \cellcolor{blue!0}{25.0} & \cellcolor{blue!1}{16.1} & \cellcolor{blue!1}{17.2} \\
 & Edit Distance & \cellcolor{blue!10}{45.0} & \cellcolor{blue!13}{61.6} & \cellcolor{blue!16}{57.8} & \cellcolor{blue!14}{63.9} & \cellcolor{blue!11}{50.0} & \cellcolor{blue!11}{54.5} & \cellcolor{blue!14}{49.4} & \cellcolor{blue!17}{70.6} & \cellcolor{blue!8}{32.2} & \cellcolor{blue!15}{62.0} & \cellcolor{blue!6}{23.9} & \cellcolor{blue!7}{31.7} \\
 & \Toketive (ours) & \cellcolor{blue!17}{60.0} & \cellcolor{blue!19}{76.6} & \cellcolor{blue!25}{72.2} & \cellcolor{blue!25}{85.0} & \cellcolor{blue!18}{68.9} & \cellcolor{blue!17}{70.0} & \cellcolor{blue!20}{59.7} & \cellcolor{blue!23}{83.5} & \cellcolor{blue!14}{44.5} & \cellcolor{blue!15}{61.7} & \cellcolor{blue!20}{47.1} & \cellcolor{blue!22}{63.7} \\
\cmidrule(lr){1-14}
\multirow{3}{*}{\OlmoTwo} & Random-sampling & \cellcolor{red!13}{51.6} & \cellcolor{red!14}{63.3} & \cellcolor{red!5}{37.8} & \cellcolor{red!5}{45.0} & \cellcolor{red!21}{77.8} & \cellcolor{red!21}{82.2} & \cellcolor{red!7}{37.2} & \cellcolor{red!9}{52.2} & \cellcolor{red!23}{62.2} & \cellcolor{red!19}{70.5} & \cellcolor{red!20}{47.8} & \cellcolor{red!15}{48.4} \\
 & Edit Distance & \cellcolor{red!19}{64.4} & \cellcolor{red!22}{82.4} & \cellcolor{red!15}{55.4} & \cellcolor{red!18}{71.3} & \cellcolor{red!20}{75.6} & \cellcolor{red!20}{80.4} & \cellcolor{red!12}{45.5} & \cellcolor{red!18}{71.6} & \cellcolor{red!23}{61.9} & \cellcolor{red!18}{69.0} & \cellcolor{red!20}{48.4} & \cellcolor{red!17}{51.7} \\
 & \Toketive (ours) & \cellcolor{red!23}{71.1} & \cellcolor{red!25}{89.5} & \cellcolor{red!17}{60.0} & \cellcolor{red!22}{80.0} & \cellcolor{red!25}{87.8} & \cellcolor{red!25}{92.7} & \cellcolor{red!21}{62.1} & \cellcolor{red!24}{85.2} & \cellcolor{red!25}{64.9} & \cellcolor{red!25}{83.3} & \cellcolor{red!25}{55.0} & \cellcolor{red!20}{59.4} \\
\cmidrule(lr){1-14}
\multirow{3}{*}{\TuluThree} & Random-sampling & \cellcolor{darkorange!2}{29.4} & \cellcolor{darkorange!2}{33.9} & \cellcolor{darkorange!1}{32.2} & \cellcolor{darkorange!1}{38.3} & \cellcolor{darkorange!0}{19.4} & \cellcolor{darkorange!0}{19.4} & \cellcolor{darkorange!4}{32.2} & \cellcolor{darkorange!3}{40.5} & \cellcolor{darkorange!8}{32.8} & \cellcolor{darkorange!5}{38.4} & \cellcolor{darkorange!0}{13.9} & \cellcolor{darkorange!0}{15.0} \\
 & Edit Distance & \cellcolor{darkorange!17}{60.0} & \cellcolor{darkorange!15}{67.1} & \cellcolor{darkorange!11}{49.4} & \cellcolor{darkorange!14}{64.5} & \cellcolor{darkorange!10}{49.4} & \cellcolor{darkorange!10}{50.0} & \cellcolor{darkorange!14}{48.9} & \cellcolor{darkorange!16}{67.6} & \cellcolor{darkorange!16}{47.3} & \cellcolor{darkorange!16}{62.6} & \cellcolor{darkorange!3}{19.4} & \cellcolor{darkorange!3}{22.8} \\
 & \Toketive (ours) & \cellcolor{darkorange!25}{75.0} & \cellcolor{darkorange!22}{82.8} & \cellcolor{darkorange!19}{63.5} & \cellcolor{darkorange!21}{77.7} & \cellcolor{darkorange!16}{63.3} & \cellcolor{darkorange!18}{72.7} & \cellcolor{darkorange!19}{58.9} & \cellcolor{darkorange!22}{81.9} & \cellcolor{darkorange!20}{55.4} & \cellcolor{darkorange!19}{70.9} & \cellcolor{darkorange!14}{38.0} & \cellcolor{darkorange!11}{39.1} \\
\cmidrule(lr){1-14}
\multirow{3}{*}{\TuluThreeOne} & Random-sampling & \cellcolor{violet!0}{23.9} & \cellcolor{violet!0}{28.9} & \cellcolor{violet!0}{28.9} & \cellcolor{violet!1}{37.2} & \cellcolor{violet!0}{19.4} & \cellcolor{violet!0}{19.4} & \cellcolor{violet!0}{23.9} & \cellcolor{violet!0}{32.2} & \cellcolor{violet!9}{33.4} & \cellcolor{violet!5}{38.9} & \cellcolor{violet!0}{13.9} & \cellcolor{violet!0}{14.4} \\
 & Edit Distance & \cellcolor{violet!11}{47.2} & \cellcolor{violet!13}{60.9} & \cellcolor{violet!11}{49.3} & \cellcolor{violet!15}{65.4} & \cellcolor{violet!11}{50.0} & \cellcolor{violet!11}{52.2} & \cellcolor{violet!13}{47.7} & \cellcolor{violet!12}{60.0} & \cellcolor{violet!15}{45.0} & \cellcolor{violet!14}{59.6} & \cellcolor{violet!2}{18.3} & \cellcolor{violet!3}{21.7} \\
 & \Toketive (ours) & \cellcolor{violet!19}{64.5} & \cellcolor{violet!21}{81.4} & \cellcolor{violet!21}{65.4} & \cellcolor{violet!21}{77.1} & \cellcolor{violet!18}{69.5} & \cellcolor{violet!19}{76.5} & \cellcolor{violet!21}{61.8} & \cellcolor{violet!22}{80.5} & \cellcolor{violet!18}{51.6} & \cellcolor{violet!19}{71.4} & \cellcolor{violet!17}{42.9} & \cellcolor{violet!14}{45.0} \\
\cmidrule(lr){1-14}
\bottomrule
\end{tabular}
}
\end{table*}

\section{Related Work}
\label{sec:related}

\textbf{Model Editing and Machine Unlearning.}
Model editing aims to update specific factual associations in pretrained language models without retraining from scratch. These techniques typically follow a locate-then-edit paradigm, using causal tracing to identify feed-forward MLP layers responsible for storing a target fact, and targeted weight updates modify the stored association~\cite{meng2022locating, meng2022mass, Fang2025_AlphaEdit, gu2401model, ma2024perturbation, zhang2025locatethenedit, zhao-etal-2025-fedleke, lirethinking, zhao2025fleke, li2026rethinking, pan2025precise, wang2026same, wang2025microedit, jiang2025anyedit, lyu2026evoedit, gupta2025lifelong, park2025context, shen2026genrecedit}. There are also parameter-preserving editing techniques which augment the models with external modules rather than modifying weights directly~\cite{qi2025incontext, huang2023transformer, hartvigsen2023aging, mitchell2022memory, zheng2023can, qi2025incontext}. Machine unlearning pursues the complementary goal of removing the influence of specific training data such that the model behaves as if it was never trained on that information~\cite{sinha2024unstar, chundawat2023can, ji2024reversing, liu2025rethinking, ren2025general, jia2024soul}. Recent work has unified editing and unlearning under the same framework, treating unlearning as a special case of targeted editing~\cite{li2026editing, liang2024locate, jung-etal-2025-come}. Alternative approaches to knowledge modification exist, but operate at different levels of granularity and cost. For instance, retraining from scratch provides the strongest guarantees but is infeasible at deployment scale~\cite{hu2024duty}. Fine-tuning offers a more practical option, but it can also become prohibitively expensive~\cite{Fang2025_AlphaEdit}. Retrieval-augmented generation (RAG) shifts knowledge control outside the model by utilizing external data, but recent works~\cite{an-etal-2025-rag, cho2024typos} show it can introduce new safety failures.

\textbf{Evaluation of Editing and Unlearning.}
Various benchmarks for model editing exist, assessing the reliability, generalization and locality of model editing techniques~\cite{meng2022locating, akyurek-etal-2023-dune, hoelscher2023detecting}. MQuAKE~\cite{zhong2023mquake} and ThinkEval~\cite{baserthinkeval} extend these by evaluating indirect knowledge leakage through multi-hop and multi-step querying respectively. A well-documented failure mode of editing and unlearning is the ripple-effect propagation of edits to other facts~\cite{cohen2024rippleedits, qin2024does, baser2026clare, rinberg2025ripplebench}. For unlearning, TOFU~\cite{maini2025tofu} WMDP~\cite{pmlr-v235-li24bc}, RWKU~\cite{jin2024rwku} have become standard evaluation suites, evaluating unlearning efficacy, model utility response quality, and hallucination avoidance.

\textbf{Security Implications of Editing and Unlearning.}
While the impact of editing and unlearning on model alignment and safety has received significant attention~\cite{10.1145/3698590, spohn2025alignthenunlearn, jiang-etal-2024-learning, shi-etal-2025-safety, wang-etal-2025-delman, zhang2025from, betley2025emergent}, the security implications of these techniques are important as well. Membership inference attacks have been applied to unlearned models to assess whether forget-set membership can be inferred post-hoc~\cite{naderloui2025rectifying, hayes2025inexact}. FUMA~\cite{deepak2025identifying} shows that gradient signals can identify what was unlearned in LLMs. Training-data extraction attacks show that PII removal through editing is often incomplete~\cite{cheng2025effective}. Reverse engineering of weight updates recovers editing subjects from weight differentials~\cite{sun2026reverse}, and TULA-DR~\cite{du2025textual} reconstructs unlearned text through constrained optimization over weight differences. 
However, all of these evaluate editing and unlearning exclusively under the canonical tokenization, implicitly assuming that canonical evaluation is sufficient to characterize model behavior. Recent data-centric metrics~\cite{lu2026waterdrum} improve the measurement of unlearning, but don't account for tokenization-aware adversaries as well. While UnUnlearning~\cite{shumailov2024ununlearning} shows that unlearned knowledge can be reintroduced via in-context learning, we show that even under standard querying, tokenization-aware adversaries can recover suppressed knowledge.

\textbf{Adversarial tokenization.}
Modern tokenizers map each string to a canonical token sequence, but the same string can be represented by many noncanonical tokenizations using the model vocabulary~\cite{wang2024tokenization}. Recent works show that models retain substantial semantic competence under such tokenizations, and that this flexibility can be weaponized for safety bypasses without changing the visible text of a request~\cite{zheng2025broken, geh2025adversarial}.

\section{Discussion: Toward tokenization-robust editing and unlearning}
Our results reveal that editing and unlearning techniques lack tokenization invariance, enabling adversaries to recover suppressed knowledge through alternative tokenizations. This raises the question: {\em how can such vulnerabilities be mitigated?}

A natural direction is to enforce tokenization-robust updates, ensuring that modifications generalize across multiple valid tokenizations of a given input. We investigate this direction through an adaptive unlearning experiment as presented in \hyperref[app:adaptive_editing]{Appendix~A}, where MEMIT iteratively incorporates tokenizations that bypass the preceding update. Our results show that this strategy can substantially reduce the bypass rate, from 0.32 for MEMIT to 0.04 after four adaptive iterations. However, this increased robustness comes at a substantial cost to locality. The corresponding ripple effect increases from $1.0\times$ to $8.6\times$, indicating that the broader updates significantly alter neighboring facts beyond the target being unlearned. These results reveal a robustness-locality trade-off, where greater resistance to tokenization-based attacks comes at the cost of broader changes to neighboring knowledge.

More generally, enforcing tokenization invariance is challenging because the space of valid tokenizations grows with input length, and different tokenizations can induce distinct internal computational trajectories. Explicitly covering this space through repeated editing may therefore require increasingly broad parameter modifications, as observed in our adaptive experiment. This suggests that simply extending localized updates to additional tokenizations may not provide a scalable solution while preserving the locality that motivates targeted editing and unlearning~\cite{cohen2024rippleedits, qin2024does, baser2026clare}.

Finally, architectural or representational approaches, such as learning tokenization-agnostic representations, may offer a more principled solution, but would require rethinking how subword tokenization interacts with model internals.

Overall, our adaptive-editing results suggest that improving tokenization robustness through increasingly broad parameter updates is possible, but comes at the expense of the locality that motivates targeted editing and unlearning. Developing more principled approaches that provide tokenization-robust knowledge modification without substantially expanding collateral effects remains an open problem.

\section{Conclusion}

We introduced \Toketive, a reference-free adversarial attack that exposes  tokenization dependence as a previously overlooked vulnerability in model editing and machine unlearning. By treating tokenization as a side channel rather than a fixed preprocessing step, \Toketive demonstrates that state-of-the-art editing and unlearning techniques produce superficial updates that fail to generalize across alternative tokenizations of the same input string. Across five models, six datasets, and six editing and unlearning techniques, we observe an average bypass rate of 38.6\%, indicating that suppressed knowledge remains broadly accessible to a tokenization-aware adversary. \Toketive unifies edit detection and pre-edit response reconstruction without requiring access to the original model, shadow models, or auxiliary classifiers, thereby avoiding assumptions that limit the practicality of prior approaches in their settings. Empirically, \Toketive achieves strong performance in both detection and reconstruction, attaining an F1 score of 84.2\% and a top-5 reconstruction accuracy of 74.5\%, while outperforming the strongest baselines by 26.2\% and 21.6\%, respectively. Our results show that current editing and unlearning techniques lack robustness to tokenization-level variation, allowing tokenization-aware adversaries like \Toketive to recover suppressed knowledge from open-weight models, highlighting a gap between their intended behavior and their actual robustness in practice.

\section*{Ethics Considerations}

This work studies security vulnerabilities in model editing and machine unlearning techniques for LLMs. In particular, we demonstrate that tokenization-aware adversaries can recover knowledge that was intended to be removed or modified. While such capabilities could be misused to extract sensitive or suppressed information, they also address a critical concern, i.e., current editing and unlearning techniques may provide a false sense of security in open-weight models.

Our goal is to advance understanding of the limitations of post-release knowledge control and to inform the design of more robust and reliable defenses. We carefully balance risks and benefits throughout this study. First, our threat model is limited to querying publicly available models using alternative tokenizations of the same input string; we do not study or enable attacks that require access to private training data, proprietary systems, or hidden interfaces. Second, all of our experiments are conducted on publicly available models and benchmark datasets, and no sensitive real-world data are involved.

We further mitigate potential harm by focusing on structural vulnerabilities (specifically, tokenization-induced variations in internal representations) rather than providing deployment-ready tools for large-scale exploitation. Our evaluation highlights fundamental limitations of existing techniques and emphasizes the need for improved evaluation protocols and defenses, rather than prescribing immediate attack deployment. We will include a clear usage statement in our repository, indicating that our work is intended solely for research and evaluation purposes, and not for misuse or unauthorized data extraction.

Consistent with prior work in machine learning security, we view this research as contributing to defensive understanding by exposing blind spots in current approaches. We follow established ethical guidelines for security research, scope our claims conservatively, and avoid releasing artifacts that could directly facilitate misuse without appropriate safeguards.

Our research involves no human subjects and did not require institutional review board approval. We explicitly consider potential downstream misuse, scope our claims to realistic open-weight settings, and frame our contribution as exposing vulnerabilities to inform the design of more robust defenses rather than facilitating attacks.


\bibliographystyle{IEEEtran}
\bibliography{sample-base}


\section*{Open Science and Responsible Disclosure}

This paper supports reproducibility while accounting for safety and responsible disclosure concerns.

\noindent\textbf{Artifacts.}
We provide an anonymous artifact repository containing:
(i) the source code for \Toketive,
(ii) scripts and configuration files for all experiments (RQ-1, RQ-2, and RQ-3),
(iii) documentation describing the experimental workflow and environment.

\noindent\textbf{Access During Review.}
All artifacts are hosted in an anonymous repository accessible to the program committee
during double-blind review. The repository contains no identifying information and is
available at: \url{https://anonymous.4open.science/r/Toketive}.

\noindent\textbf{Reproducibility.}
All core claims in the paper can be independently evaluated using the provided artifacts. To support reproducibility, we provide code, detailed documentation and dataset information.

\noindent\textbf{Models and Datasets.} 
All five models evaluated in this work are publicly available open-weight models accessible via HuggingFace. All datasets are publicly available benchmarks.

\section*{Acknowledgments}

This paper was edited for grammar, spelling, and light style polishing using ChatGPT and Claude.


\begin{table*}
\caption{Expected Calibration Error (ECE) of representational entanglement and edit distance as predictors of tokenization convergence across models and datasets. Lower ECE indicates better calibration. Bold indicates the better signal.}
\label{tab:ece}
\centering
\begin{tabular}{llrrrrrr}
\toprule
Model & Method & Authors & CounterFact & Known-1000 & MQuAKE & RippleEdits & WorldFacts \\
\midrule
\multirow[t]{2}{*}{\LlamaThree} & Repr. Entanglement & \textbf{0.174} & \textbf{0.102} & \textbf{0.120} & \textbf{0.089} & \textbf{0.118} & \textbf{0.112} \\
 & Edit Distance & 0.398 & 0.283 & 0.215 & 0.336 & 0.235 & 0.424 \\
\cline{1-8}
\multirow[t]{2}{*}{\LlamaThreeOne} & Repr. Entanglement & \textbf{0.168} & \textbf{0.121} & \textbf{0.108} & \textbf{0.043} & \textbf{0.133} & \textbf{0.133} \\
 & Edit Distance & 0.404 & 0.317 & 0.324 & 0.266 & 0.254 & 0.461 \\
\cline{1-8}
\multirow[t]{2}{*}{\OlmoTwo} & Repr. Entanglement & \textbf{0.180} & \textbf{0.097} & \textbf{0.090} & \textbf{0.037} & \textbf{0.097} & \textbf{0.174} \\
 & Edit Distance & 0.298 & 0.197 & 0.225 & 0.156 & 0.200 & 0.442 \\
\cline{1-8}
\multirow[t]{2}{*}{\TuluThreeOne} & Repr. Entanglement & \textbf{0.233} & \textbf{0.104} & \textbf{0.145} & \textbf{0.051} & \textbf{0.061} & \textbf{0.216} \\
 & Edit Distance & 0.466 & 0.288 & 0.375 & 0.266 & 0.346 & 0.534 \\
\cline{1-8}
\multirow[t]{2}{*}{\TuluThree} & Repr. Entanglement & \textbf{0.228} & \textbf{0.132} & \textbf{0.133} & \textbf{0.077} & \textbf{0.077} & \textbf{0.192} \\
 & Edit Distance & 0.439 & 0.333 & 0.348 & 0.306 & 0.356 & 0.483 \\
\cline{1-8}
\bottomrule
\end{tabular}
\end{table*}

\begin{table*}
\caption{Maximum Calibration Error (MCE) of representational entanglement and edit distance as predictors of tokenization convergence across models and datasets. MCE measures the worst-case calibration gap across score bins. Lower MCE indicates better worst-case calibration. Bold indicates the better signal per model-dataset pair.}
\label{tab:mce}
\centering
\begin{tabular}{llrrrrrr}
\toprule
 & Dataset & Authors & CounterFact & Known-1000 & MQuAKE & RippleEdits & WorldFacts \\
Model & Method &  &  &  &  &  &  \\
\midrule
\multirow[t]{2}{*}{Llama3-8B} & Repr. Entanglement & \textbf{0.293} & \textbf{0.214} & \textbf{0.269} & \textbf{0.207} & \textbf{0.202} & \textbf{0.193} \\
 & Edit Distance & 0.473 & 0.380 & 0.373 & 0.427 & 0.441 & 0.498 \\
\cline{1-8}
\multirow[t]{2}{*}{Llama3.1-8B} & Repr. Entanglement & \textbf{0.273} & \textbf{0.340} & \textbf{0.163} & \textbf{0.093} & \textbf{0.190} & \textbf{0.280} \\
 & Edit Distance & 0.446 & 0.380 & 0.504 & 0.381 & 0.546 & 0.577 \\
\cline{1-8}
\multirow[t]{2}{*}{OLMo2-13B} & Repr. Entanglement & 0.436 & 0.263 & \textbf{0.166} & \textbf{0.116} & \textbf{0.203} & \textbf{0.252} \\
 & Edit Distance & \textbf{0.330} & \textbf{0.250} & 0.392 & 0.201 & 0.394 & 0.513 \\
\cline{1-8}
\multirow[t]{2}{*}{Tulu3.1-8B} & Repr. Entanglement & \textbf{0.366} & \textbf{0.258} & \textbf{0.281} & \textbf{0.161} & \textbf{0.095} & \textbf{0.363} \\
 & Edit Distance & 0.537 & 0.380 & 0.539 & 0.379 & 0.578 & 0.681 \\
\cline{1-8}
\multirow[t]{2}{*}{Tulu3-8B} & Repr. Entanglement & \textbf{0.384} & 0.393 & \textbf{0.295} & \textbf{0.220} & \textbf{0.102} & \textbf{0.329} \\
 & Edit Distance & 0.495 & \textbf{0.393} & 0.510 & 0.363 & 0.605 & 0.620 \\
\cline{1-8}
\bottomrule
\end{tabular}
\end{table*}

\section*{Appendix A\\Adaptive Editing and Unlearning against Tokenization-Based Bypasses}
\label{app:adaptive_editing}

To evaluate whether tokenization-aware editing and unlearning can mitigate the vulnerability exposed by \Toketive, we additionally consider an adaptive variant of MEMIT. Unlike the standard setting, where the editing procedure operates only on the canonical tokenization of the target prompt, the adaptive procedure explicitly incorporates bypass tokenizations discovered after each editing iteration. This experiment evaluates whether a developer who is aware of the tokenization-based side channel, and can eliminate the residual factual associations exposed by alternative tokenizations.

\begin{algorithm}
\caption{Adaptive MEMIT Unlearning}
\label{alg:adaptive_memit}
\begin{algorithmic}[1]
\Require Pretrained model $M$, canonical prompt $p_c$, original response $o_{\mathrm{old}}$, target response $o_{\mathrm{new}}$, maximum iterations $T$
\Ensure Updated model $M$
\State $M \gets \mathrm{MEMIT}(M, p_c \rightarrow o_{\mathrm{new}})$
\For{$t=2$ to $T$}
    \State $S \gets \mathrm{GenerateNoncanonicalTokenizations}(p_c)$
    \State $b \gets \emptyset$
    \For{each $p \in S$}
        \If{$M(p)$ contains $o_{\mathrm{old}}$}
            \State $b \gets p$
            \State \textbf{break}
        \EndIf
    \EndFor
    \If{$b = \emptyset$}
        \State \textbf{break}
    \EndIf
    \State $M \gets \mathrm{MEMIT}(M, b \rightarrow o_{\mathrm{new}})$
\EndFor
\State \Return $M$
\end{algorithmic}
\end{algorithm}

\begin{figure}
\centering

    \includegraphics[width=\linewidth]{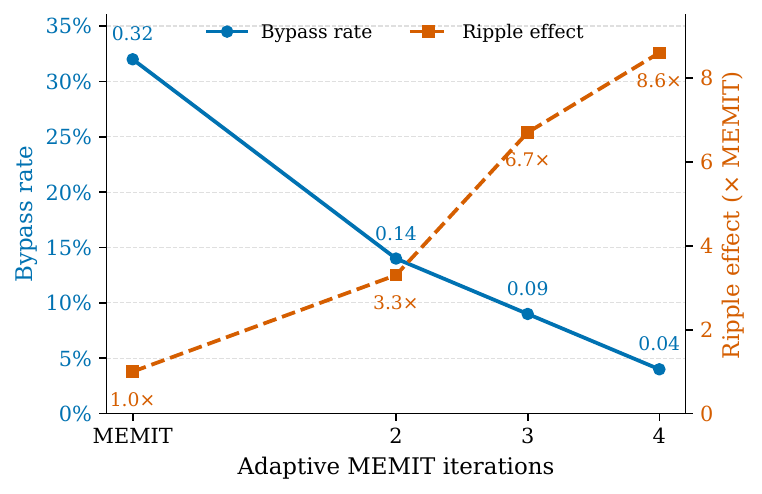}
    \caption{Increasing the number of adaptive MEMIT iterations progressively reduces the bypass rate, but substantially increases the ripple effect on neighboring knowledge. Bypass rate and ripple effect are measured relative to standard MEMIT.}
    \label{fig:tradeoff}
\end{figure}

\textbf{Experimental Setup.} We conduct the experiment on \LlamaThree using the same highly entangled edit-control fact pairs described in \hyperref[app:finetuning]{Appendix~B}. Specifically, we use 1,000 edit-control pairs with representational entanglement greater than 0.9. We evaluate standard MEMIT against adaptive MEMIT under the same unlearning objective and measure both the tokenization bypass rate and the resulting ripple effects.

Starting from the canonical prompt, standard MEMIT is first applied to suppress the target fact. We then use \Toketive to search for noncanonical tokenizations that continue to recover the original response. When a bypass tokenization is identified, it is added as an additional unlearning target and MEMIT is applied again. This process is repeated for up to four adaptive iterations. Thus, each iteration expands the set of tokenizations explicitly covered by the unlearning procedure.

\textbf{Adaptive Editing.} Algorithm~\ref{alg:adaptive_memit} summarizes the adaptive procedure. Given a canonical prompt $p_c$, its original response $o_{\mathrm{old}}$, and the desired unlearning response $o_{\mathrm{new}}$, we first perform standard MEMIT unlearning on $p_c$. At each subsequent iteration, we generate a set of noncanonical tokenizations of $p_c$ and identify a tokenization whose model response still contains $o_{\mathrm{old}}$. If such a bypass is found, we apply MEMIT to that tokenization using the same unlearning target $o_{\mathrm{new}}$. The procedure terminates when no bypass is found or when the maximum number of adaptive iterations is reached.

\begin{table}[t]
\centering
\caption{Effect of adaptive MEMIT on tokenization-based bypasses and collateral changes. Ripple effects are normalized to standard MEMIT.}
\label{tab:adaptive_memit}
\begin{tabular}{lcc}
\toprule
Method & Bypass Rate $\downarrow$ & Ripple Effect $\downarrow$ \\
\midrule
MEMIT & 0.32 & $1.0\times$ \\
Adaptive MEMIT (2 iter.) & 0.14 & $3.3\times$ \\
Adaptive MEMIT (3 iter.) & 0.09 & $6.7\times$ \\
Adaptive MEMIT (4 iter.) & 0.04 & $8.6\times$ \\
\bottomrule
\end{tabular}
\end{table}

\textbf{Results.}
As shown in Fig.~\ref{fig:tradeoff}, Adaptive MEMIT substantially reduces the effectiveness of tokenization-based bypasses. The bypass rate decreases from 0.32 for standard MEMIT to 0.14 after two adaptive iterations, 0.09 after three iterations, and 0.04 after four iterations. Thus, explicitly incorporating discovered bypass tokenizations into subsequent unlearning updates can substantially improve robustness against tokenization-aware adversaries. However, this increased robustness comes at a progressively larger cost to locality. The ripple effect increases from $1.0\times$ for standard MEMIT to $3.3\times$, $6.7\times$, and $8.6\times$ after two, three, and four adaptive iterations, respectively. Each additional adaptive iteration therefore reduces the remaining bypasses while simultaneously increasing collateral changes to neighboring knowledge.

These results reveal a clear trade-off between tokenization robustness and update locality. While a developer can the mitigate tokenization-based side-channel by explicitly incorporating discovered bypass tokenizations into the unlearning procedure, doing so progressively expands the region of the model affected by the update. In particular, reducing the bypass rate from 0.32 to 0.04 requires an $8.6\times$ increase in ripple effects relative to standard MEMIT. This suggests that tokenization-aware defenses can improve robustness, but may undermine the locality that motivates targeted editing and unlearning.

\section*{Appendix B\\Fine-Tuning for Targeted Unlearning and Editing}
\label{app:finetuning}

We evaluate vanilla fine-tuning when used for targeted unlearning to examine whether its broader parameter updates provide greater robustness to tokenization-based bypasses. We evaluate fine-tuning in terms of both tokenization bypass rate and collateral changes to neighboring knowledge. We conduct this experiment on \LlamaThree using 1,000 edit-control fact pairs selected from the factual entanglement artifacts of~\cite{baserthinkeval}. We restrict the evaluation to highly entangled pairs, with representational entanglement greater than 0.9, providing a stringent setting in which modifications to the target fact are likely to affect neighboring knowledge. We measure the bypass rate using the same protocol as our main experiments and quantify collateral changes using the original-answer log-probability shift following~\cite{baserthinkeval}.

\begin{table}
\centering
\caption{Comparison of MEMIT and vanilla fine-tuning for targeted unlearning. Bypass rate measures the fraction of alternative tokenizations that recover the pre-update response. Ripple effect is normalized to MEMIT.}
\label{tab:finetuning}
\begin{tabular}{lcc}
\toprule
Method & Bypass Rate $\downarrow$ & Ripple Effect $\downarrow$ \\
\midrule
MEMIT & 0.32 & $1.0\times$ \\
Fine-tuning & 0.08 & $5.2\times$ \\
\bottomrule
\end{tabular}
\end{table}

\textbf{Results.}
Fine-tuning substantially reduces the bypass rate, from 0.32 for MEMIT to 0.08, indicating that its broader parameter updates provide greater robustness to alternative tokenizations. However, this increased robustness comes at a substantial cost to locality. Fine-tuning produces a $5.2\times$ larger ripple effect than MEMIT, indicating that suppressing a broader set of tokenization variants also causes substantially greater changes to neighboring knowledge. Thus, although fine-tuning provides stronger robustness against tokenization-based bypasses, its collateral effects make it less suitable for targeted unlearning and editing, where localized modification is a primary objective. The fine-tuning configuration and evaluation protocol are provided in Table~\ref{tab:ft-config}.

\begin{table}
\centering
\caption{Configuration of the fine-tuning baseline and the shared evaluation
protocol. All values are the defaults in the released code; no per-fact or
per-model tuning was performed.}
\label{tab:ft-config}
\small
\begin{tabular}{@{}ll@{}}
\toprule
\textbf{Parameter} & \textbf{Value} \\
\midrule
\multicolumn{2}{@{}l}{\textit{Fine-tuning optimization}}\\
Trainable parameters      & All (full model, no freezing) \\
Optimizer                 & SGD, momentum $=0$ \\
Learning rate             & $5\times10^{-5}$ \\
Maximum steps             & 25 \\
Early-stopping loss       & $5\times10^{-2}$ \\
Batch size                & 1 (single prompt--target pair) \\
Objective                 & Causal-LM cross-entropy \\
Model precision           & fp32 \\
\midrule
\multicolumn{2}{@{}l}{\textit{Protocol}}\\
Model                     & \LlamaThree \\
Editing baseline          & MEMIT \\
\bottomrule
\end{tabular}
\end{table}

\section*{Appendix C\\Tokenization Sampling Procedure}
\label{app:sampler}

To facilitate reproducibility, we provide the complete tokenization sampling procedure used by \Toketive. \Toketive does not rely on a formal theory of knowledge localization, nor do we claim that representational entanglement or residual computational paths constitute the definitive mechanism by which factual knowledge is stored in LLMs. Algorithm~\ref{alg:sampler} specifies the candidate-generation process, adaptive controller update, tokenizer constraints, duplicate handling, and attempt budget used to generate alternative tokenizations. The sampler operates without tokenizer-specific heuristics: candidate segments are retained only when they satisfy the tokenizer round-trip constraint $\tau^{-1}(\tau(c))=c$, ensuring that each sampled tokenization decodes exactly to the original input. The adaptive controller updates $\alpha$ using the cosine similarities of the most recent $B$ scored proposals, while all scored proposals, including those outside the target similarity band, are retained in the history. We use a maximum of $nM$ sampling attempts for $n$ requested tokenizations, with $M=300$ attempts per sample.

The sampler uses a fixed budget of $M=300$ proposals per requested sample, yielding a maximum of $nM$ attempts for a quota of $n$: $n_{\mathrm{det}}=20$ for detection and $n_{\mathrm{rec}}=50$ for reconstruction. The loop terminates as soon as the quota is met, so these values represent worst-case rather than typical sampling costs. If the budget is exhausted before the quota is reached, we fill the shortfall using validated but out-of-band proposals in ascending order of their distance to the target band. This maintains a fixed sample size and ensures comparable detection and reconstruction statistics across facts. Thus, the band boundaries serve as sampling targets rather than hard guarantees; if the fallback pool is also insufficient, the procedure returns fewer than $n$ samples. Algorithm~\ref{alg:sampler} (Lines 3 and 15) specifies both behaviors.

\begin{algorithm*}
\small
\caption{Adaptive tokenization sampler used by \Toketive. $\tau,\tau^{-1}$ are the tokenizer's encode/decode maps with special tokens disabled on all calls; $v^*=\tau(p)$ is the canonical tokenization; $\textsc{Hidden}$ extracts the subject-final hidden state at probe layer $\mathcal{L}$. Constants: maximum segment length $W=19$ characters, attempt budget $M=300$ per requested sample, step $\eta=0.08$, buffer $B=10$, $\alpha_0=0.5$. The round-trip test $\tau^{-1}(\tau(c))=c$ is what keeps segments byte- and Unicode-safe without tokenizer-specific rules; $D$ records every scored proposal, including rejected candidates, so the controller adapts using the full observed proposal history. Here $s$ denotes the subject whose final-token position is probed, and $h^{*}=\textsc{Hidden}(f_{\theta'},v^{*},\mathcal{L},s)$ the canonical reference representation against which candidate tokenizations are scored.}

\label{alg:sampler}
\begin{algorithmic}[1]

\Procedure{Sample}{$f_{\theta'},p,s,h^*,[\beta_l,\beta_h),n$}
  \State $\mathcal{C},\mathcal{F},\mathcal{S}\gets\emptyset$;\;
         $D\gets[\,]$;\; $\alpha\gets\alpha_0$;\; $k\gets0$
  \While{$|\mathcal{C}|<n$ \textbf{and} $k<nM$}
    \State $k\gets k+1$;\;
           $\alpha\gets$\Call{UpdateAlpha}{$\alpha,D,\beta_l,\beta_h$};\;
           $u\gets$\Call{GenTok}{$p,\alpha$}
    \State \textbf{if} $u=\bot$ \textbf{or} $u\in\mathcal{S}$ \textbf{then
           continue} \Comment{invalid or duplicate}
    \State $\mathcal{S}\gets\mathcal{S}\cup\{u\}$;\;
           $e\gets\cos\!\big(\Call{Hidden}{f_{\theta'},u,\mathcal{L},s},h^*\big)$;\;
           append $e$ to $D$
    \If{$\beta_l\le e<\beta_h$}
      \State $\mathcal{C}\gets\mathcal{C}\cup\{u\}$
    \Else
      \State $d\gets\min(|e-\beta_l|,\,|e-\beta_h|)$;\;
             $\mathcal{F}\gets\mathcal{F}\cup\{(u,d)\}$
             \Comment{distance to band}
    \EndIf
  \EndWhile
  \State \textbf{if} $|\mathcal{C}|<n$ \textbf{then} move $(u,d)$ from
         $\mathcal{F}$ into $\mathcal{C}$ in ascending $d$ until
         $|\mathcal{C}|=n$ or $\mathcal{F}=\emptyset$
  \State \Return $\mathcal{C}$
\EndProcedure

\Statex
\Procedure{GenTok}{$p,\alpha$}
  \State $m\gets1$ \textbf{if} $\alpha<0.3$; $2$ \textbf{if} $\alpha<0.5$;
         $3$ \textbf{if} $\alpha<0.7$; \textbf{else} $4$
         \Comment{max tokens per segment}
  \State $u\gets[\,]$;\; $i\gets0$
  \While{$i<|p|$}
    \State $\mathcal{A}\gets\big\{(\tau(c_{ij}),j)\;:\;c_{ij}=p[i{:}j],\;
           0<j-i\le W,\; j\le|p|,\;|\tau(c_{ij})|\le m,\;
           \tau^{-1}(\tau(c_{ij}))=c_{ij}\big\}$
    \If{$\mathcal{A}=\emptyset$}
      \State $u\gets u\Vert\tau(p[i])$;\; $i\gets i+1$
             \Comment{single-character fallback ensures progress}
    \Else
      \ForAll{$(t,j)\in\mathcal{A}$, with $\ell\gets j-i$}
        \State \textbf{if} $\alpha\le0.5$ \textbf{then} $w(t,j)\gets\ell/2$
               \Comment{favor longer segments}
        \State \textbf{else} $w(t,j)\gets1/\max(1,\ell/2)$, doubled if $|t|>1$
               \Comment{favor short, multi-token splits}
      \EndFor
      \State sample $(t,j)\!\sim\!\mathcal{A}$ with
             $P(t,j)=w(t,j)/\!\!\sum_{(t',j')\in\mathcal{A}}\!\!w(t',j')$;\;
             $u\gets u\Vert t$;\; $i\gets j$
    \EndIf
  \EndWhile
  \State \Return $\bot$ \textbf{if} $u=v^*$ \textbf{or} $\tau^{-1}(u)\ne p$,
         \textbf{else} $u$
\EndProcedure

\Statex
\Procedure{UpdateAlpha}{$\alpha,D,\beta_l,\beta_h$}
  \State \textbf{if} $|D|<B$ \textbf{then return} $\alpha$
  \State $\bar{s}\gets\mathrm{mean}\big(D[|D|-B:|D|]\big)$
  \State \textbf{if} $\bar{s}>\beta_h$ \textbf{then return}
         $\min(0.95,\alpha+\eta)$
  \State \textbf{if} $\bar{s}<\beta_l$ \textbf{then return}
         $\max(0.05,\alpha-\eta)$
  \State \Return $\mathrm{clip}(\alpha+\varepsilon,0.05,0.95)$,\;
         $\varepsilon\sim\mathcal{U}(-\eta/2,\eta/2)$
         \Comment{in-band jitter avoids stagnation}
\EndProcedure

\end{algorithmic}
\end{algorithm*}

\section*{Appendix D\\Cost and Efficiency Analysis of \Toketive}
\label{app:rq3}

The dominant cost of \Toketive for each edited fact is the number of forward passes through $f_{\theta'}$. The canonical reference step requires a single forward pass to extract hidden states. 

The detection and reconstruction samplers each perform at most $M \cdot n_{\text{det}}$ and $M \cdot n_{\text{rec}}$ tokenization attempts, respectively, where $M$ is the per-sample attempt budget. Each attempt requires a forward pass to compute hidden representations for the entanglement score (cost $\mathcal{O}(F)$), while each accepted sample additionally incurs an autoregressive generation cost (cost $\mathcal{O}(G)$).

The total cost per fact is therefore:
\[
\mathcal{O}\!\left(
M(n_{\text{det}} + n_{\text{rec}})\cdot F \;+\;
(n_{\text{det}} + n_{\text{rec}})\cdot G
\right),
\]
where $F$ denotes the cost of a single forward pass, and $G$ denotes the cost of autoregressive generation, which scales with output length and typically satisfies $G \gg F$.

The random and edit-distance baselines avoid hidden-state extraction, eliminating the $F$-term associated with scoring. Their cost is dominated by generation:
\[
\mathcal{O}\!\left(
(n_{\text{det}} + n_{\text{rec}})\cdot G
\right),
\]
with negligible additional overhead from tokenization and edit-distance computation. In the worst case, \Toketive performs $M(n_{\text{det}} + n_{\text{rec}})$ forward passes for scoring; in practice, the sampler terminates early once band quotas are satisfied.

\begin{figure*}
    \centering
    \begin{subfigure}[b]{0.48\linewidth}
        \includegraphics[width=\linewidth]{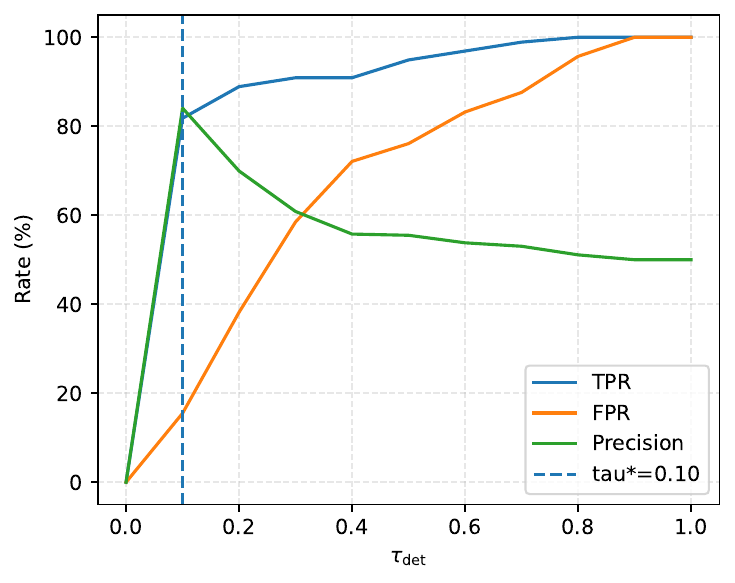}
        \caption{TPR, FPR, and Precision}
        \label{fig:tau_rates}
    \end{subfigure}
    \hfill
    \begin{subfigure}[b]{0.48\linewidth}
        \includegraphics[width=\linewidth]{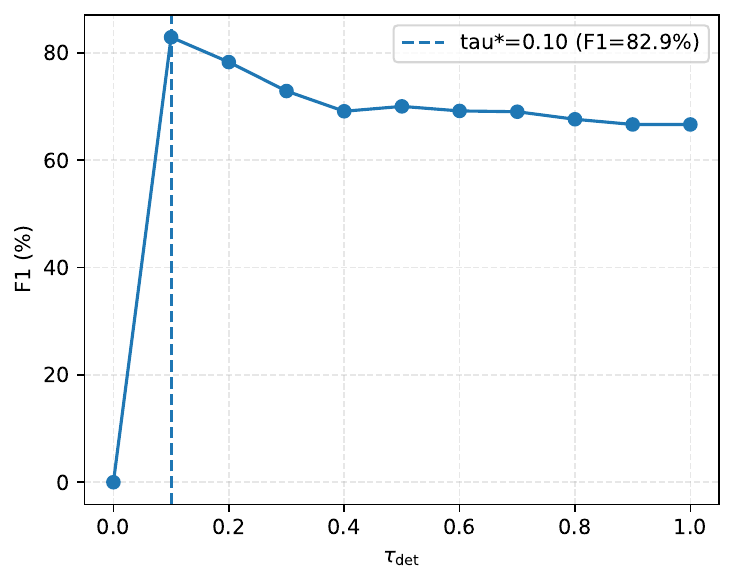}
        \caption{F1 score}
        \label{fig:tau_f1}
    \end{subfigure}
    \caption{Sensitivity of edit detection performance to the 
    detection threshold $\tau_{\text{det}}$, averaged across 
    all models, datasets, and techniques. \textbf{(a)} TPR, 
    FPR, and Precision as a function of $\tau_{\text{det}}$. 
    At $\tau^* = 0.10$, TPR and Precision are both near 
    $83\%$ while FPR remains low, marking the optimal 
    precision-recall balance. \textbf{(b)} F1 score as a 
    function of $\tau_{\text{det}}$. F1 peaks sharply at 
    $\tau^* = 0.10$ ($82.9\%$) and degrades monotonically 
    for larger values, confirming this as a robust operating 
    point. The plateau near $67\%$ at 
    $\tau_{\text{det}} \rightarrow 1.0$ corresponds to the 
    degenerate case where all facts are labeled edited 
    (TPR $= 100\%$, FPR $= 100\%$, Precision $\approx 50\%$).}
    \label{fig:tau_det_sensitivity}
\end{figure*}

\section*{Appendix E\\Hyperparameter Calibration}
\label{app:hyperparams}

 They are selected once using a single development configuration and are kept fixed for all remaining $179$ experiments. No model-, dataset-, or editing-technique-specific retuning is performed. We intentionally avoid per-target tuning to demonstrate that \Toketive doesn’t depend on target-specific calibration.
 
\paragraph{\textbf{Detection threshold $\tau_{\text{det}}$.}}
The detection threshold $\tau_{\text{det}}$ was calibrated 
on a single held-out configuration: \LlamaThree with MEMIT 
on Known-1000. Fig.~\ref{fig:tau_det_sensitivity} shows 
F1 as a function of $\tau_{\text{det}}$ for this 
configuration. F1 peaks sharply at $\tau^* = 0.10$ 
(F1 $= 82.9\%$) and degrades monotonically for larger 
values. We adopt $\tau_{\text{det}} = 0.10$ as a fixed 
threshold and apply it uniformly across all remaining 
$179$ model-dataset-technique combinations without 
further tuning. The consistent detection performance 
reported in Table~\ref{tab:rq3_detection_main} across 
diverse models and datasets suggests that this threshold 
generalizes well beyond the calibration setting.

\paragraph{\textbf{Band boundaries.}}
Band boundaries for the representational entanglement were determined by sweeping 
$\beta^{\text{det}}_h$ (with $\beta^{\text{det}}_l = 0$ 
fixed) for detection and $\beta^{\text{rec}}_l$ (with 
$\beta^{\text{rec}}_h = 1.0$ fixed) for reconstruction 
on the same held-out configuration, using F1 and top-1 
accuracy as the respective criteria. The values used are as follows.

For the representational entanglement sampler:
\begin{itemize}
    \item Detection band: 
          $[\beta^{\text{det}}_l,\, \beta^{\text{det}}_h) 
          = [0.3,\, 0.5)$
    \item Reconstruction band: 
          $[\beta^{\text{rec}}_l,\, \beta^{\text{rec}}_h) 
          = [0.7,\, 1.0)$
\end{itemize}

For the edit distance sampler, we utilise the following values:
\begin{itemize}
    \item Detection band: 
          $[\beta^{\text{det}}_l,\, \beta^{\text{det}}_h) 
          = [0.80,\, 1.00)$
    \item Reconstruction band: 
          $[\beta^{\text{rec}}_l,\, \beta^{\text{rec}}_h) 
          = [0.00,\, 0.50)$
\end{itemize}

The detection band for entanglement ($[0.30, 0.50)$) 
captures tokenizations that are meaningfully divergent 
from canonical internally but retain sufficient semantic 
signal to produce coherent completions. The reconstruction 
band ($[0.70, 1.00)$) targets high-similarity tokenizations 
that are likely to draw on the same factual subspace as 
the canonical path while being just divergent enough to 
escape patched circuits, consistent with the design 
motivation in Section~\ref{sec:sampler}.

For edit distance, the detection band ($[0.80, 1.00)$) 
selects highly divergent tokenizations at the lexical 
level, while the reconstruction band ($[0.00, 0.50)$) 
selects tokenizations that are lexically close to 
canonical. These choices mirror the entanglement band 
logic but operate on token-sequence surface similarity 
rather than hidden-state similarity.

All band boundaries and $\tau_{\text{det}}$ are fixed 
at the calibrated values for all $180$ experiments 
reported in this paper. No per-dataset or per-model 
tuning was performed.

\begin{table*}
\caption{Mean representational entanglement and edit distance scores for convergent (Conv) and non-convergent (Non-conv) noncanonical tokenizations, and their difference (Conv $-$ Non-conv), across all models and datasets. 
Positive differences indicate higher scores for convergent 
tokenizations. For representational entanglement, a positive 
difference reflects stronger internal similarity to the 
canonical hidden state for convergent tokenizations.}
\label{tab:app_rq1_means}
\resizebox{0.99\textwidth}{!}{

\begin{tabular}{ll|ccc|ccc|ccc|ccc|ccc|ccc}
\toprule
 \multirow{2}{*}{\centering Model} &
\multirow{2}{*}{\centering Method} & 
\multicolumn{3}{c}{Real Authors} & 
\multicolumn{3}{c}{CounterFact} & 
\multicolumn{3}{c}{Known-1000} & 
\multicolumn{3}{c}{MQuAKE} & 
\multicolumn{3}{c}{RippleEdits} & 
\multicolumn{3}{c}{World Facts} \\
 &  & Conv & Non-conv & Diff & Conv & Non-conv & Diff & Conv & Non-conv & Diff & Conv & Non-conv & Diff & Conv & Non-conv & Diff & Conv & Non-conv & Diff \\
\midrule
\multirow[t]{2}{*}{\LlamaThree} & Repr. Ent. & 0.669 & 0.436 & 0.233 & 0.717 & 0.593 & 0.124 & 0.737 & 0.549 & 0.188 & 0.750 & 0.548 & 0.202 & 0.793 & 0.669 & 0.124 & 0.775 & 0.635 & 0.140 \\
 & Edit distance & 0.736 & 0.807 & -0.071 & 0.682 & 0.752 & -0.070 & 0.708 & 0.771 & -0.064 & 0.717 & 0.774 & -0.057 & 0.740 & 0.772 & -0.032 & 0.739 & 0.797 & -0.058 \\
\cline{1-20}
\multirow[t]{2}{*}{\LlamaThreeOne} & Repr. Ent. & 0.691 & 0.442 & 0.249 & 0.732 & 0.592 & 0.141 & 0.733 & 0.565 & 0.169 & 0.765 & 0.559 & 0.206 & 0.812 & 0.680 & 0.132 & 0.773 & 0.642 & 0.131 \\
 & Edit distance & 0.738 & 0.810 & -0.072 & 0.671 & 0.758 & -0.087 & 0.724 & 0.769 & -0.045 & 0.708 & 0.772 & -0.065 & 0.749 & 0.768 & -0.019 & 0.746 & 0.789 & -0.043 \\
\cline{1-20}
\multirow[t]{2}{*}{\OlmoTwo} & Repr. Ent. & 0.598 & 0.332 & 0.266 & 0.677 & 0.539 & 0.139 & 0.675 & 0.465 & 0.210 & 0.731 & 0.465 & 0.267 & 0.754 & 0.609 & 0.146 & 0.737 & 0.530 & 0.207 \\
 & Edit distance & 0.728 & 0.803 & -0.075 & 0.681 & 0.748 & -0.067 & 0.720 & 0.756 & -0.037 & 0.673 & 0.767 & -0.094 & 0.741 & 0.771 & -0.030 & 0.737 & 0.811 & -0.073 \\
\cline{1-20}
\multirow[t]{2}{*}{\TuluThree} & Repr. Ent. & 0.649 & 0.371 & 0.278 & 0.712 & 0.566 & 0.146 & 0.708 & 0.533 & 0.175 & 0.743 & 0.508 & 0.235 & 0.784 & 0.655 & 0.129 & 0.744 & 0.574 & 0.170 \\
 & Edit distance & 0.739 & 0.805 & -0.066 & 0.699 & 0.754 & -0.056 & 0.730 & 0.771 & -0.041 & 0.710 & 0.776 & -0.066 & 0.748 & 0.759 & -0.011 & 0.745 & 0.788 & -0.043 \\
\cline{1-20}
\multirow[t]{2}{*}{\TuluThreeOne} & Repr. Ent. & 0.648 & 0.374 & 0.274 & 0.703 & 0.575 & 0.129 & 0.715 & 0.524 & 0.191 & 0.745 & 0.512 & 0.233 & 0.788 & 0.656 & 0.132 & 0.750 & 0.583 & 0.166 \\
 & Edit distance & 0.738 & 0.809 & -0.071 & 0.698 & 0.765 & -0.067 & 0.728 & 0.771 & -0.044 & 0.711 & 0.779 & -0.068 & 0.755 & 0.767 & -0.012 & 0.739 & 0.780 & -0.042 \\
\cline{1-20}
\bottomrule
\end{tabular}
}
\end{table*}

\section*{Appendix F\\Mean Score Analysis}
\label{app:rq1_means}

Table~\ref{tab:app_rq1_means} reports the mean representational 
entanglement and edit distance scores for convergent and 
non-convergent noncanonical tokenizations, along with their 
difference (Conv $-$ Non-conv), across all models and datasets.

For representational entanglement, convergent tokenizations 
consistently achieve higher mean scores than non-convergent 
ones across all model-dataset combinations, with differences 
ranging from $0.124$ to $0.278$. The direction of this 
difference is stable, as convergent tokenizations are always 
more similar to the canonical hidden state than non-convergent 
ones, indicating that representational entanglement scores 
carry a consistent directional signal about factual retrieval 
behavior. The largest gaps appear on Real Authors and MQuAKE, 
where the mean difference exceeds $0.20$ for most models, 
while RippleEdits and CounterFact show smaller but still 
consistent separations.

For edit distance, the pattern is reversed and substantially 
weaker. Non-convergent tokenizations have slightly higher 
mean edit distances than convergent ones in all settings, 
yielding negative differences ranging from $-0.011$ to 
$-0.094$. The magnitude of this separation is small (typically $0.03$-$0.07$) indicating that convergent and 
non-convergent tokenizations differ only marginally in their 
lexical distance from the canonical sequence. This indicates 
that while edit distance captures some surface-level 
variation, it does not reliably separate the two groups, 
consistent with the low AUC and $|r_b|$ values reported in 
Table~\ref{tab:predictability}.

Taken together, these mean score differences provide 
additional support that representational entanglement 
captures a meaningful internal signal about factual retrieval, 
while edit distance does not. The consistent positive 
direction of the entanglement difference across all 
$5 \times 6 = 30$ model-dataset combinations, with no 
exceptions, further validates its use as the routing 
criterion in \Toketive's adaptive sampler.

\begin{table*}
\centering
\caption{Examples of \Toketive{} outcomes on \LlamaThree{} edited with AlphaEdit.
All responses are generated after editing. The edited target answer is
\textit{George R.R.~Martin}, while the pre-edit answer is
\textit{George Orwell}. Alternative tokenizations can therefore recover the
suppressed answer even though the canonical tokenization produces the edited
response.}
\label{tab:qualitative_examples}
\small
\begin{tabular}{p{0.18\linewidth} p{0.25\linewidth} p{0.43\linewidth} p{0.09\linewidth}}
\toprule
\textbf{Tokenization} & \textbf{Token sequence} & \textbf{Post-edit response}\\
\midrule

\textbf{Noncanonical} 
&
\texttt{['T','h','e',' ','n','ove','l',' ','198','4',...]}
&
The novel \textit{1984} was written by \textbf{George Orwell}, a British
author, and published in 1949. The book is a \ldots
\\[3pt]

\textbf{Noncanonical}
&
\texttt{['Th','e',' ','nov','el',' ','198','4',' ',...]}
&
The novel \textit{1984} was written by \textbf{Michael Moorcock} in 1948.
It was a satire \ldots
\\[3pt]

\textbf{Canonical}
&
\texttt{['The',' novel',' ','198','4',' was',' written',' by']}
&
The novel \textit{1984} was written by \textbf{George R.R.~Martin}, and it
was published in 1979. The novel \ldots
\\

\bottomrule
\end{tabular}
\end{table*}

\section*{Appendix G\\Distributional Visualizations}
\label{app:rq1_dist}

\begin{figure*}
\centering
\begin{subfigure}{0.19\linewidth}
    \includegraphics[width=\linewidth]{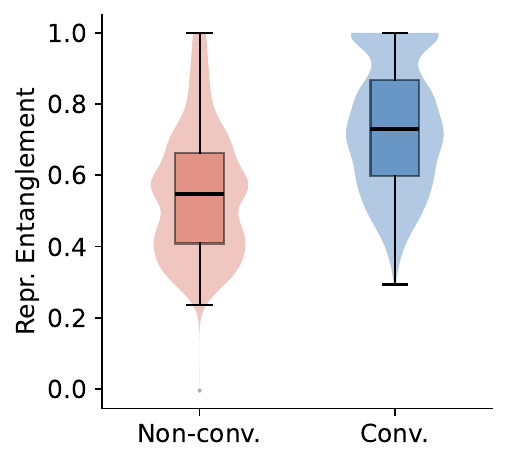}
    \caption{\LlamaThree}
\end{subfigure}
\hfill
\begin{subfigure}{0.19\linewidth}
    \includegraphics[width=\linewidth]{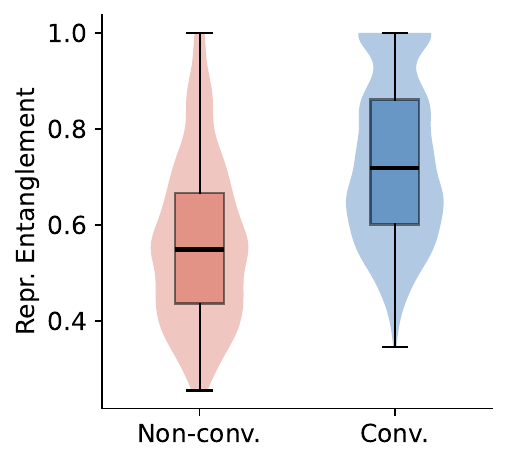}
    \caption{\LlamaThreeOne}
\end{subfigure}
\hfill
\begin{subfigure}{0.19\linewidth}
    \includegraphics[width=\linewidth]{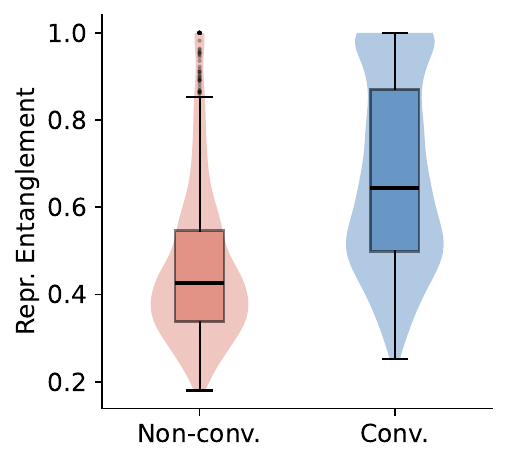}
    \caption{\OlmoTwo}
\end{subfigure}
\hfill
\begin{subfigure}{0.19\linewidth}
    \includegraphics[width=\linewidth]{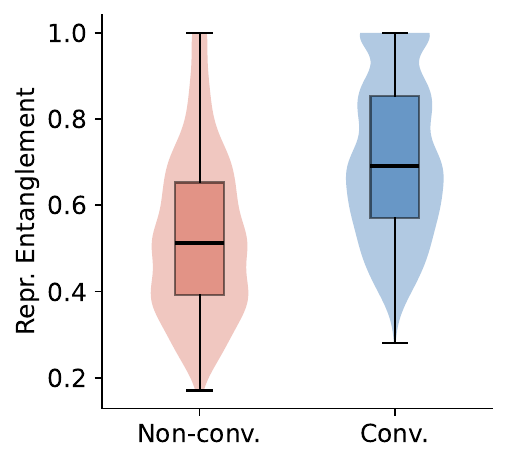}
    \caption{\TuluThree}
\end{subfigure}
\hfill
\begin{subfigure}{0.19\linewidth}
    \includegraphics[width=\linewidth]{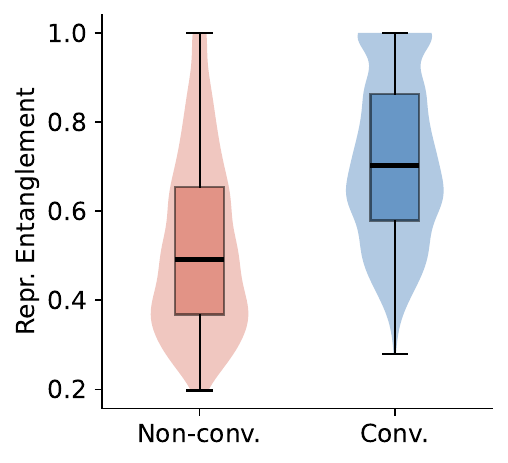}
    \caption{\TuluThreeOne}
\end{subfigure}

\caption{Representational entanglement score distributions for
Known-1000 across all five models. Convergent tokenizations
shown in blue, non-convergent in red.}
\label{fig:app_clare_known_1000}
\end{figure*}

\begin{figure*}
\centering
\begin{subfigure}{0.19\linewidth}
    \includegraphics[width=\linewidth]{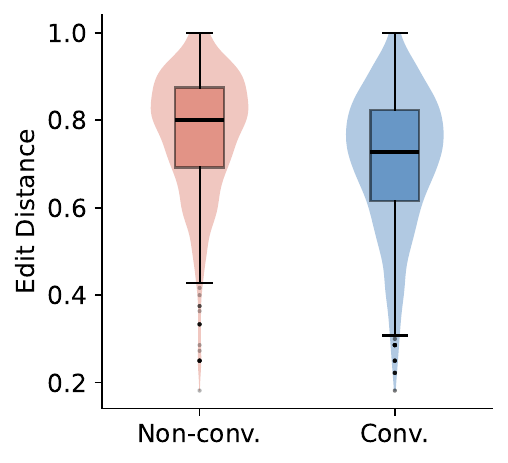}
    \caption{\LlamaThree}
\end{subfigure}
\hfill
\begin{subfigure}{0.19\linewidth}
    \includegraphics[width=\linewidth]{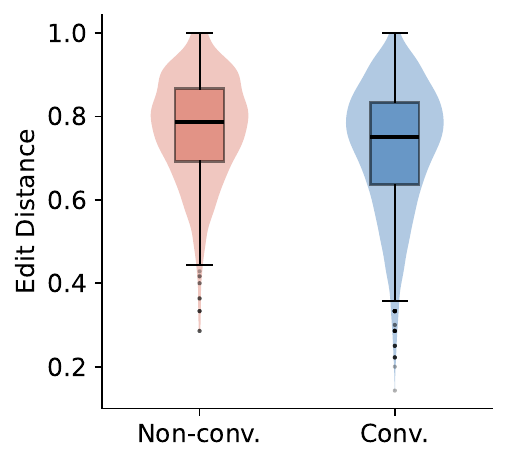}
    \caption{\LlamaThreeOne}
\end{subfigure}
\hfill
\begin{subfigure}{0.19\linewidth}
    \includegraphics[width=\linewidth]{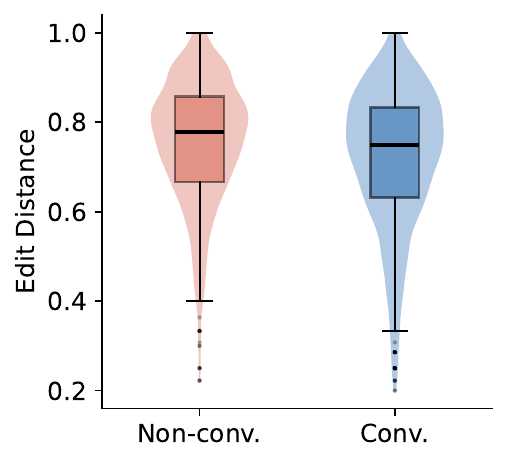}
    \caption{\OlmoTwo}
\end{subfigure}
\hfill
\begin{subfigure}{0.19\linewidth}
    \includegraphics[width=\linewidth]{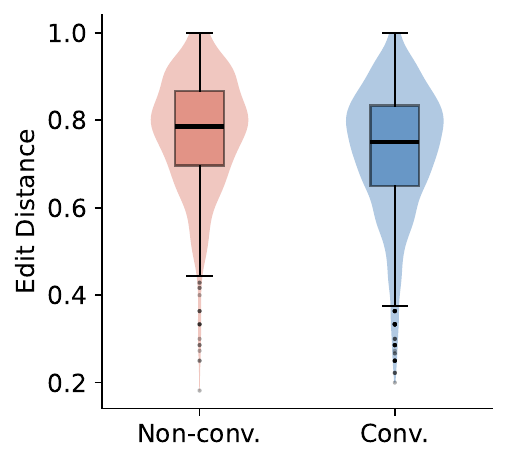}
    \caption{\TuluThree}
\end{subfigure}
\hfill
\begin{subfigure}{0.19\linewidth}
    \includegraphics[width=\linewidth]{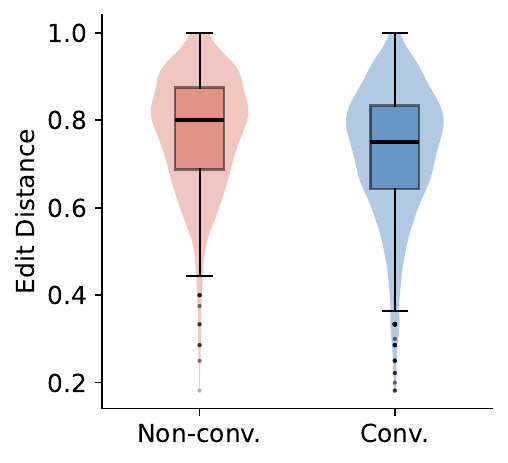}
    \caption{\TuluThreeOne}
\end{subfigure}

\caption{Edit distance score distributions for Known-1000 across
all five models. Convergent tokenizations shown in blue,
non-convergent in red.}
\label{fig:app_editdist_known_1000}
\end{figure*}

\begin{figure*}
\centering
\begin{subfigure}{0.19\linewidth}
    \includegraphics[width=\linewidth]{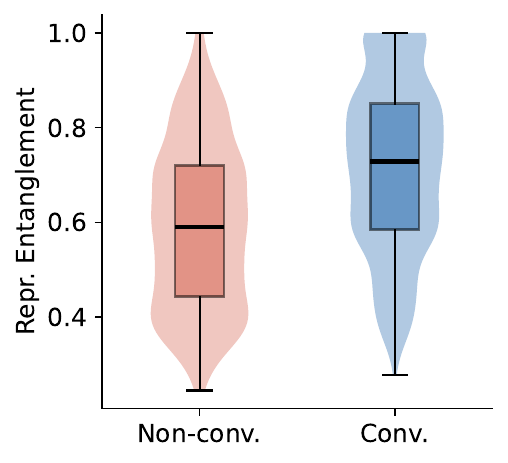}
    \caption{\LlamaThree}
\end{subfigure}
\hfill
\begin{subfigure}{0.19\linewidth}
    \includegraphics[width=\linewidth]{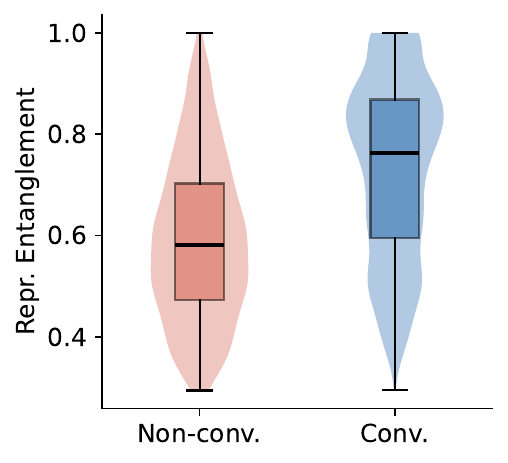}
    \caption{\LlamaThreeOne}
\end{subfigure}
\hfill
\begin{subfigure}{0.19\linewidth}
    \includegraphics[width=\linewidth]{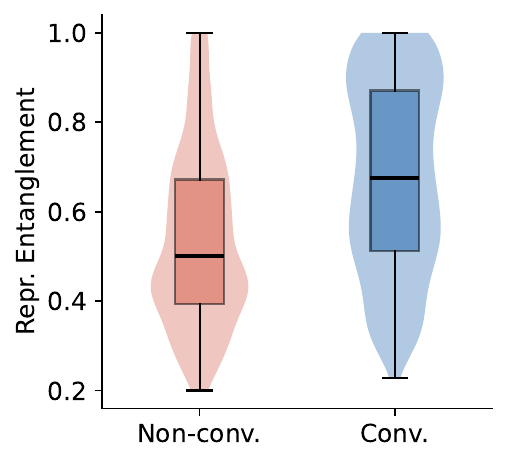}
    \caption{\OlmoTwo}
\end{subfigure}
\hfill
\begin{subfigure}{0.19\linewidth}
    \includegraphics[width=\linewidth]{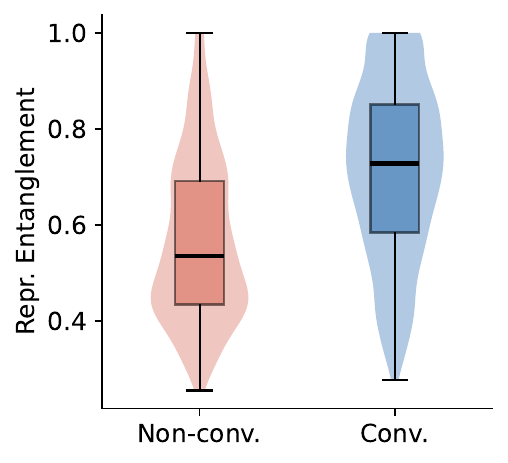}
    \caption{\TuluThree}
\end{subfigure}
\hfill
\begin{subfigure}{0.19\linewidth}
    \includegraphics[width=\linewidth]{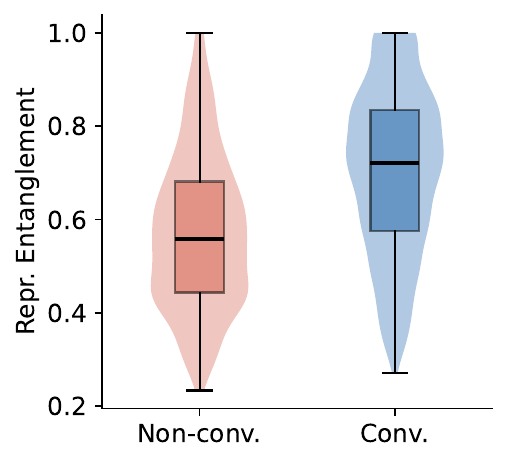}
    \caption{\TuluThreeOne}
\end{subfigure}

\caption{Representational entanglement score distributions for
CounterFact across all five models.}
\label{fig:app_clare_counterfact}
\end{figure*}

\begin{figure*}
\centering
\begin{subfigure}{0.19\linewidth}
    \includegraphics[width=\linewidth]{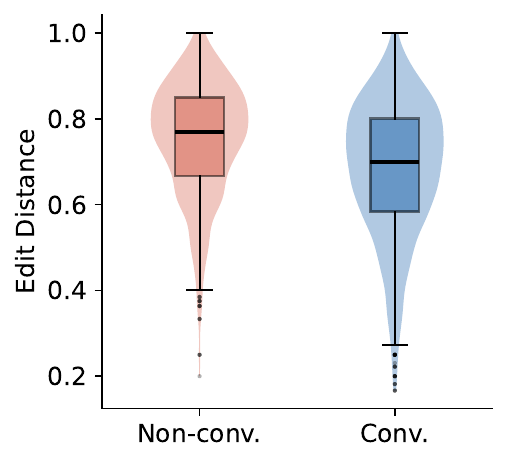}
    \caption{\LlamaThree}
\end{subfigure}
\hfill
\begin{subfigure}{0.19\linewidth}
    \includegraphics[width=\linewidth]{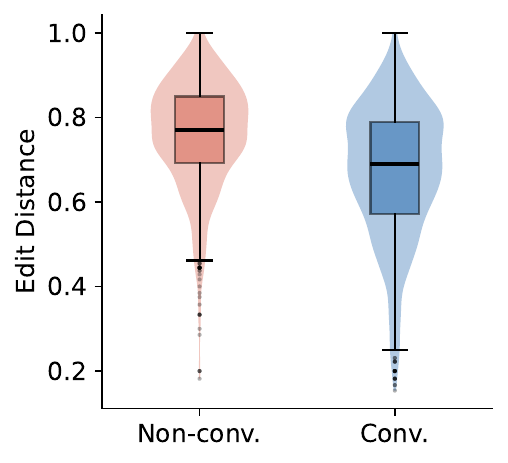}
    \caption{\LlamaThreeOne}
\end{subfigure}
\hfill
\begin{subfigure}{0.19\linewidth}
    \includegraphics[width=\linewidth]{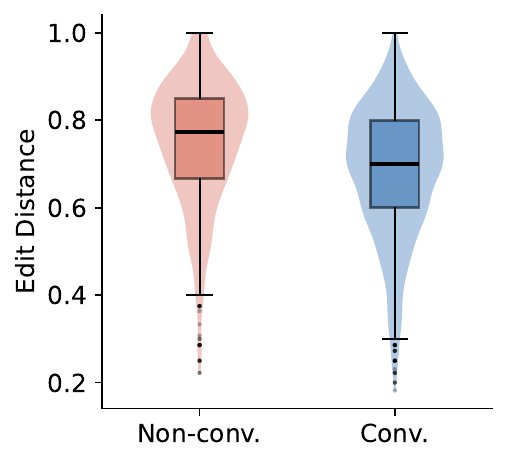}
    \caption{\OlmoTwo}
\end{subfigure}
\hfill
\begin{subfigure}{0.19\linewidth}
    \includegraphics[width=\linewidth]{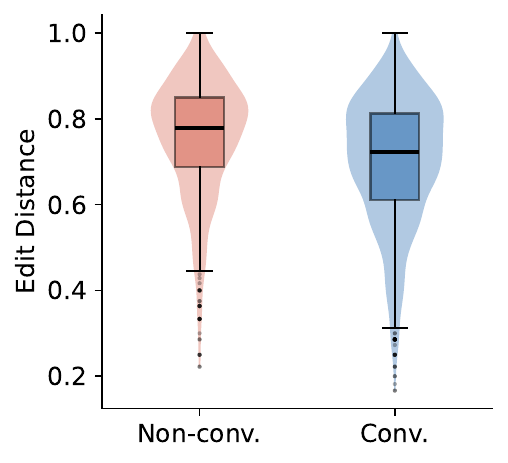}
    \caption{\TuluThree}
\end{subfigure}
\hfill
\begin{subfigure}{0.19\linewidth}
    \includegraphics[width=\linewidth]{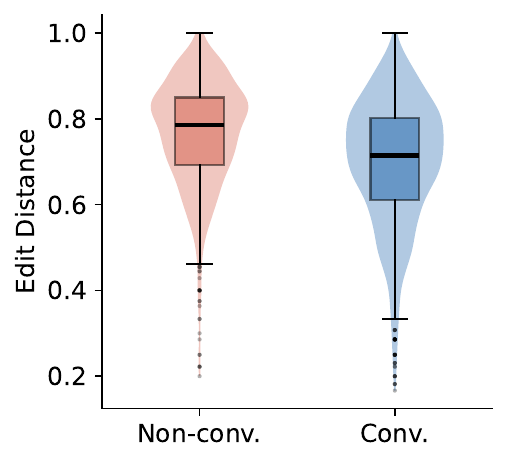}
    \caption{\TuluThreeOne}
\end{subfigure}

\caption{Edit distance score distributions for CounterFact
across all five models.}
\label{fig:app_editdist_counterfact}
\end{figure*}

\begin{figure*}
\centering
\begin{subfigure}{0.19\linewidth}
    \includegraphics[width=\linewidth]{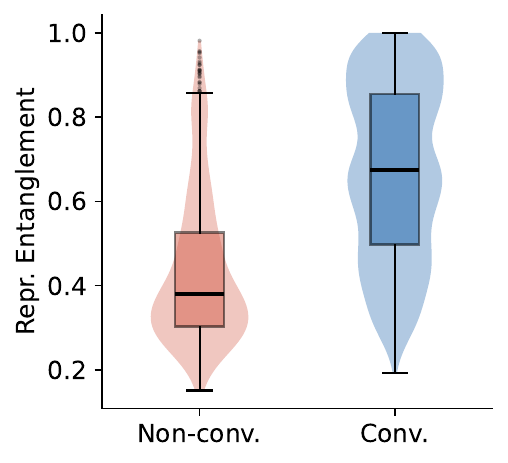}
    \caption{\LlamaThree}
\end{subfigure}
\hfill
\begin{subfigure}{0.19\linewidth}
    \includegraphics[width=\linewidth]{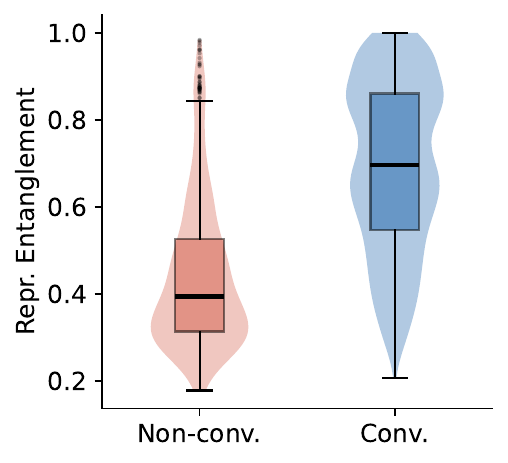}
    \caption{\LlamaThreeOne}
\end{subfigure}
\hfill
\begin{subfigure}{0.19\linewidth}
    \includegraphics[width=\linewidth]{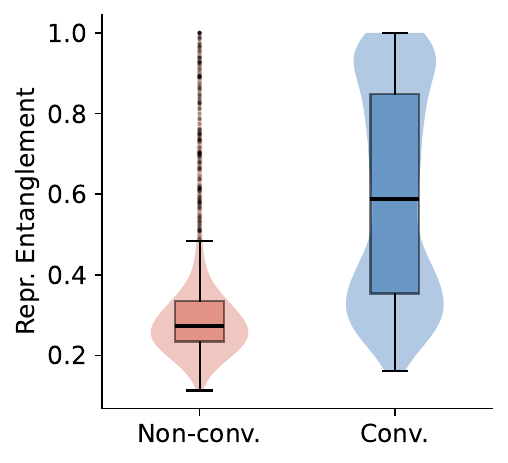}
    \caption{\OlmoTwo}
\end{subfigure}
\hfill
\begin{subfigure}{0.19\linewidth}
    \includegraphics[width=\linewidth]{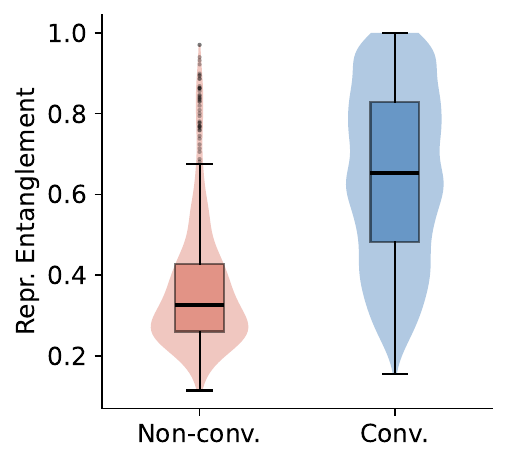}
    \caption{\TuluThree}
\end{subfigure}
\hfill
\begin{subfigure}{0.19\linewidth}
    \includegraphics[width=\linewidth]{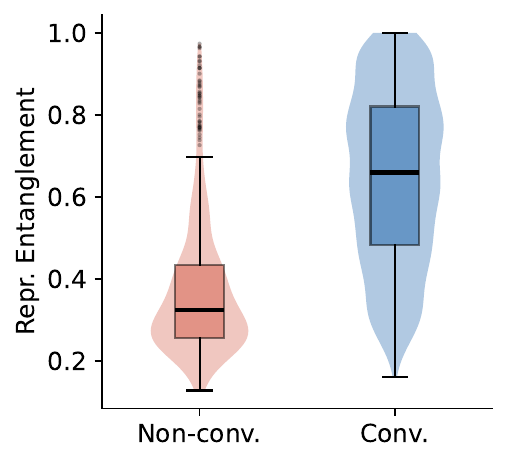}
    \caption{\TuluThreeOne}
\end{subfigure}

\caption{Representational entanglement score distributions for
Real Authors across all five models.}
\label{fig:app_clare_authors}
\end{figure*}

\begin{figure*}
\centering
\begin{subfigure}{0.19\linewidth}
    \includegraphics[width=\linewidth]{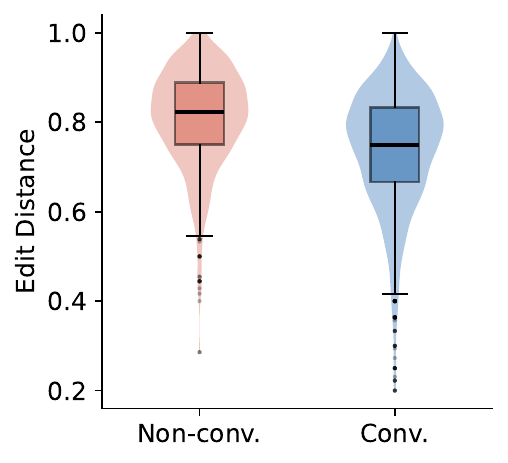}
    \caption{\LlamaThree}
\end{subfigure}
\hfill
\begin{subfigure}{0.19\linewidth}
    \includegraphics[width=\linewidth]{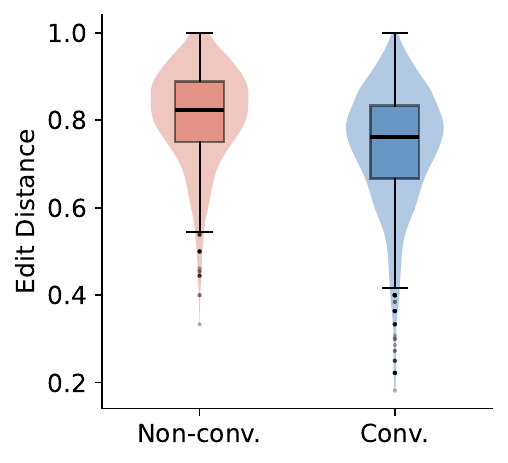}
    \caption{\LlamaThreeOne}
\end{subfigure}
\hfill
\begin{subfigure}{0.19\linewidth}
    \includegraphics[width=\linewidth]{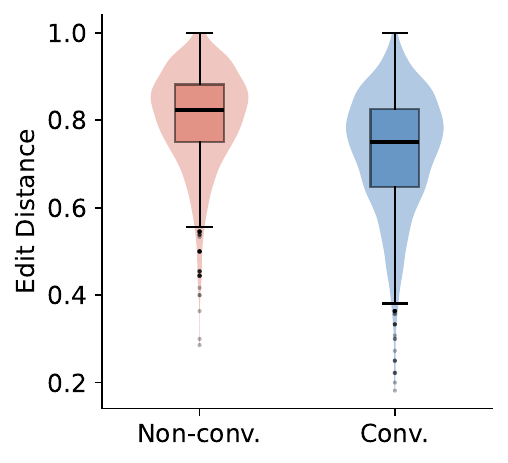}
    \caption{\OlmoTwo}
\end{subfigure}
\hfill
\begin{subfigure}{0.19\linewidth}
    \includegraphics[width=\linewidth]{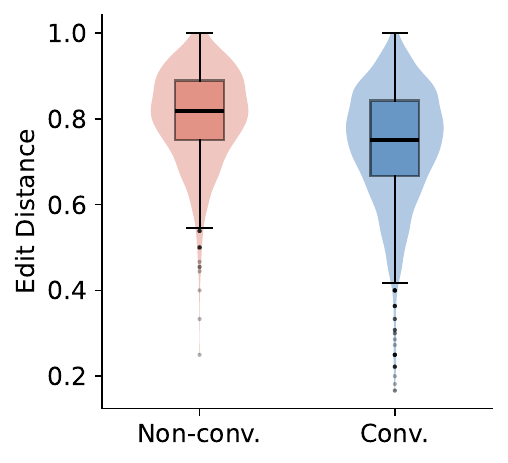}
    \caption{\TuluThree}
\end{subfigure}
\hfill
\begin{subfigure}{0.19\linewidth}
    \includegraphics[width=\linewidth]{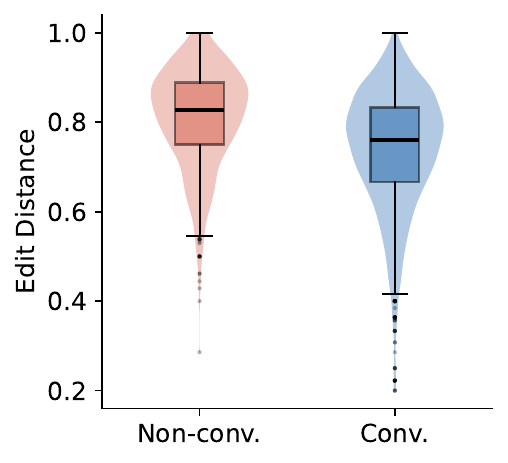}
    \caption{\TuluThreeOne}
\end{subfigure}

\caption{Edit distance score distributions for Real Authors
across all five models.}
\label{fig:app_editdist_authors}
\end{figure*}

\begin{figure*}
\centering
\begin{subfigure}{0.19\linewidth}
    \includegraphics[width=\linewidth]{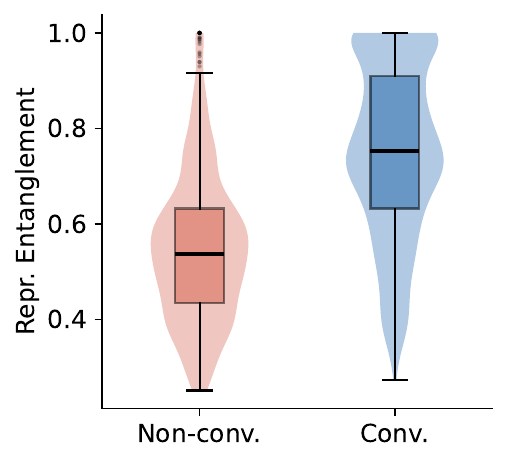}
    \caption{\LlamaThree}
\end{subfigure}
\hfill
\begin{subfigure}{0.19\linewidth}
    \includegraphics[width=\linewidth]{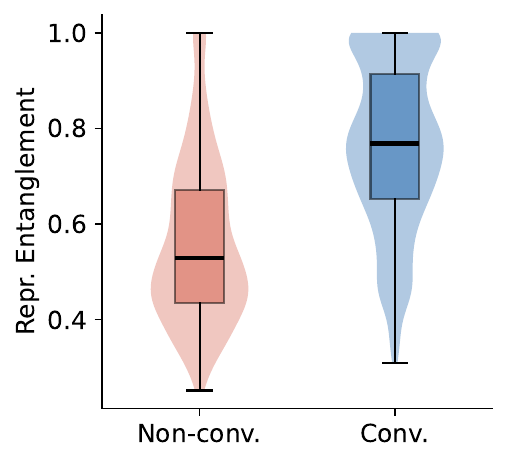}
    \caption{\LlamaThreeOne}
\end{subfigure}
\hfill
\begin{subfigure}{0.19\linewidth}
    \includegraphics[width=\linewidth]{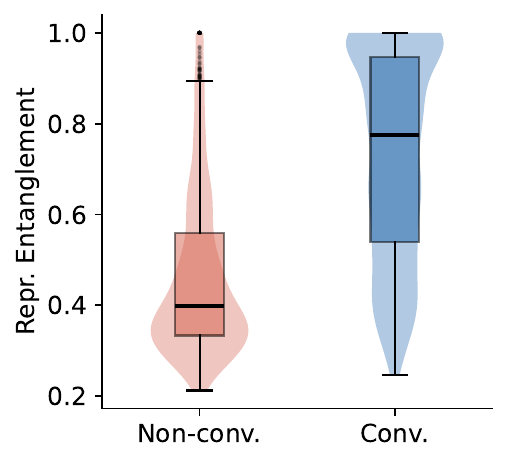}
    \caption{\OlmoTwo}
\end{subfigure}
\hfill
\begin{subfigure}{0.19\linewidth}
    \includegraphics[width=\linewidth]{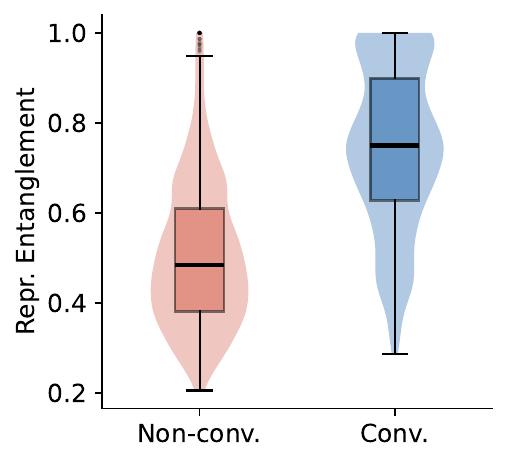}
    \caption{\TuluThree}
\end{subfigure}
\hfill
\begin{subfigure}{0.19\linewidth}
    \includegraphics[width=\linewidth]{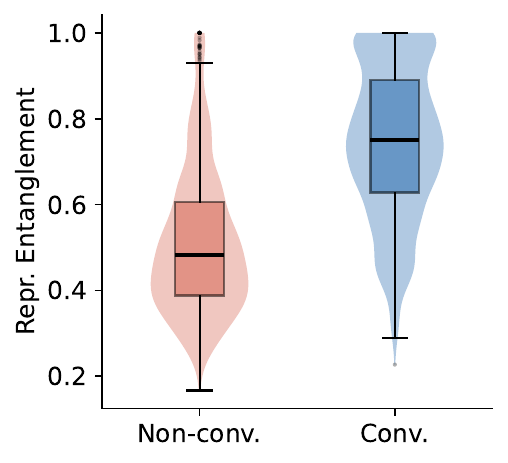}
    \caption{\TuluThreeOne}
\end{subfigure}

\caption{Representational entanglement score distributions for
MQuAKE across all five models.}
\label{fig:app_clare_mquake}
\end{figure*}

\begin{figure*}
\centering
\begin{subfigure}{0.19\linewidth}
    \includegraphics[width=\linewidth]{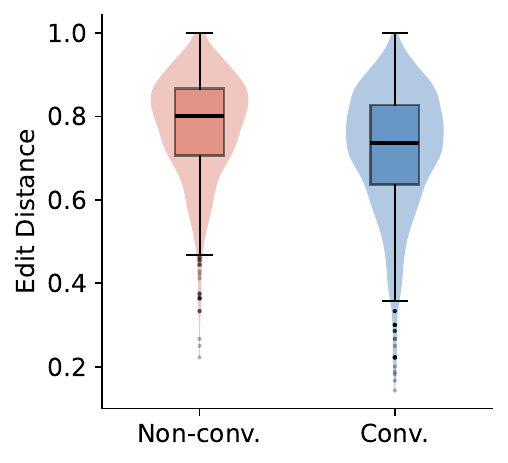}
    \caption{\LlamaThree}
\end{subfigure}
\hfill
\begin{subfigure}{0.19\linewidth}
    \includegraphics[width=\linewidth]{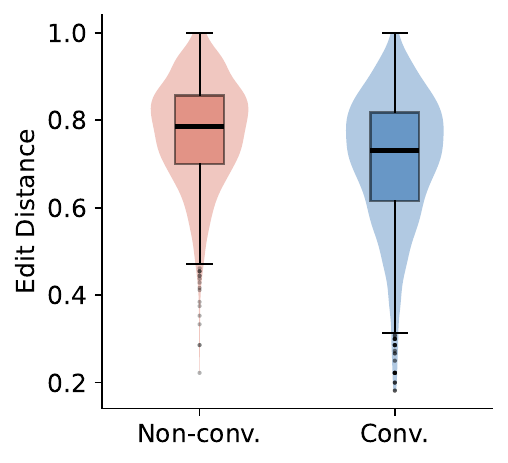}
    \caption{\LlamaThreeOne}
\end{subfigure}
\hfill
\begin{subfigure}{0.19\linewidth}
    \includegraphics[width=\linewidth]{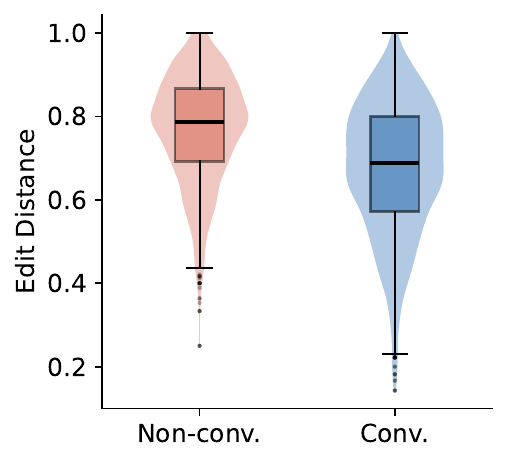}
    \caption{\OlmoTwo}
\end{subfigure}
\hfill
\begin{subfigure}{0.19\linewidth}
    \includegraphics[width=\linewidth]{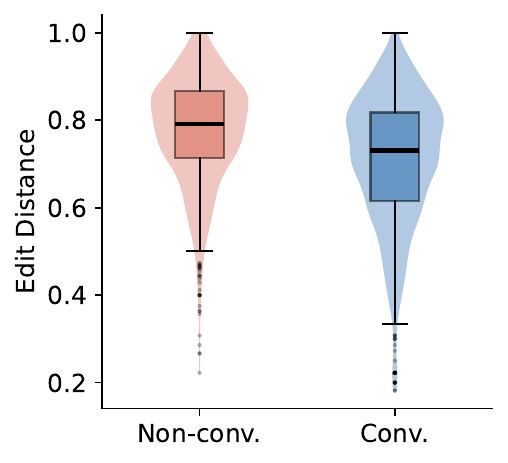}
    \caption{\TuluThree}
\end{subfigure}
\hfill
\begin{subfigure}{0.19\linewidth}
    \includegraphics[width=\linewidth]{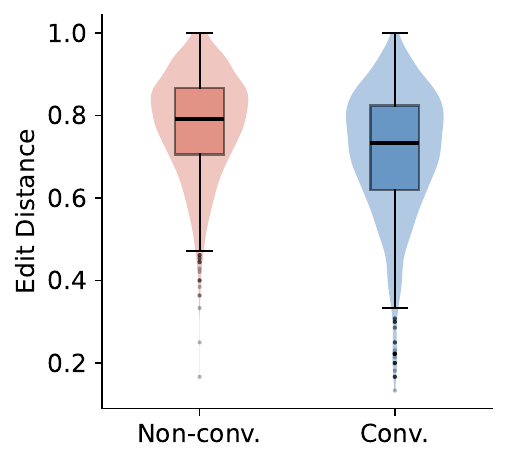}
    \caption{\TuluThreeOne}
\end{subfigure}

\caption{Edit distance score distributions for MQuAKE across
all five models.}
\label{fig:app_editdist_mquake}
\end{figure*}

\begin{figure*}
\centering
\begin{subfigure}{0.19\linewidth}
    \includegraphics[width=\linewidth]{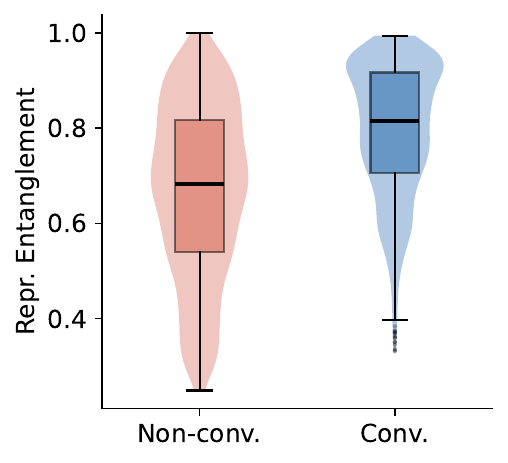}
    \caption{\LlamaThree}
\end{subfigure}
\hfill
\begin{subfigure}{0.19\linewidth}
    \includegraphics[width=\linewidth]{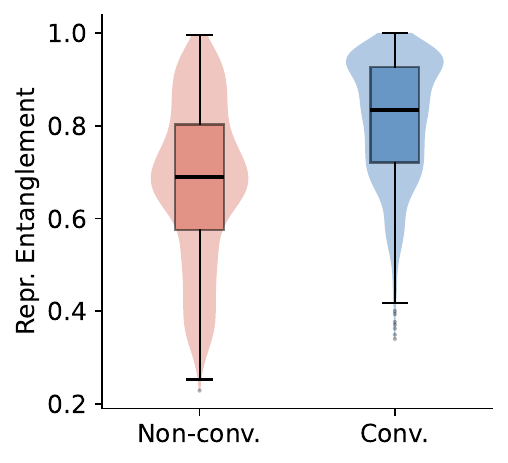}
    \caption{\LlamaThreeOne}
\end{subfigure}
\hfill
\begin{subfigure}{0.19\linewidth}
    \includegraphics[width=\linewidth]{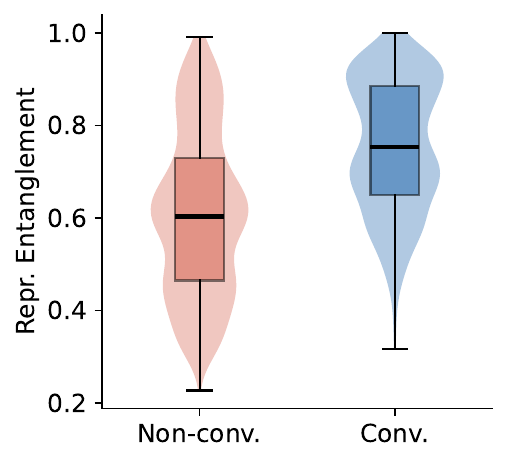}
    \caption{\OlmoTwo}
\end{subfigure}
\hfill
\begin{subfigure}{0.19\linewidth}
    \includegraphics[width=\linewidth]{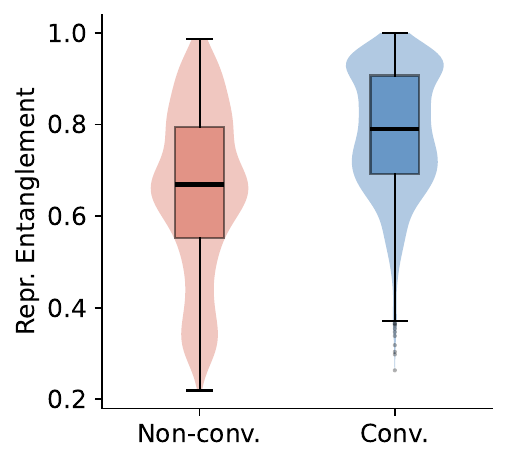}
    \caption{\TuluThree}
\end{subfigure}
\hfill
\begin{subfigure}{0.19\linewidth}
    \includegraphics[width=\linewidth]{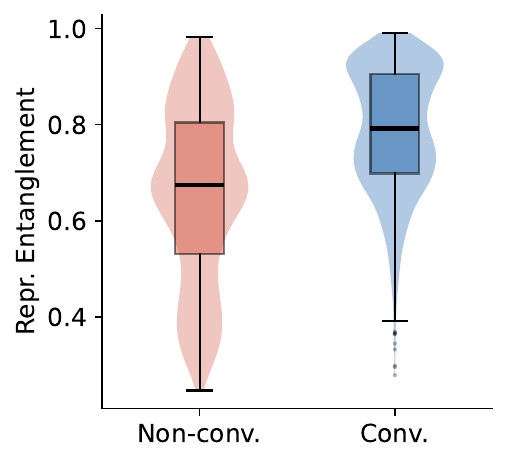}
    \caption{\TuluThreeOne}
\end{subfigure}

\caption{Representational entanglement score distributions for
RippleEdits across all five models.}
\label{fig:app_clare_rippleedits}
\end{figure*}

\begin{figure*}
\centering
\begin{subfigure}{0.19\linewidth}
    \includegraphics[width=\linewidth]{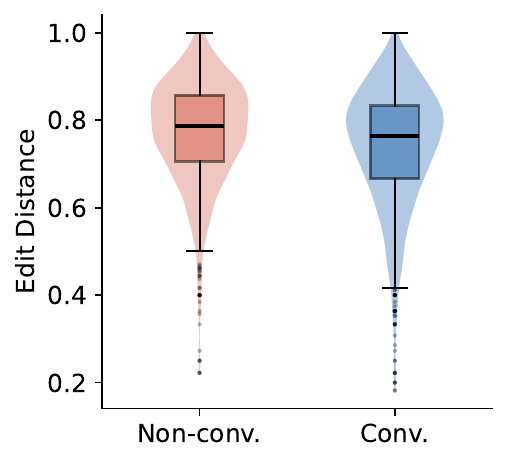}
    \caption{\LlamaThree}
\end{subfigure}
\hfill
\begin{subfigure}{0.19\linewidth}
    \includegraphics[width=\linewidth]{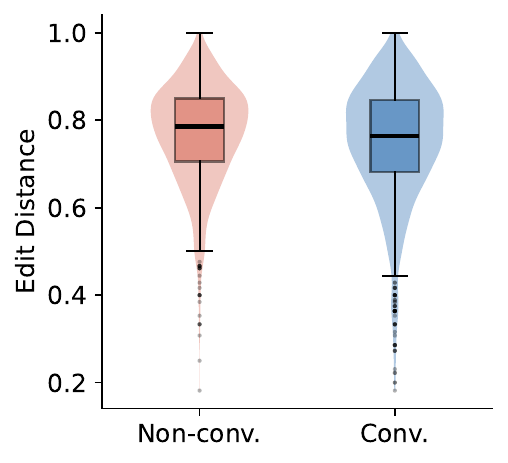}
    \caption{\LlamaThreeOne}
\end{subfigure}
\hfill
\begin{subfigure}{0.19\linewidth}
    \includegraphics[width=\linewidth]{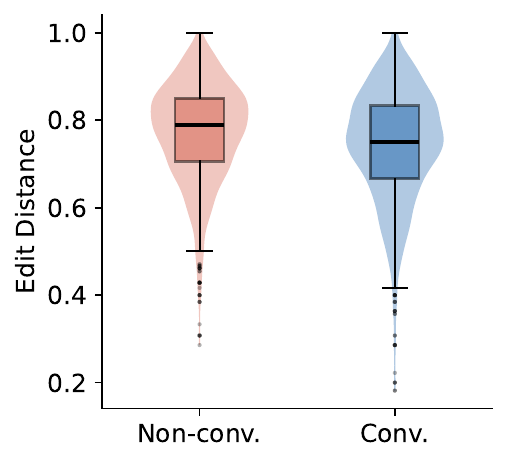}
    \caption{\OlmoTwo}
\end{subfigure}
\hfill
\begin{subfigure}{0.19\linewidth}
    \includegraphics[width=\linewidth]{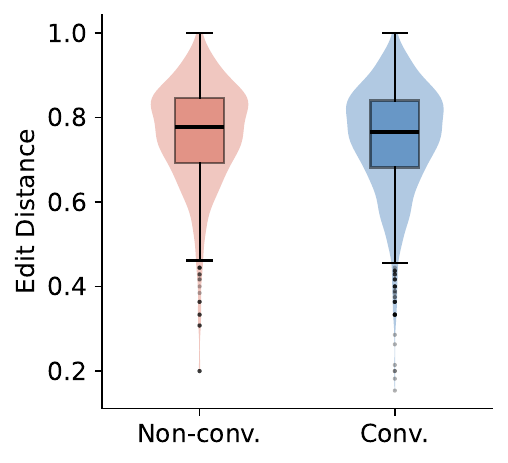}
    \caption{\TuluThree}
\end{subfigure}
\hfill
\begin{subfigure}{0.19\linewidth}
    \includegraphics[width=\linewidth]{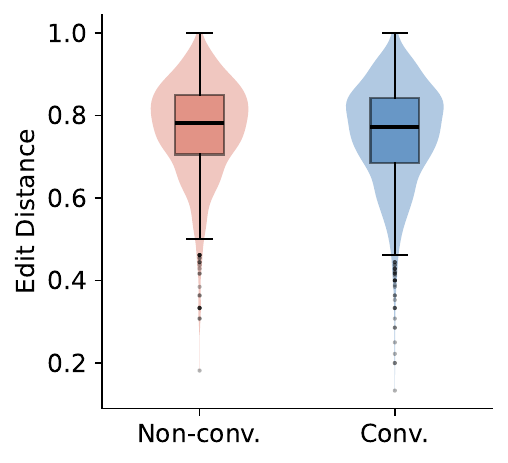}
    \caption{\TuluThreeOne}
\end{subfigure}

\caption{Edit distance score distributions for RippleEdits
across all five models.}
\label{fig:app_editdist_rippleedits}
\end{figure*}

\begin{figure*}
\centering
\begin{subfigure}{0.19\linewidth}
    \includegraphics[width=\linewidth]{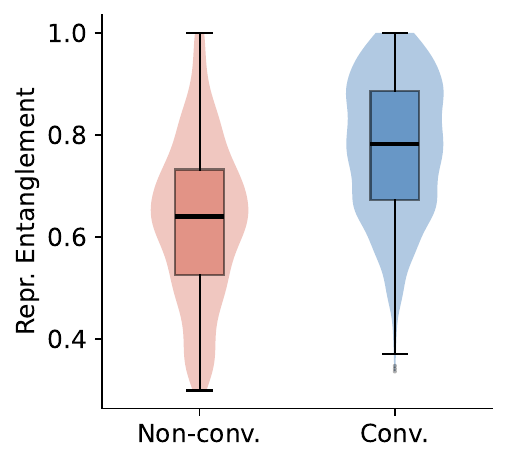}
    \caption{\LlamaThree}
\end{subfigure}
\hfill
\begin{subfigure}{0.19\linewidth}
    \includegraphics[width=\linewidth]{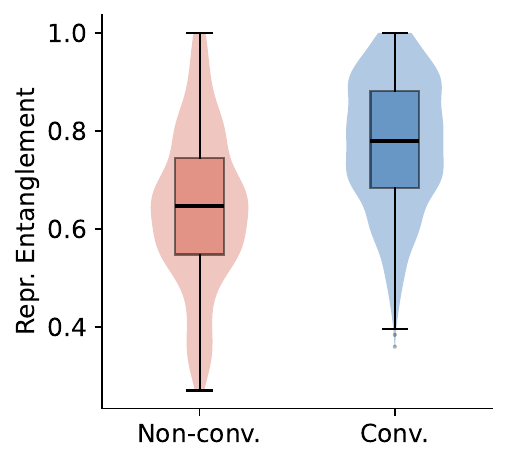}
    \caption{\LlamaThreeOne}
\end{subfigure}
\hfill
\begin{subfigure}{0.19\linewidth}
    \includegraphics[width=\linewidth]{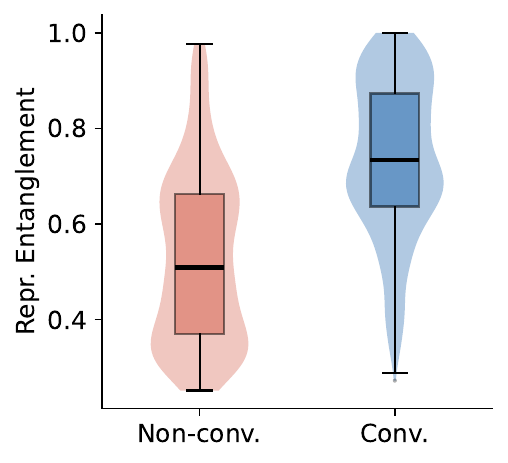}
    \caption{\OlmoTwo}
\end{subfigure}
\hfill
\begin{subfigure}{0.19\linewidth}
    \includegraphics[width=\linewidth]{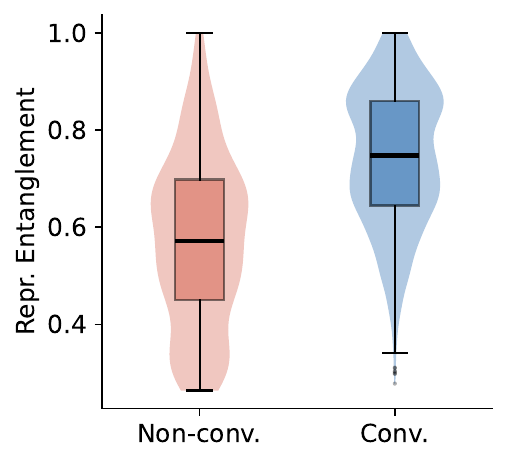}
    \caption{\TuluThree}
\end{subfigure}
\hfill
\begin{subfigure}{0.19\linewidth}
    \includegraphics[width=\linewidth]{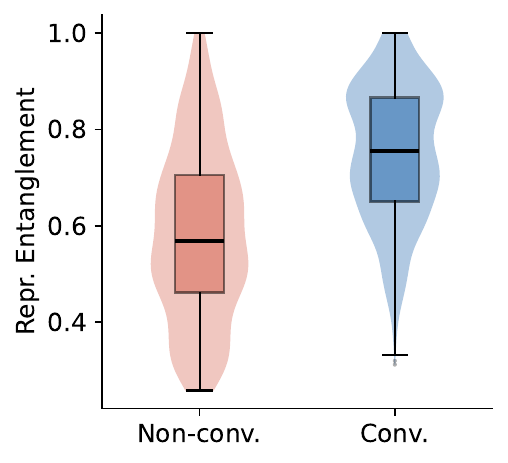}
    \caption{\TuluThreeOne}
\end{subfigure}

\caption{Representational entanglement score distributions for
World Facts across all five models.}
\label{fig:app_clare_world_facts}
\end{figure*}

\begin{figure*}
\centering
\begin{subfigure}{0.19\linewidth}
    \includegraphics[width=\linewidth]{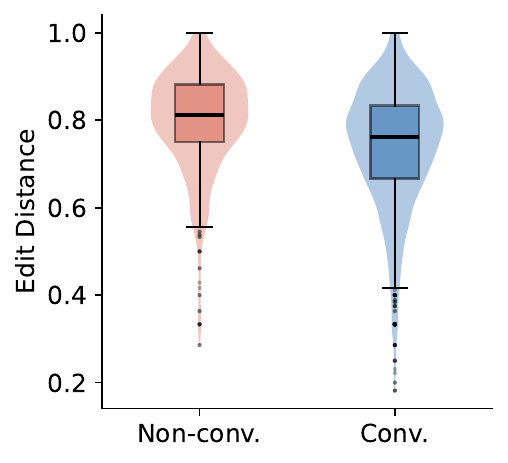}
    \caption{\LlamaThree}
\end{subfigure}
\hfill
\begin{subfigure}{0.19\linewidth}
    \includegraphics[width=\linewidth]{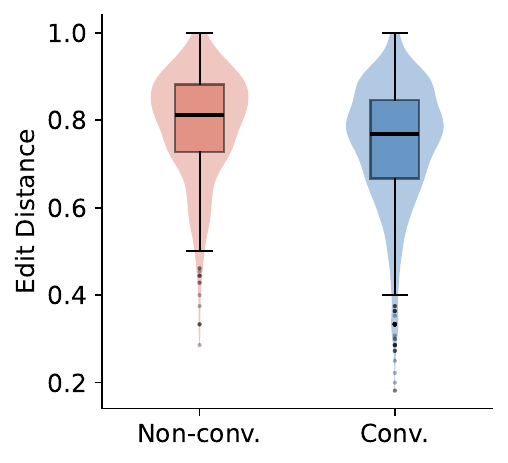}
    \caption{\LlamaThreeOne}
\end{subfigure}
\hfill
\begin{subfigure}{0.19\linewidth}
    \includegraphics[width=\linewidth]{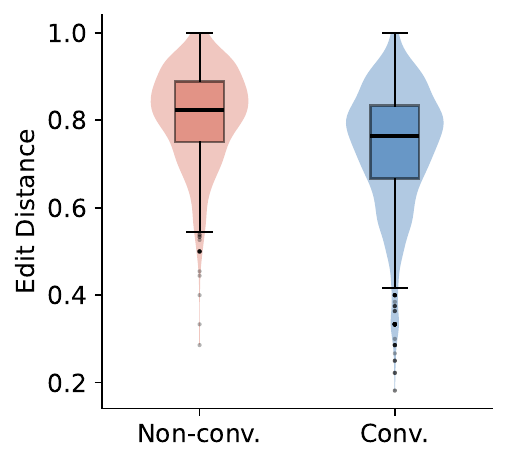}
    \caption{\OlmoTwo}
\end{subfigure}
\hfill
\begin{subfigure}{0.19\linewidth}
    \includegraphics[width=\linewidth]{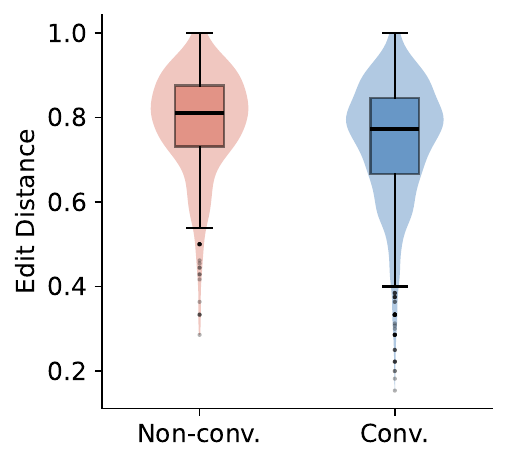}
    \caption{\TuluThree}
\end{subfigure}
\hfill
\begin{subfigure}{0.19\linewidth}
    \includegraphics[width=\linewidth]{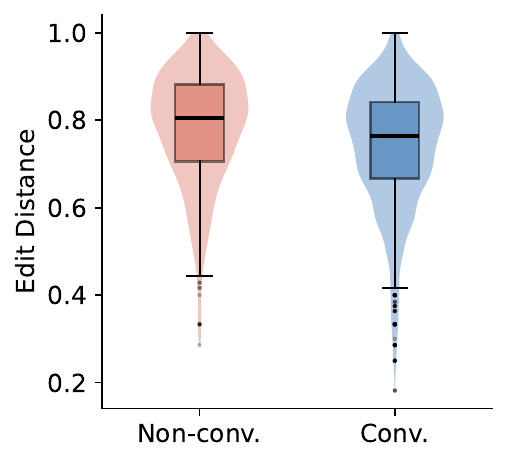}
    \caption{\TuluThreeOne}
\end{subfigure}

\caption{Edit distance score distributions for World Facts
across all five models.}
\label{fig:app_editdist_world_facts}
\end{figure*}

\begin{figure*}
\centering
\begin{subfigure}{0.19\linewidth}
    \includegraphics[width=\linewidth]{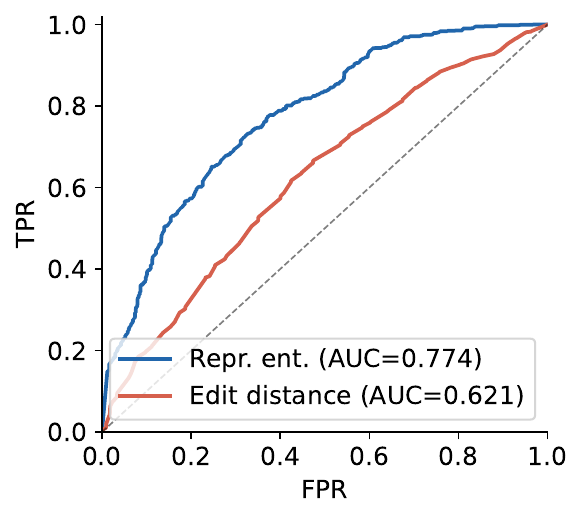}
    \caption{\LlamaThree}
\end{subfigure}
\hfill
\begin{subfigure}{0.19\linewidth}
    \includegraphics[width=\linewidth]{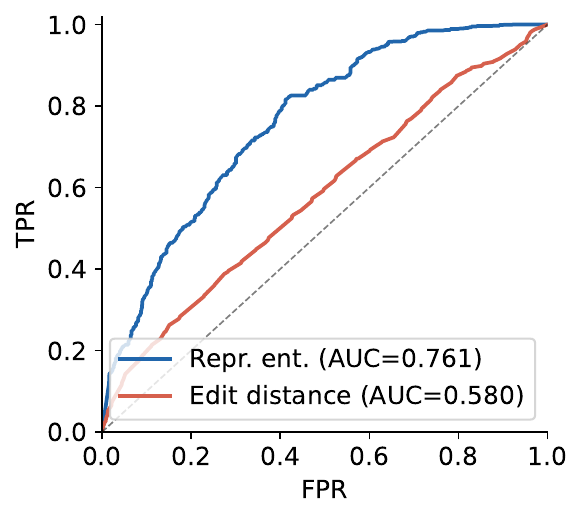}
    \caption{\LlamaThreeOne}
\end{subfigure}
\hfill
\begin{subfigure}{0.19\linewidth}
    \includegraphics[width=\linewidth]{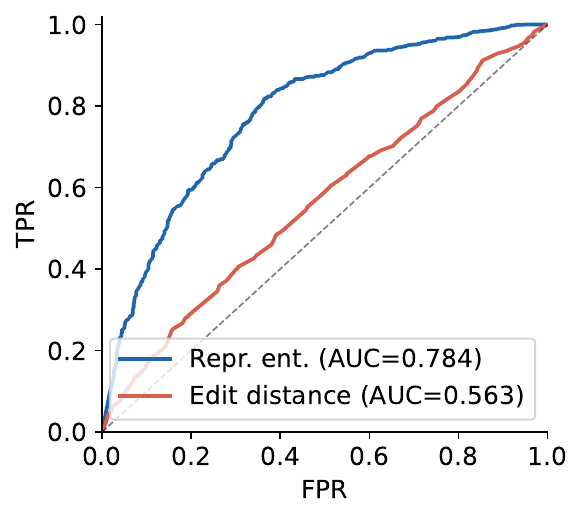}
    \caption{\OlmoTwo}
\end{subfigure}
\hfill
\begin{subfigure}{0.19\linewidth}
    \includegraphics[width=\linewidth]{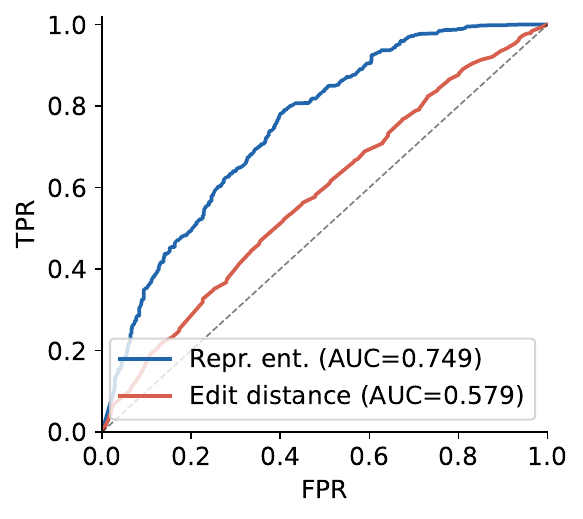}
    \caption{\TuluThree}
\end{subfigure}
\hfill
\begin{subfigure}{0.19\linewidth}
    \includegraphics[width=\linewidth]{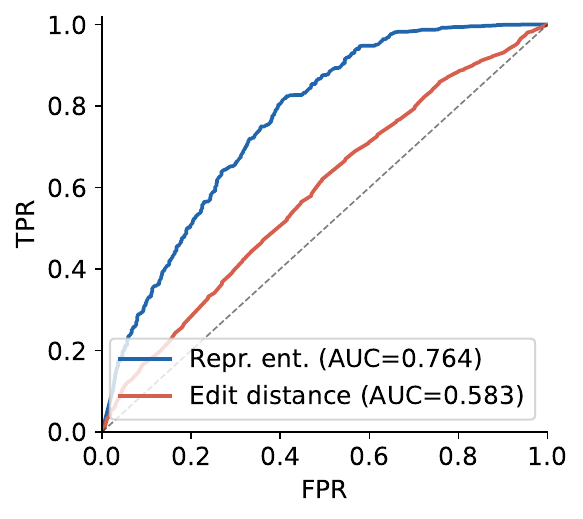}
    \caption{\TuluThreeOne}
\end{subfigure}

\caption{ROC curves for Known-1000 across all five models.
Representational entanglement (blue) vs.\ edit distance (red). Dashed line indicates
random baseline.}
\label{fig:app_roc_known_1000}
\end{figure*}

\begin{figure*}
\centering
\begin{subfigure}{0.19\linewidth}
    \includegraphics[width=\linewidth]{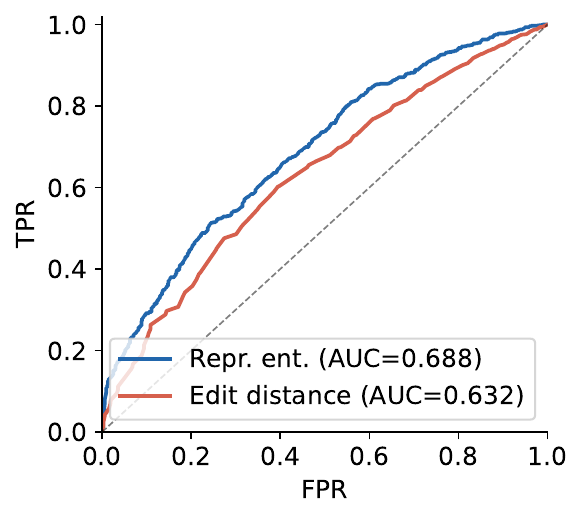}
    \caption{\LlamaThree}
\end{subfigure}
\hfill
\begin{subfigure}{0.19\linewidth}
    \includegraphics[width=\linewidth]{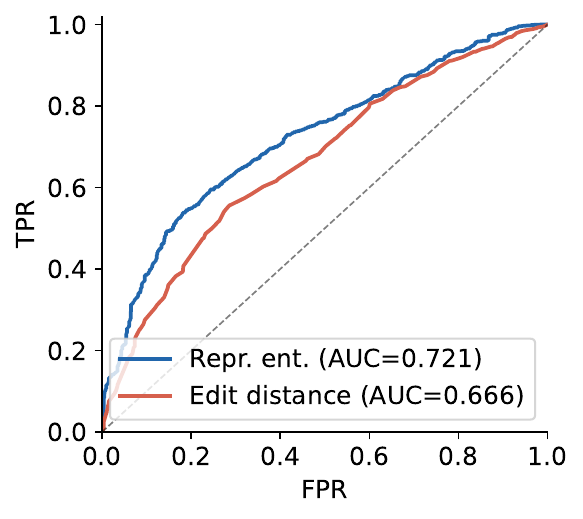}
    \caption{\LlamaThreeOne}
\end{subfigure}
\hfill
\begin{subfigure}{0.19\linewidth}
    \includegraphics[width=\linewidth]{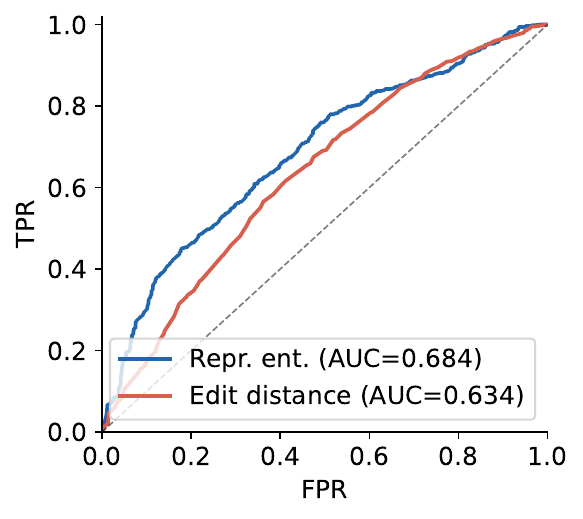}
    \caption{\OlmoTwo}
\end{subfigure}
\hfill
\begin{subfigure}{0.19\linewidth}
    \includegraphics[width=\linewidth]{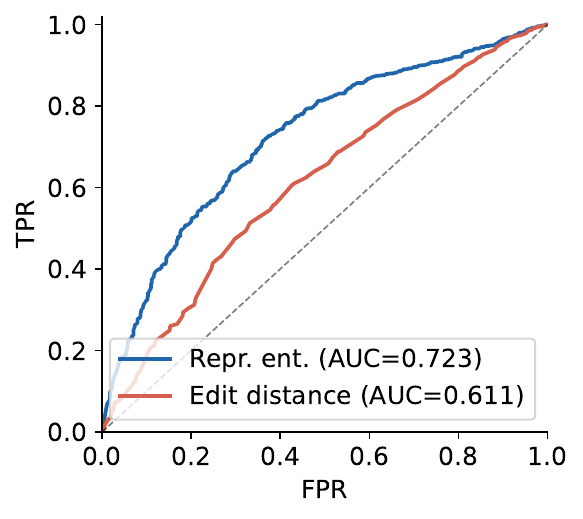}
    \caption{\TuluThree}
\end{subfigure}
\hfill
\begin{subfigure}{0.19\linewidth}
    \includegraphics[width=\linewidth]{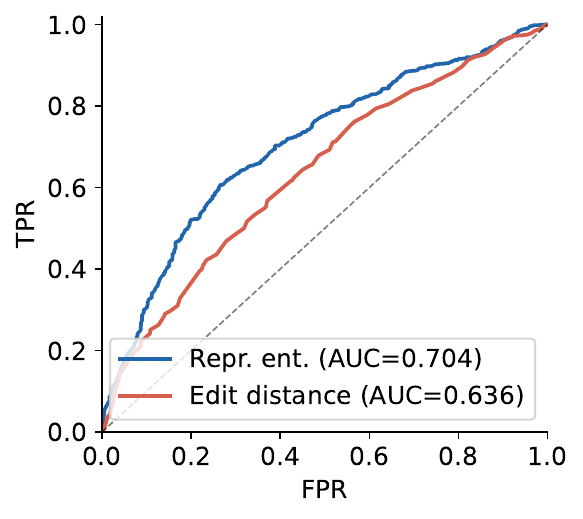}
    \caption{\TuluThreeOne}
\end{subfigure}
\caption{ROC curves for CounterFact across all five models.}
\label{fig:app_roc_counterfact}
\end{figure*}

\begin{figure*}
\centering
\begin{subfigure}{0.19\linewidth}
    \includegraphics[width=\linewidth]{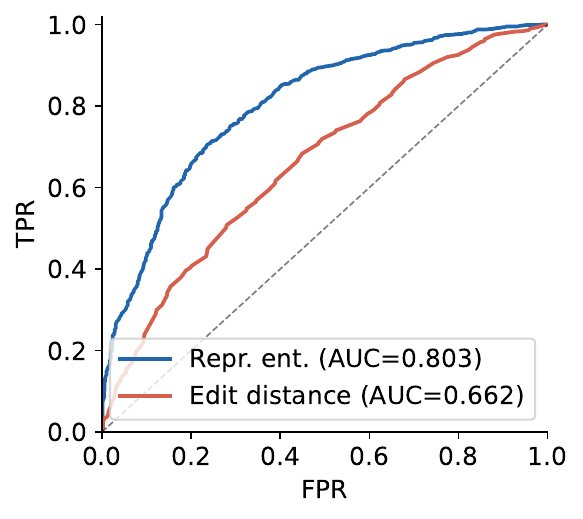}
    \caption{\LlamaThree}
\end{subfigure}
\hfill
\begin{subfigure}{0.19\linewidth}
    \includegraphics[width=\linewidth]{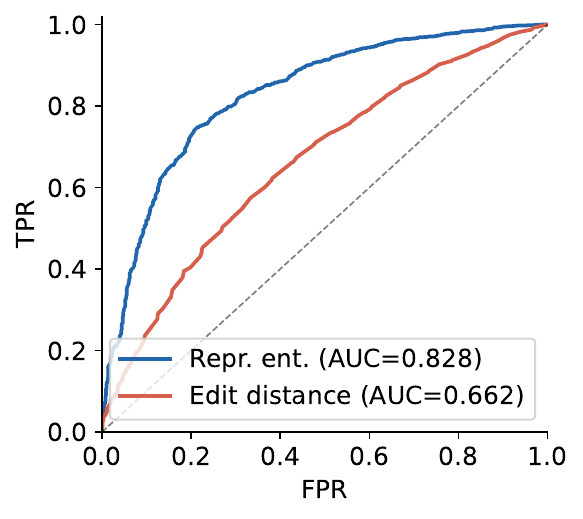}
    \caption{\LlamaThreeOne}
\end{subfigure}
\hfill
\begin{subfigure}{0.19\linewidth}
    \includegraphics[width=\linewidth]{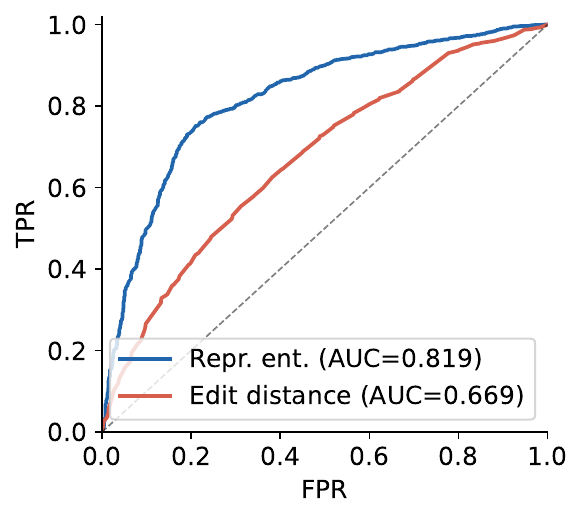}
    \caption{\OlmoTwo}
\end{subfigure}
\hfill
\begin{subfigure}{0.19\linewidth}
    \includegraphics[width=\linewidth]{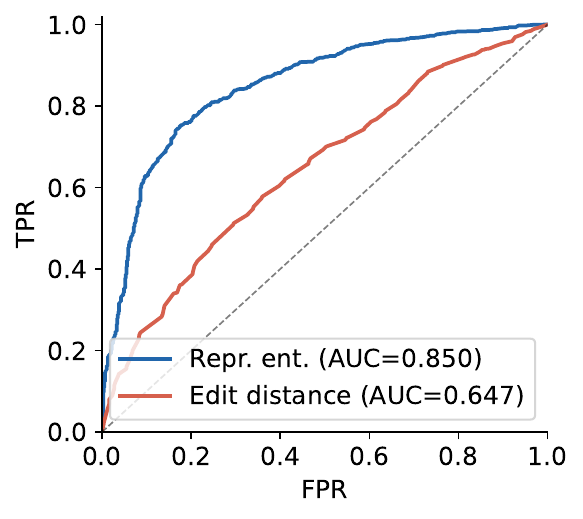}
    \caption{\TuluThree}
\end{subfigure}
\hfill
\begin{subfigure}{0.19\linewidth}
    \includegraphics[width=\linewidth]{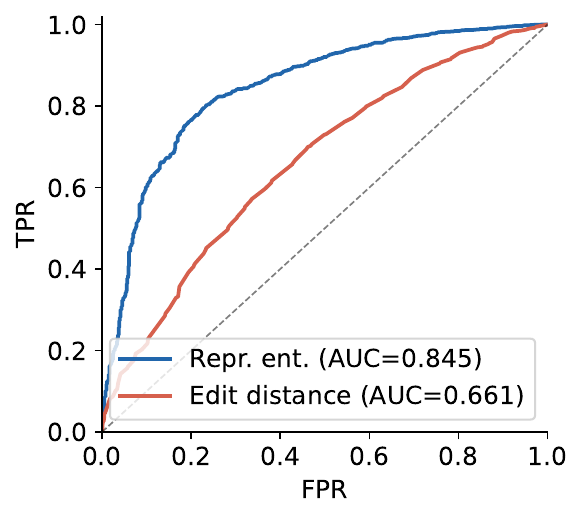}
    \caption{\TuluThreeOne}
\end{subfigure}

\caption{ROC curves for Real Authors across all five models.}
\label{fig:app_roc_authors}
\end{figure*}

\begin{figure*}
\centering
\begin{subfigure}{0.19\linewidth}
    \includegraphics[width=\linewidth]{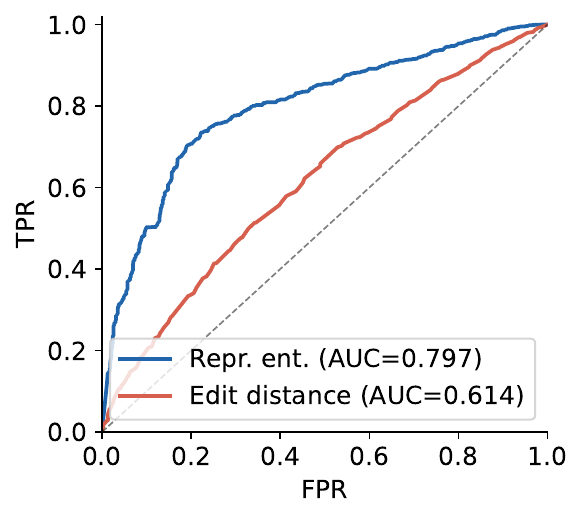}
    \caption{\LlamaThree}
\end{subfigure}
\hfill
\begin{subfigure}{0.19\linewidth}
    \includegraphics[width=\linewidth]{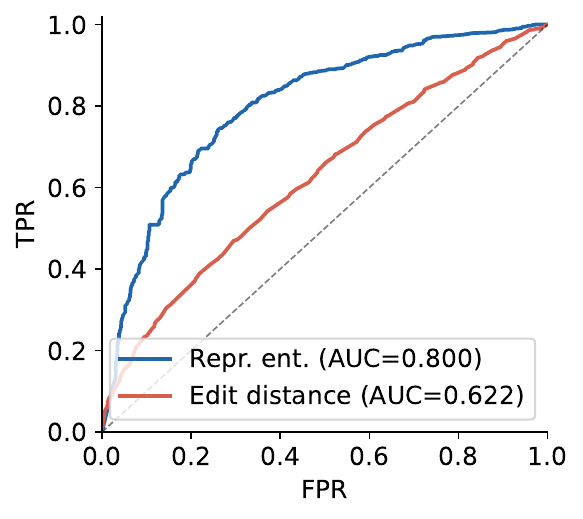}
    \caption{\LlamaThreeOne}
\end{subfigure}
\hfill
\begin{subfigure}{0.19\linewidth}
    \includegraphics[width=\linewidth]{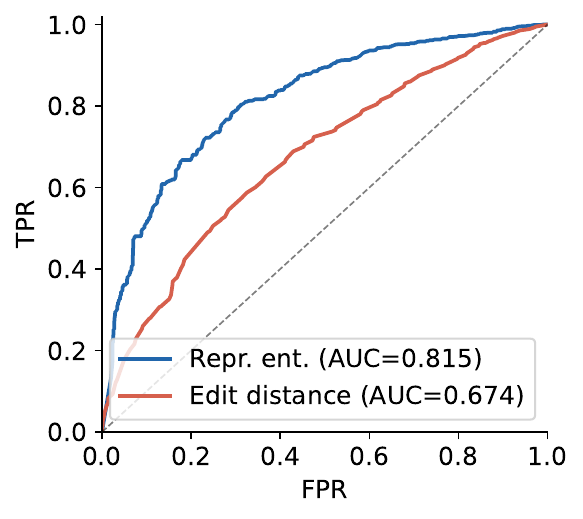}
    \caption{\OlmoTwo}
\end{subfigure}
\hfill
\begin{subfigure}{0.19\linewidth}
    \includegraphics[width=\linewidth]{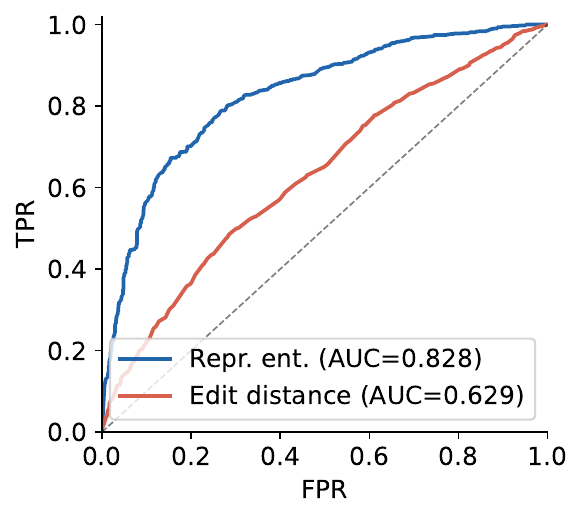}
    \caption{\TuluThree}
\end{subfigure}
\hfill
\begin{subfigure}{0.19\linewidth}
    \includegraphics[width=\linewidth]{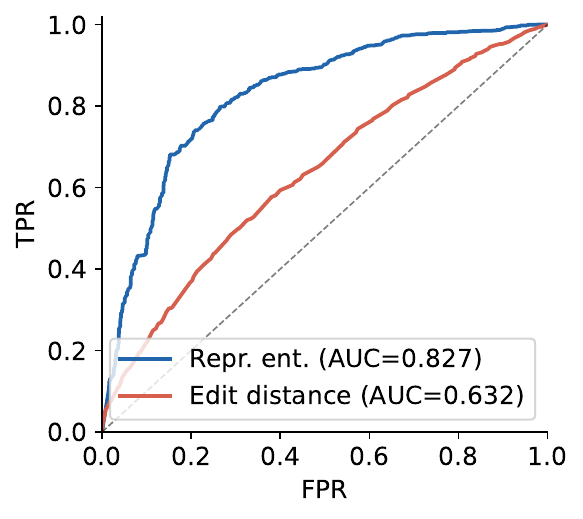}
    \caption{\TuluThreeOne}
\end{subfigure}

\caption{ROC curves for MQuAKE across all five models.}
\label{fig:app_roc_mquake}
\end{figure*}

\begin{figure*}
\centering
\begin{subfigure}{0.19\linewidth}
    \includegraphics[width=\linewidth]{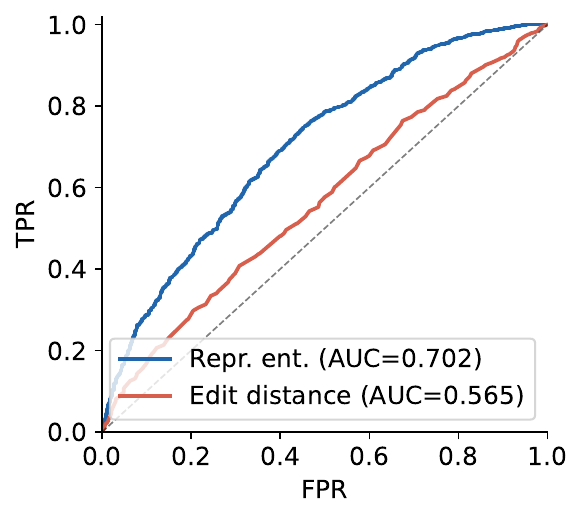}
    \caption{\LlamaThree}
\end{subfigure}
\hfill
\begin{subfigure}{0.19\linewidth}
    \includegraphics[width=\linewidth]{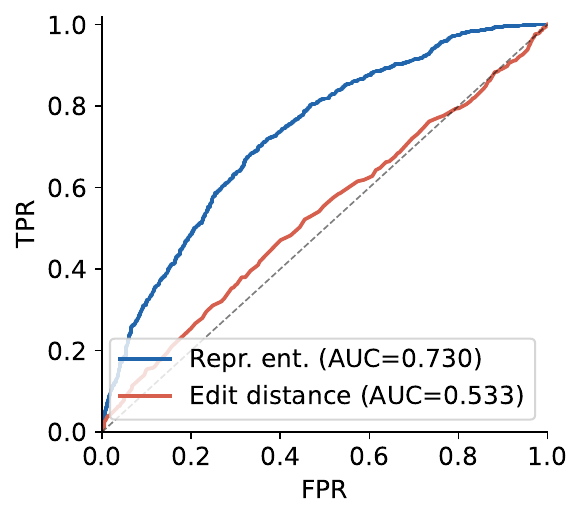}
    \caption{\LlamaThreeOne}
\end{subfigure}
\hfill
\begin{subfigure}{0.19\linewidth}
    \includegraphics[width=\linewidth]{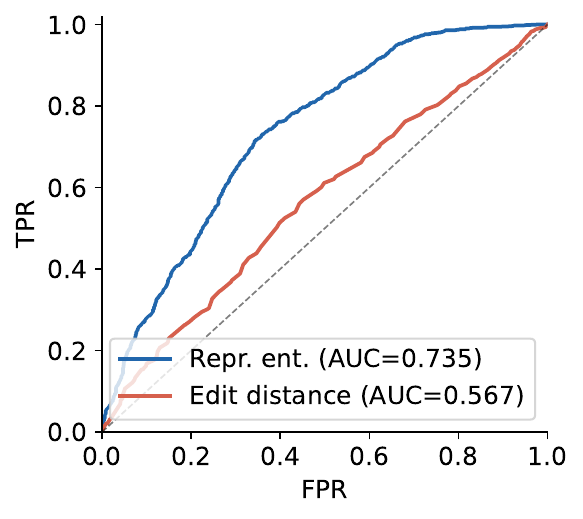}
    \caption{\OlmoTwo}
\end{subfigure}
\hfill
\begin{subfigure}{0.19\linewidth}
    \includegraphics[width=\linewidth]{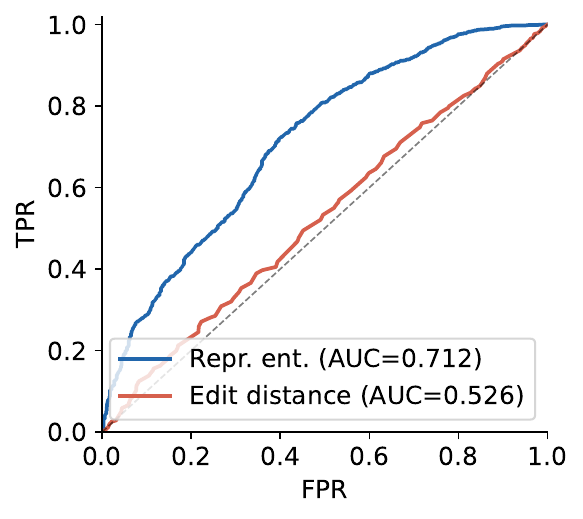}
    \caption{\TuluThree}
\end{subfigure}
\hfill
\begin{subfigure}{0.19\linewidth}
    \includegraphics[width=\linewidth]{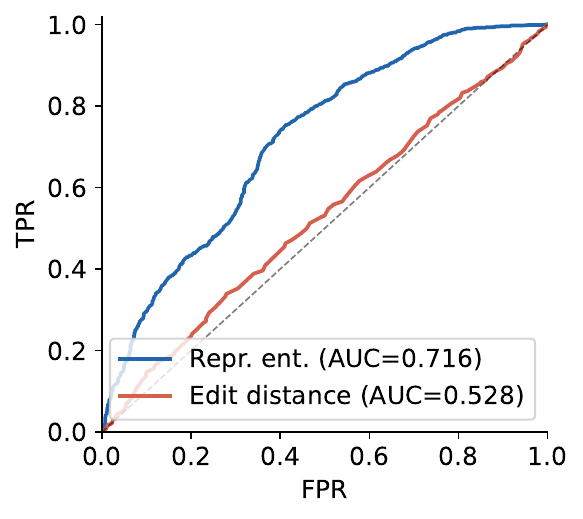}
    \caption{\TuluThreeOne}
\end{subfigure}

\caption{ROC curves for RippleEdits across all five models.}
\label{fig:app_roc_rippleedits}
\end{figure*}

\begin{figure*}
\centering
\begin{subfigure}{0.19\linewidth}
    \includegraphics[width=\linewidth]{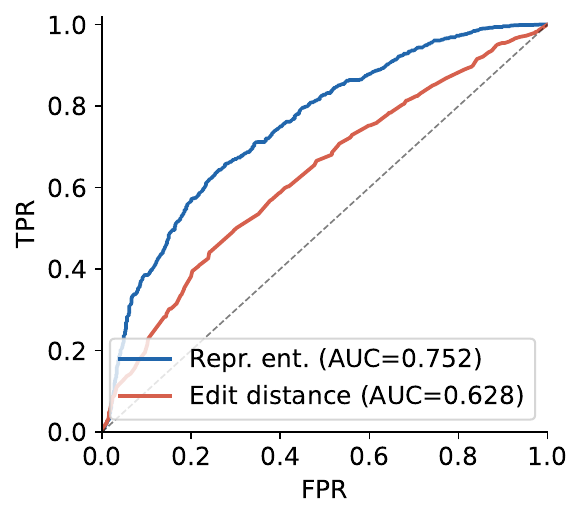}
    \caption{\LlamaThree}
\end{subfigure}
\hfill
\begin{subfigure}{0.19\linewidth}
    \includegraphics[width=\linewidth]{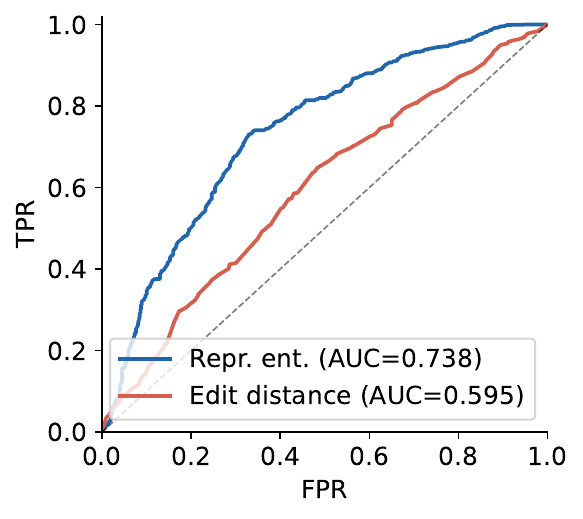}
    \caption{\LlamaThreeOne}
\end{subfigure}
\hfill
\begin{subfigure}{0.19\linewidth}
    \includegraphics[width=\linewidth]{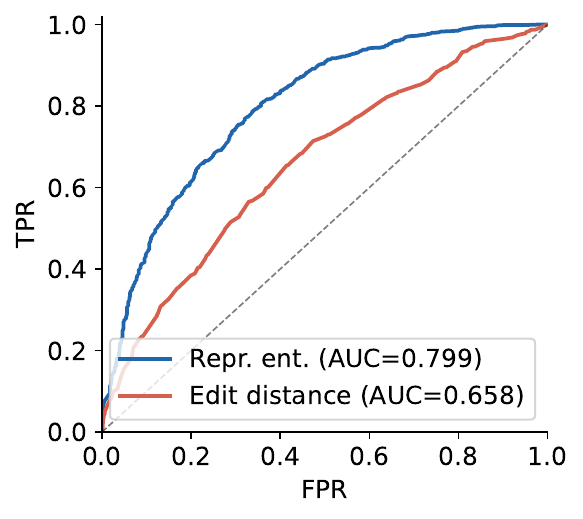}
    \caption{\OlmoTwo}
\end{subfigure}
\hfill
\begin{subfigure}{0.19\linewidth}
    \includegraphics[width=\linewidth]{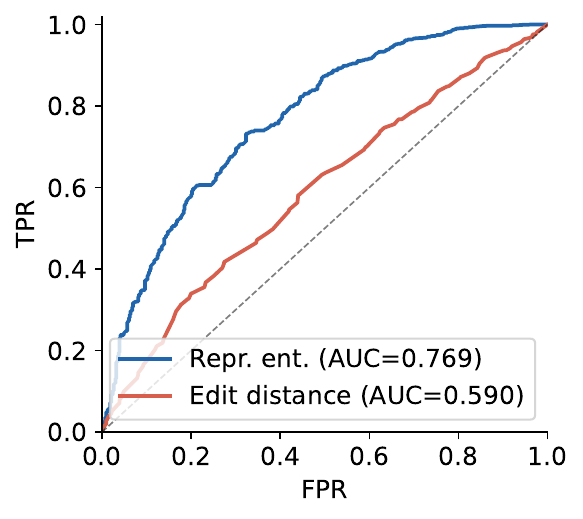}
    \caption{\TuluThree}
\end{subfigure}
\hfill
\begin{subfigure}{0.19\linewidth}
    \includegraphics[width=\linewidth]{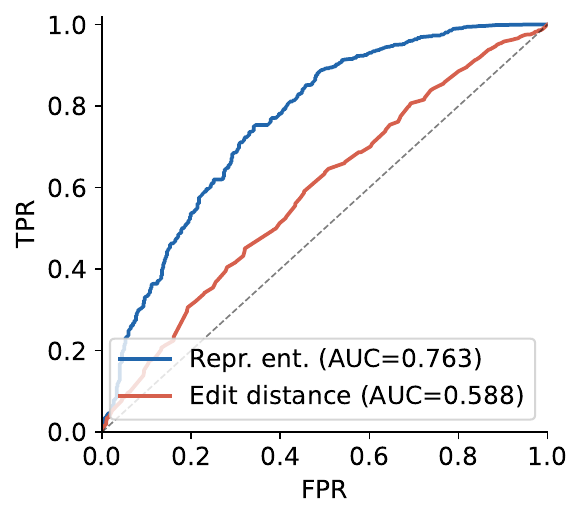}
    \caption{\TuluThreeOne}
\end{subfigure}

\caption{ROC curves for World Facts across all five models.}
\label{fig:app_roc_world_facts}
\end{figure*}

Figures~\ref{fig:app_clare_known_1000}-\ref{fig:app_editdist_world_facts} present distributional visualizations of representational entanglement and edit distance as predictors of tokenization convergence. Figures~\ref{fig:app_roc_known_1000}-\ref{fig:app_roc_world_facts} shows ROC curves for all models across all six datasets, illustrating the discrimination advantage of representational entanglement over edit distance at all operating thresholds. Across all combinations, representational entanglement consistently exhibits a rightward shift for convergent tokenizations relative to non-convergent ones, with clearly separated interquartile ranges in most settings. Edit distance distributions overlap substantially between convergent and non-convergent groups across all datasets, with near-identical medians and largely overlapping interquartile ranges. As visible in the violin plots, the edit distance distributions for convergent and non-convergent tokenizations share similar shapes and central tendencies, and representational entanglement distributions show a pronounced mass shift toward higher scores for convergent tokenizations.

\section*{Appendix H\\Calibration Analysis}
\label{sec:calibration}
Beyond discrimination (AUC) and effect size ($|r_b|$) reported in Section~\ref{sec:rq1}, we evaluate the calibration of representational entanglement and edit distance as predictors of tokenization convergence. A well-calibrated signal produces confidence scores whose magnitude reliably reflects the actual probability of convergence, not just their relative ordering. We measure calibration using Expected Calibration Error (ECE), computed by partitioning scores into ten equal-width bins and measuring the weighted average gap between mean predicted score and observed convergence rate per bin. Lower ECE indicates better calibration.

Table~\ref{tab:ece} reports ECE for representational entanglement and edit distance across all five models and six datasets. Representational entanglement is substantially better calibrated (lower valyes) than edit distance across all settings. The gap is most pronounced on TOFU-World Facts, where edit distance ECE exceeds $0.4$ across all models while representational entanglement remains below $0.25$. This indicates that edit distance scores are not only less discriminative (as shown by AUC and $|r_b|$ in Table~\ref{tab:predictability}) but also poorly calibrated. Representational entanglement, by contrast, produces scores whose magnitude is meaningfully aligned with convergence probability, justifying its use as a routing signal in \Toketive's adaptive sampler rather than merely as a ranking criterion.

Table~\ref{tab:mce} reports Maximum Calibration Error (MCE), which measures the worst-case calibration gap across bins rather than the average. MCE is more sensitive to localized miscalibration and complements ECE by identifying settings where a signal fails badly in a specific score range even if its average calibration is acceptable. Representational entanglement achieves lower MCE than edit distance in the majority of model-dataset combinations. \OlmoTwo is a partial exception: edit distance achieves lower MCE on Real Authors ($0.330$ vs.\ $0.436$) and CounterFact ($0.250$ vs.\ $0.263$), suggesting that the worst-case calibration of entanglement scores is slightly degraded for this model on these datasets. However, representational entanglement retains lower MCE on the remaining four datasets for \OlmoTwo, and its ECE advantage holds uniformly(Table~\ref{tab:ece}). Taken together, ECE and MCE show that representational entanglement produces more reliable confidence estimates than edit distance across the full score range and across all models, further supporting its use as the routing signal in \Toketive's adaptive sampler.

\begin{table*}
\centering
\caption{Full breakdown of noncanonical tokenization responses 
across models, techniques, and datasets. New (\%) reports the 
fraction producing the post-edit answer $o^*$; Old (\%) 
reports the fraction recovering the pre-update answer $o$ 
(bypass rate); Other (\%) reports the fraction producing 
neither.}
\label{tab:app_rq2_full}
\setlength{\tabcolsep}{3pt}
\resizebox{0.99\textwidth}{!}{
\begin{tabular}{ll 
    rrr rrr rrr rrr rrr rrr}
\toprule
& & \multicolumn{3}{c}{Known-1000} 
  & \multicolumn{3}{c}{CounterFact} 
  & \multicolumn{3}{c}{TOFU-Real Authors} 
  & \multicolumn{3}{c}{MQuAKE} 
  & \multicolumn{3}{c}{RippleEdits} 
  & \multicolumn{3}{c}{TOFU-World Facts} \\
\cmidrule(lr){3-5}\cmidrule(lr){6-8}\cmidrule(lr){9-11}
\cmidrule(lr){12-14}\cmidrule(lr){15-17}\cmidrule(lr){18-20}
Model & Technique 
    & New & Old & Other 
    & New & Old & Other 
    & New & Old & Other 
    & New & Old & Other 
    & New & Old & Other 
    & New & Old & Other \\
\midrule
\multirow{6}{*}{\rotatebox[origin=c]{90}{\LlamaThree}}
    & MEMIT     & 21.5 & 36.5 & 42.0 & 16.8 & 50.3 & 32.9 
               & 29.7 & 43.8 & 26.5 & 15.6 & 46.9 & 37.5 
               & 21.0 & 37.5 & 41.5 & 48.0 & 27.0 & 25.0 \\
    & RECT      & 19.3 & 39.1 & 41.5 & 15.2 & 51.6 & 33.2 
               & 26.5 & 46.4 & 27.1 & 11.7 & 49.7 & 38.5 
               & 16.9 & 42.3 & 40.8 & 44.0 & 30.1 & 25.9 \\
    & PRUNE     & 21.0 & 35.8 & 43.2 & 17.8 & 49.3 & 32.9 
               & 31.1 & 42.6 & 26.3 & 12.6 & 49.5 & 37.9 
               & 23.0 & 37.7 & 39.3 & 44.9 & 28.7 & 26.4 \\
    & AlphaEdit & 24.6 & 33.8 & 41.6 & 20.7 & 47.3 & 32.0 
               & 30.0 & 42.6 & 27.4 & 17.1 & 43.1 & 39.7 
               & 24.9 & 35.6 & 39.5 & 54.3 & 23.2 & 22.5 \\
    & CoME      & 25.5 & 32.9 & 41.6 & 19.8 & 48.3 & 32.0 
               & 29.0 & 44.1 & 26.9 & 18.5 & 42.7 & 38.8 
               & 29.8 & 33.1 & 37.1 & 56.1 & 21.1 & 22.8 \\
    & LTU       &  4.5 & 32.1 & 63.4 &  6.4 & 39.1 & 54.5 
               &  1.0 & 57.1 & 41.9 &  1.3 & 47.5 & 51.2 
               &  1.6 & 26.9 & 71.5 & 12.2 & 51.8 & 36.0 \\
\cmidrule(lr){1-20}
\multirow{6}{*}{\rotatebox[origin=c]{90}{\LlamaThreeOne}}
    & MEMIT     & 21.9 & 42.0 & 36.1 & 19.5 & 49.1 & 31.4 
               & 30.4 & 42.1 & 27.5 & 15.5 & 39.7 & 44.8 
               & 25.0 & 32.3 & 42.7 & 49.3 & 30.9 & 19.7 \\
    & RECT      & 19.6 & 43.9 & 36.5 & 16.0 & 51.5 & 32.5 
               & 27.0 & 45.2 & 27.7 & 12.7 & 42.0 & 45.2 
               & 21.3 & 38.3 & 40.5 & 44.5 & 34.5 & 20.9 \\
    & PRUNE     & 23.1 & 39.9 & 37.1 & 21.2 & 48.3 & 30.5 
               & 29.3 & 43.4 & 27.3 & 15.4 & 39.5 & 45.1 
               & 29.1 & 30.2 & 40.7 & 51.3 & 29.7 & 18.9 \\
    & AlphaEdit & 28.0 & 36.9 & 35.1 & 26.1 & 45.8 & 28.1 
               & 31.6 & 40.5 & 27.9 & 21.4 & 31.2 & 47.4 
               & 31.0 & 30.2 & 38.8 & 59.3 & 22.1 & 18.7 \\
    & CoME      & 27.0 & 38.3 & 34.7 & 27.0 & 44.3 & 28.7 
               & 34.8 & 37.8 & 27.4 & 19.7 & 35.2 & 45.0 
               & 34.3 & 25.3 & 40.4 & 57.3 & 23.9 & 18.8 \\
    & LTU       &  5.8 & 42.8 & 51.4 &  4.9 & 46.0 & 49.1 
               &  1.2 & 56.4 & 42.4 &  0.3 & 42.2 & 57.5 
               &  1.9 & 27.8 & 70.3 & 13.7 & 48.6 & 37.6 \\
\cmidrule(lr){1-20}
\multirow{6}{*}{\rotatebox[origin=c]{90}{\OlmoTwo}}
    & MEMIT     & 15.1 & 36.1 & 48.8 & 18.7 & 29.0 & 52.3 
               & 11.4 & 38.5 & 50.1 & 17.3 & 24.6 & 58.2 
               & 11.5 & 35.8 & 52.7 & 28.1 & 36.6 & 35.3 \\
    & RECT      & 13.1 & 38.0 & 49.0 & 18.2 & 30.0 & 51.8 
               & 11.3 & 38.6 & 50.1 & 17.5 & 24.2 & 58.3 
               & 11.6 & 36.1 & 52.3 & 29.2 & 35.7 & 35.2 \\
    & PRUNE     & 13.9 & 36.0 & 50.1 & 20.9 & 27.8 & 51.3 
               & 11.6 & 38.0 & 50.3 & 17.9 & 24.0 & 58.1 
               & 12.2 & 35.4 & 52.4 & 31.6 & 32.9 & 35.4 \\
    & AlphaEdit & 18.0 & 29.7 & 52.3 & 22.2 & 26.7 & 51.1 
               & 12.7 & 37.2 & 50.1 & 21.6 & 19.7 & 58.7 
               & 15.1 & 31.9 & 53.1 & 37.7 & 25.6 & 36.7 \\
    & CoME      & 15.8 & 33.8 & 50.4 & 20.9 & 27.7 & 51.4 
               & 12.5 & 37.1 & 50.4 & 19.6 & 21.7 & 58.7 
               & 12.3 & 35.3 & 52.4 & 31.6 & 32.3 & 36.1 \\
    & LTU       &  6.8 & 38.7 & 54.5 &  5.6 & 36.5 & 57.9 
               &  6.1 & 37.5 & 56.4 &  6.9 & 31.9 & 61.2 
               &  5.2 & 39.6 & 55.2 & 13.1 & 36.9 & 30.0 \\
\cmidrule(lr){1-20}
\multirow{6}{*}{\rotatebox[origin=c]{90}{\TuluThreeOne}}
    & MEMIT     & 22.4 & 43.5 & 34.1 & 19.0 & 49.7 & 31.2 
               & 27.4 & 49.0 & 23.6 & 20.1 & 42.8 & 37.1 
               & 25.9 & 38.6 & 35.4 & 55.6 & 30.5 & 13.9 \\
    & RECT      & 19.7 & 46.3 & 33.9 & 17.6 & 50.5 & 31.9 
               & 24.9 & 51.1 & 24.0 & 18.8 & 44.2 & 37.0 
               & 20.1 & 43.7 & 36.2 & 51.6 & 35.0 & 13.4 \\
    & PRUNE     & 24.0 & 43.2 & 32.8 & 21.8 & 47.2 & 31.0 
               & 28.4 & 47.8 & 23.8 & 22.7 & 40.6 & 36.7 
               & 27.1 & 37.6 & 35.2 & 53.5 & 32.3 & 14.2 \\
    & AlphaEdit & 31.7 & 37.0 & 31.3 & 27.5 & 43.0 & 29.5 
               & 31.1 & 45.3 & 23.6 & 25.9 & 38.4 & 35.7 
               & 31.8 & 32.2 & 36.0 & 67.6 & 20.8 & 11.6 \\
    & CoME      & 26.9 & 40.1 & 33.0 & 24.4 & 45.5 & 30.1 
               & 31.1 & 45.5 & 23.4 & 25.1 & 39.7 & 35.2 
               & 33.2 & 30.7 & 36.1 & 60.7 & 26.6 & 12.7 \\
    & LTU       &  2.4 & 39.6 & 58.0 &  6.3 & 44.0 & 49.7 
               &  2.7 & 58.9 & 38.4 &  5.8 & 45.1 & 49.1 
               &  5.2 & 39.0 & 55.8 & 13.8 & 54.0 & 32.2 \\
\cmidrule(lr){1-20}
\multirow{6}{*}{\rotatebox[origin=c]{90}{\TuluThree}}
    & MEMIT     & 21.9 & 39.8 & 38.2 & 18.3 & 45.8 & 35.9 
               & 30.2 & 37.1 & 32.7 & 14.1 & 42.2 & 43.7 
               & 19.0 & 45.5 & 35.4 & 55.7 & 32.3 & 12.0 \\
    & RECT      & 19.4 & 42.1 & 38.4 & 16.2 & 47.3 & 36.5 
               & 27.5 & 39.8 & 32.7 & 13.8 & 43.1 & 43.1 
               & 16.8 & 47.4 & 35.8 & 51.3 & 37.4 & 11.3 \\
    & PRUNE     & 21.7 & 40.3 & 38.0 & 18.6 & 45.5 & 35.9 
               & 30.4 & 36.8 & 32.8 & 15.9 & 41.4 & 42.7 
               & 22.4 & 44.8 & 32.7 & 55.1 & 32.4 & 12.5 \\
    & AlphaEdit & 28.4 & 33.8 & 37.8 & 27.0 & 39.2 & 33.8 
               & 31.7 & 35.7 & 32.6 & 25.6 & 33.0 & 41.4 
               & 27.5 & 35.7 & 36.8 & 65.5 & 23.5 & 11.0 \\
    & CoME      & 24.1 & 38.0 & 37.9 & 23.8 & 42.1 & 34.1 
               & 33.3 & 34.6 & 32.2 & 22.0 & 37.2 & 40.8 
               & 24.0 & 39.8 & 36.1 & 61.2 & 28.0 & 10.8 \\
    & LTU       &  2.0 & 33.8 & 64.2 &  5.4 & 40.5 & 54.1 
               &  2.3 & 48.7 & 49.0 &  0.1 & 46.0 & 54.0 
               &  4.2 & 40.1 & 55.7 & 15.9 & 53.9 & 30.2 \\
\bottomrule
\end{tabular}
}
\end{table*}

\section*{Appendix I\\Qualitative Examples of Tokenization-Aware Recovery}
\label{app:qualitative_examples}

Table~\ref{tab:qualitative_examples} presents three representative outcomes of
\Toketive on \LlamaThree edited with AlphaEdit. All responses are generated
after the editing procedure. In the first case, a noncanonical tokenization
recovers the original pre-edit answer, \textit{George Orwell}, despite the
canonical tokenization producing the edited answer, \textit{George R.R.~Martin}.
This represents a successful bypass of the edit. In the second case, the
noncanonical tokenization produces an incorrect response, \textit{Michael
Moorcock}, rather than recovering either the pre-edit or edited answer. This
illustrates a failed recovery despite successfully deviating from the edited
behavior. Finally, the canonical tokenization produces the intended edited
answer, \textit{George R.R.~Martin}, demonstrating that the edit remains
effective under the representation on which it was originally applied.
Together, these examples illustrate that tokenization-aware sampling can
separate tokenizations that preserve the original factual association from
those that either lose the association or remain affected by the edit.

\section*{Appendix J\\Additional Analysis for Q2: Tokenization Invariance of Editing and Unlearning}
\label{app:rq2}

Table~\ref{tab:app_rq2_full} reports the full breakdown of 
noncanonical tokenization responses across all five models, 
six datasets, and six techniques. For each combination, we 
report the fraction of noncanonical tokenizations that produce 
the new answer (New\%), the pre-update answer (Old\%), or 
neither (Other\%). The bypass rate reported in the main paper 
(Table~\ref{tab:bypass_rates}) corresponds directly to the 
Old\% column here.

Several patterns are consistent across all models and 
datasets. The Old\% column shows that a substantial 
fraction of noncanonical tokenizations recover the pre-update 
response in every setting, with no technique achieving 
near-zero bypass rates. The New\% column shows that the 
post-edit answer is produced by a minority of noncanonical 
tokenizations in most settings, indicating that the edit 
generalizes poorly beyond the canonical path. The Other\% 
column (tokenizations producing neither the old nor the 
new answer) is largest for LTU across most settings.


\end{document}